\DocumentMetadata{}
\documentclass[sigconf,dvipsnames,anonymous=false,authorversion]{acmart}

\AtBeginDocument{%
  \providecommand\BibTeX{{%
    \normalfont B\kern-0.5em{\scshape i\kern-0.25em b}\kern-0.8em\TeX}}}

\acmConference[SIGMOD '25]{The 2025 International Conference on Management of Data}{June 22--27, 2025}{Berlin, Germany}
\usepackage{listings}
\usepackage{color}
\usepackage{xcolor}
\usepackage[plain]{algorithm2e}
\usepackage[noend]{algorithmic}

\usepackage{colortbl}
\usepackage{subcaption}

\usepackage{lipsum}
\usepackage{verbatim}
\usepackage{colortbl}
\usepackage{multirow}
\usepackage{soul}

\usepackage{enumitem}

\newlist{questions}{enumerate}{2}
\setlist[questions,1]{label=\textbf{(Q\arabic*)},ref=\textbf{(Q\arabic*)}}
\setlist[questions,2]{label=(\alph*),ref=\thequestionsi(\alph*)}

\NewDocumentCommand{\codeword}{v}{%
    \texttt{\textcolor{violet}{#1}}%
}

\NewDocumentCommand{\colorcodeword}{vv}{%
    \textbf{\texttt{\textcolor{#1}{#2}}}%
}

\makeatletter
\def\NAT@def@citea{\def\@citea{\NAT@separator}}
\makeatother

\definecolor{lightgray}{gray}{0.95}

\begin{document}

\title{PDX: A Data Layout for Vector Similarity Search}


% \iffalse
\author{Leonardo Kuffo}
\affiliation{%
  \institution{CWI}
  \city{Amsterdam}
  \country{The Netherlands}}
%\email{lxkr@cwi.nl}

\author{Elena Krippner}
\affiliation{%
  \institution{CWI}
  \city{Amsterdam}
  \country{The Netherlands}}
%\email{ekrippner@cwi.nl}

\author{Peter Boncz}
\affiliation{%
  \institution{CWI}
  \city{Amsterdam}
  \country{The Netherlands}}
%\email{boncz@cwi.nl}
% \fi

%\renewcommand{\shortauthors}{Afroozeh and Kuffó and Boncz}

\begin{abstract}

We propose Partition Dimensions Across (PDX), a data layout for vectors (e.g., embeddings) that, similar to PAX~\cite{PAX}, stores multiple vectors in one block, using a vertical layout for the dimensions (Figure \ref{fig:pdx}). PDX accelerates exact and approximate similarity search thanks to its dimension-by-dimension search strategy that operates on multiple-vectors-at-a-time in tight loops. It beats SIMD-optimized distance kernels on standard horizontal vector storage (avg 40\% faster), only relying on scalar code that gets auto-vectorized. We combined the PDX layout with recent dimension-pruning algorithms ADSampling~\cite{adsampling} and BSA~\cite{bsa} that accelerate approximate vector search. 
We found that these algorithms on the horizontal vector layout can {\em lose} to SIMD-optimized linear scans, even if they are SIMD-optimized. However, when used on PDX, their benefit is restored to 2-7x. We find that search on PDX is especially fast if a limited number of dimensions has to be scanned fully, which is what the dimension-pruning approaches do. We finally introduce PDX-BOND, an even more flexible dimension-pruning strategy, with good performance on exact search and reasonable performance on approximate search. Unlike previous pruning algorithms, it can work on vector data "as-is" without preprocessing; making it attractive for vector databases with frequent updates.

% We propose Partition Dimensions Across (PDX), a novel data layout for vectors that stores together the values of each dimension within vectorgroups (Figure \ref{fig:pdx}). PDX improves search speed in both exact and approximate Vector Similarity Search thanks to our dimension-by-dimension search framework that operates on multiple-vectors-at-a-time only relying on scalar code that can be auto-vectorized by modern compilers. Furthermore, we combined the PDX layout with recently proposed dimensions-pruning algorithms (ADSampling~\cite{adsampling}, BSA~\cite{bsa}) to speed up a vector similarity search on an IVF index search by 18x on average (more than 30x in some datasets) without any compromise of recall or preprocessing overhead. PDX improves on these algorithms thanks to the efficient data access patterns, which are cache-friendly and can adapt per query and dataset. 

\end{abstract}

%\keywords{vector similarity search, approximate nearest neighbor search, data formats, vector data management, vectorized execution}

\maketitle

\section{Introduction}

K-Nearest Neighbour Search (KNNS), also referred to nowadays as Vector Similarity Search (VSS), has rapidly become a core component of a variety of applications: information/multimedia retrieval, data pipelines, code co-piloting, LLMs pipelines, etc. The KNNS problem consists of finding the K-vectors within a collection that are the most similar to a query vector based on a distance or similarity metric (e.g., Euclidean, Cosine, Manhattan).  KNNS is computationally intensive, as providing \textit{exact} answers requires a large number of computations. The latter makes KNNS inefficient for large-scale workloads, especially at the throughput needed by LLMs and information retrieval applications. 

\begin{figure}[t!]
\centering
\includegraphics[width=0.85\columnwidth]{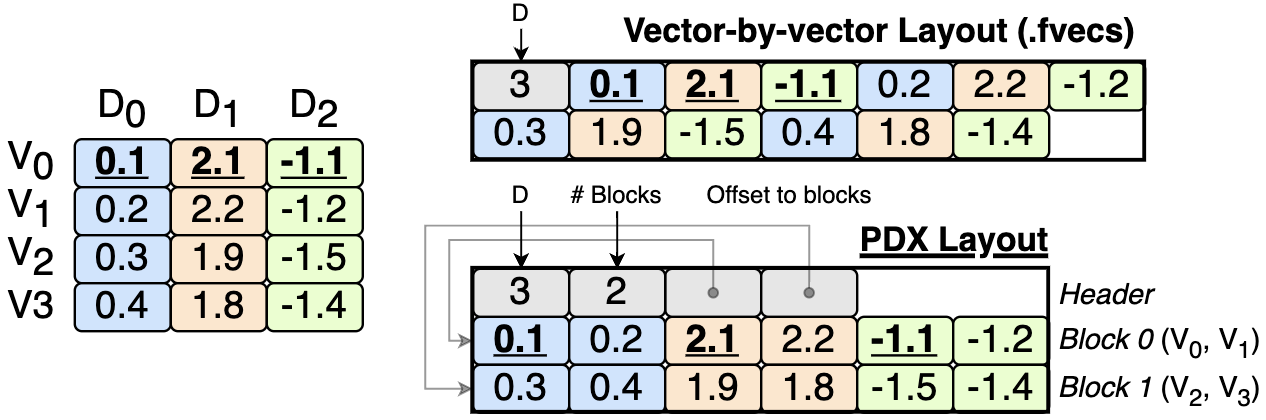}
\vspace*{-4mm}
\caption{PDX stores dimensions in a vertical layout, allowing efficient dimension-by-dimension distance calculation, more opportunities for SIMD execution, and better memory locality for search algorithms that prune dimensions.}
\label{fig:pdx}
\vspace*{-4mm}
\end{figure}

However, certain applications can tolerate \textit{approximate answers}, that is, to only obtain a subset of the actual neighbors of the query (Approximate Nearest Neighbor Search). Giving up exactness opened opportunities to develop approximate indexes based on bucketing ~\cite{pqivf, surveylsh}, trees~\cite{annoy, flann}, and graphs~\cite{hnsw, nsg, hcnng, surveygraph} that when used together with \textit{quantization} to reduce the size of the vectors~\cite{pqivf, lvq, rabitq, anisotropic, polysemous, lowquant}, achieve throughput close to $10^5$ queries per second on modestly sized datasets~\cite{annbench}. Therefore, it is no surprise that a flurry of improvements to approximate VSS have been developed in recent years, mainly focusing on improving existing index structures (better data access patterns~\cite{starling, turbolvq}, GPU optimizations~\cite{bang, ngt}, leveraging disk storage~\cite{diskann, starling, aisaq}, SIMD kernels~\cite{turbolvq, usearch}) and reducing the trade-off between losing information in the vectors (quantization) and achieving higher recalls~\cite{rabitq, lvq, turbolvq, lowquant}.

Both approximate and exact VSS share a common theme: the \textit{distance calculation} (referred to as Distance Comparison Operation (DCO) in~\cite{adsampling}). DCOs are the most time-consuming operation in a VSS~\cite{adsampling, candy, bsa}, followed by the access to the data itself (especially in a RAM-constrained environment)~\cite{sancaaccess}. Despite this, few efforts have been made to improve these. In other words, during an approximate or exact search, if one wants to determine if a vector will make it into the K-nearest neighbours of a query, all of its dimensions have to be accessed. ADSampling~\cite{adsampling} and BSA~\cite{bsa} improve on this by performing a distance approximation only by evaluating \textit{some} dimensions of a vector, an idea first explored many years back in BOND~\cite{bond} and FNN~\cite{fnn} for exact search. ADSampling randomly projects the vector collection and queries to make them suitable for a reliable distance approximation using only a few dimensions (as low as 2\% in some datasets), speeding up IVF~\cite{pqivf} and HNSW~\cite{hnsw} index search by x5.6 and x2.6, respectively, with little accuracy loss. BSA improved ADSampling speed by replacing the random projection with learned PCA projections, resulting in tighter approximations and, thus, earlier \textit{pruning} of vectors at the expense of more intensive data preprocessing. 

We believe that the core idea in ADSampling and BSA of pruning dimensions at search time is the next leap in VSS, as the DCO is performed in any VSS setting. However, the current de-facto layout to store vectors (the vector-by-vector/horizontal/N-ary layout in Figure \ref{fig:pdx}) prevents these algorithms from \textit{always} beating SIMD-optimized searches due to the latency to evaluate their pruning bounds. Furthermore, the horizontal layout implies that dimensions that are never visited are still loaded, wasting memory bandwidth~\cite{sancaaccess, bond}. A vertical layout was proposed two decades ago~\cite{bond} for column-at-a-time image search with dimensions pruning using partial distance calculations. However, methods like ADSampling and BSA need to compute full distances before starting pruning, making them incompatible with the idea of full column-at-a-time processing.

We introduce \textbf{Partition Dimensions Across (PDX)} (Figure \ref{fig:pdx}), a data layout for vectors that stores vectors dimension-at-a-time within {\small \tt blocks} (analogous to rowgroups in Parquet~\cite{parquet}). PDX allows for efficient per-dimension access, which is ideal for performing partial distance calculations such as the ones recently proposed in ADSampling and BSA. Furthermore, we introduce \textbf{PDXearch}, a search framework applicable in exact and approximate KNN that leverages the PDX layout by \textit{adaptively} scanning dimensions as required by the underlying algorithm and query. In PDXearch, a search happens dimension-by-dimension rather than vector-by-vector. The latter gets the best out of modern compilers as searches can be \textit{vectorized} by processing multiple-vectors-at-a-time~\cite{monetdb, superscalar} while at the same time improving data access patterns and cache utilization. PDXearch does not rely on SIMD instructions to be fast (its code auto-vectorizes on {\small \tt float32} vectors), making it portable to any ISA and SIMD register width. PDXearch speeds up ADSampling and BSA SIMDized versions by 4.6x and 2.3x without any loss in recall, achieving 5.3x faster searches than the FAISS ~\cite{faisspaper} {\small\tt IVF\_FLAT} index on average in high-dimensional datasets. 

Finally, we introduce \textbf{PDX-BOND}: A VSS algorithm in the same line of ADSampling and BSA that leverages PDX by first accessing the most relevant dimensions {\em relative to an incoming query}. PDX-BOND does not have \textit{any} recall trade-off (can do exact search) and does not require data transformations or parameter tuning to achieve comparable performance to ADSampling. PDX-BOND also outperforms USearch~\cite{usearch}, Milvus~\cite{milvus}, and FAISS (state-of-the-art systems) on exact search by 4.0x, 3.3x, and 2.5x on average.

\noindent Our main contributions are:
\vspace*{-1mm}
\begin{itemize}
    \item The design of \textbf{PDX}, a new data layout for vectors alongside \textbf{PDXearch}: a framework to perform pruned VSS dimension-by-dimension (Section \ref{sec:pdx}). 
    \item The insight that the ADSampling~\cite{adsampling} and BSA~\cite{bsa} dimension pruning algorithms, which were originally evaluated with scalar code, can be actually slower than SIMD-optimized searches. Thanks to PDX, they regain clear superiority.
    \item An experimental evaluation of our search framework on 10 vector datasets and 4 CPU architectures, demonstrating the versatility and effectiveness of PDX to achieve significant speedups both in exact and approximate settings (Section \ref{sec:eval}). 
    \item The design and evaluation of \textbf{PDX-BOND}: A VSS algorithm that leverages the PDX layout to visit first the most relevant dimensions relative to the incoming query. 
    \item An open-source implementation of our algorithms in C++ with Python bindings for their ease of use (\url{https://github.com/cwida/PDX}). 
\end{itemize}

\section{Preliminaries}\label{sec:preliminaries}

\subsection{K-Nearest Neighbour Search (KNNS)}\label{sec:theknnproblem}
Given a collection $V$ of $n$ multi-dimensional objects $\{v_0, v_1, \cdots, v_{n-1}\}$, defined on a $D$-dimensional space, and a $D$-dimensional query $q$, the KNNS problem tries to find a subset $R \subset V$, containing the $k$ most \textit{similar} objects to $q$. The notion of similarity between two objects $(v, q)$ is measured using a function $\delta(v, q)$. Usually, $\delta$ is a distance or similarity function defined in an Euclidean space (e.g., Euclidean Distance L2, Manhattan Distance L1, Hamming Distance, Cosine Similarity, Inner Product). The KNNS problem consists of finding the objects that minimize $\delta$. The Squared Euclidean Distance (L2) is one of the most commonly used distance metrics for KNNS, and it is defined as $\delta(v, q) = \sum_{i=0}^{D} (v_i - q_i)^2$.

To obtain $R$, $\delta$ needs to be computed for every $v\in V$, leading to a large number of operations. In modern CPUs, a KNNS within millions of vectors and hundreds of dimensions can be answered in a few hundred milliseconds~\cite{annbench, candy}. However, modern applications such as RAG pipelines often need sub-millisecond performance to cope with the increasing throughput of requests, rendering KNNS unfeasible on large-scale data at high throughput.  However, in most embedding-based applications, approximate answers are often as good as mathematically exact ones. This allowed the KNNS problem to \text{scale} by returning only \textit{approximate} results, renaming the problem to \textbf{Approximate k-Nearest Neighbor Search (ANNS)}. ANNS aims to return a result set $\hat{R}$, in which the quality of the elements of $\hat{R}$, with respect to a query, is measured using the \textit{recall@k} metric. Recall@k measures the percentage of intersection between the vectors in $R$ and $\hat{R}$ when answering the same query ($recall@k = \frac{R \bigcap \hat{R}}{k}$). There is no consensus on a "good enough" recall, as this depends on the application that uses ANNS. 

Trading off accuracy for speed has resulted in two ideas that, when used together, can achieve throughputs of up to $10^5$ queries per second (QPS) on modestly sized datasets~\cite{annbench, candy}: Approximate indexes and quantization (reducing the size of the vectors). 

\vspace*{3mm}\noindent{\bf Approximate indexes} aim to build data structures that guide the search to the most suitable place of the D-dimensional space in which the query \textit{may} find its nearest neighbours. There exist four types of indexes: graph-based~\cite{hnsw, nsg, hcnng, surveygraph}, tree-based~\cite{annoy, flann}, bucket-based~\cite{pqivf, surveylsh} and hybrids~\cite{spann, ngt}. Their common goal is only to evaluate the distance/similarity function between $q$ and a smaller set of vectors $V' \subset V$, such that $V'$ contains all or most of the elements in $R$. The approximate indexes that have seen the most adoption are HNSW~\cite{hnsw} (Hierarchical Navigable Small Worlds) and IVF~\cite{pqivf} (Inverted Files).

Graph indexes like HNSW have seen great success in achieving desirable recall in most datasets~\cite{annbench, candy}. Graph indexes organize objects into a graph where nodes represent the vectors, and edges reflect their similarity. The property of navigability and "small world"~\cite{smallworld} is desired on the graph so that a greedy search can reach the answers to a query in logarithmic complexity~\cite{surveygraph}. On the other hand, IVF is a bucket-based index that applies a non-optimized Lloyd algorithm (k-means) to the vector collection to group them into lists/buckets. At search time, the distance metric is first evaluated with the centroids of each bucket, and the vectors inside the nearest centroids buckets are chosen for evaluation. A higher number of buckets can be probed to trade off speed for more recall~\cite{faisspaper, milvus}. IVF has shown to work modestly well in most datasets~\cite{annbench, candy} while being able to scale better than graph indexes, which usually are non-feasible to compute in commodity hardware on huge datasets (in the order of a couple of gigabytes~\cite{annbench}) due to their memory requirements and long construction times~\cite{faisspaper}. A commonly used \textit{hybrid index} consists of building an HNSW index on the centroids generated by IVF to find the most promising buckets quickly~\cite{spann}. The latter is usually feasible in commodity hardware as the amount of centroids is set in the order of $\sqrt{n}$~\cite{faisspaper, milvus}. 

\begin{figure}[t!]
\centering
\includegraphics[width=0.85\columnwidth]{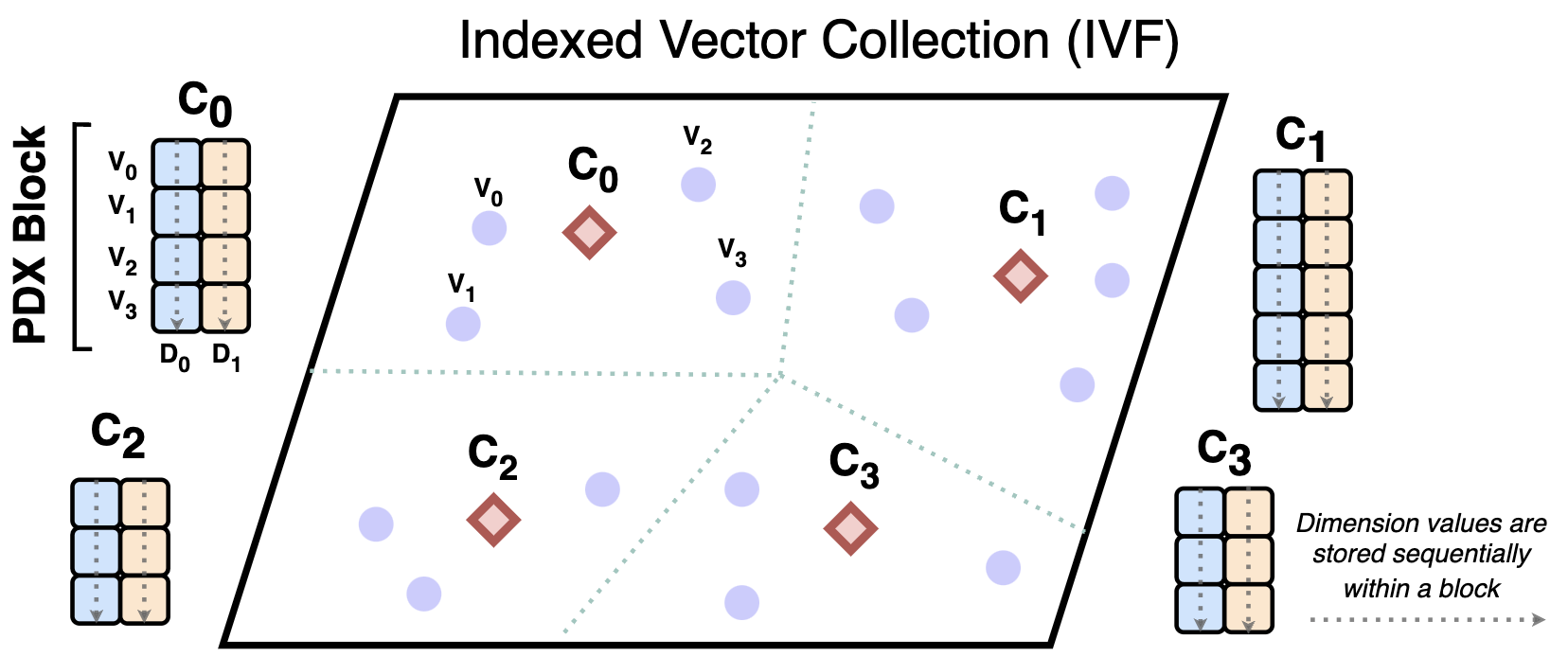}
\vspace*{-4mm}
\caption{Example of an IVF index on a collection of vectors. The IVF buckets naturally map to the concept of blocks of vectors in the PDX layout.}
\vspace*{-3mm}
\label{fig:ivf-pdx}
\end{figure}

In this study, we focus our experiments on exact search and bucketing indexes (IVF), as the PDX layout is a perfect fit for them. In fact, the notion of {\small \tt blocks} is similar to the bucketing characteristic of IVF (Figure \ref{fig:ivf-pdx}). In Section \ref{sec:discussion}, we discuss how the PDX layout can be used in the near future on graph indexes like HNSW. 

\begin{table}[]
\centering
\caption{Vector Datasets}\vspace*{-4mm}
\label{tab:datasets}
\resizebox{0.98\columnwidth}{!}{%
\begin{tabular}{|l|l|c|c|c|c|}
\hline
\multicolumn{1}{|c|}{\textbf{Dataset}} & \multicolumn{1}{c|}{\textbf{Semantics}} & \textbf{Size}       & \textbf{N. Queries}       &  \textbf{Dim.$\downarrow$} & \multicolumn{1}{c|}{\textbf{Distribution}} \\ \hline \hline
NYTimes                       & TF-IDF Features                & 290,000    & 10,000  & 16        & \raisebox{-.4\height}{\includegraphics[width=0.08\textwidth]{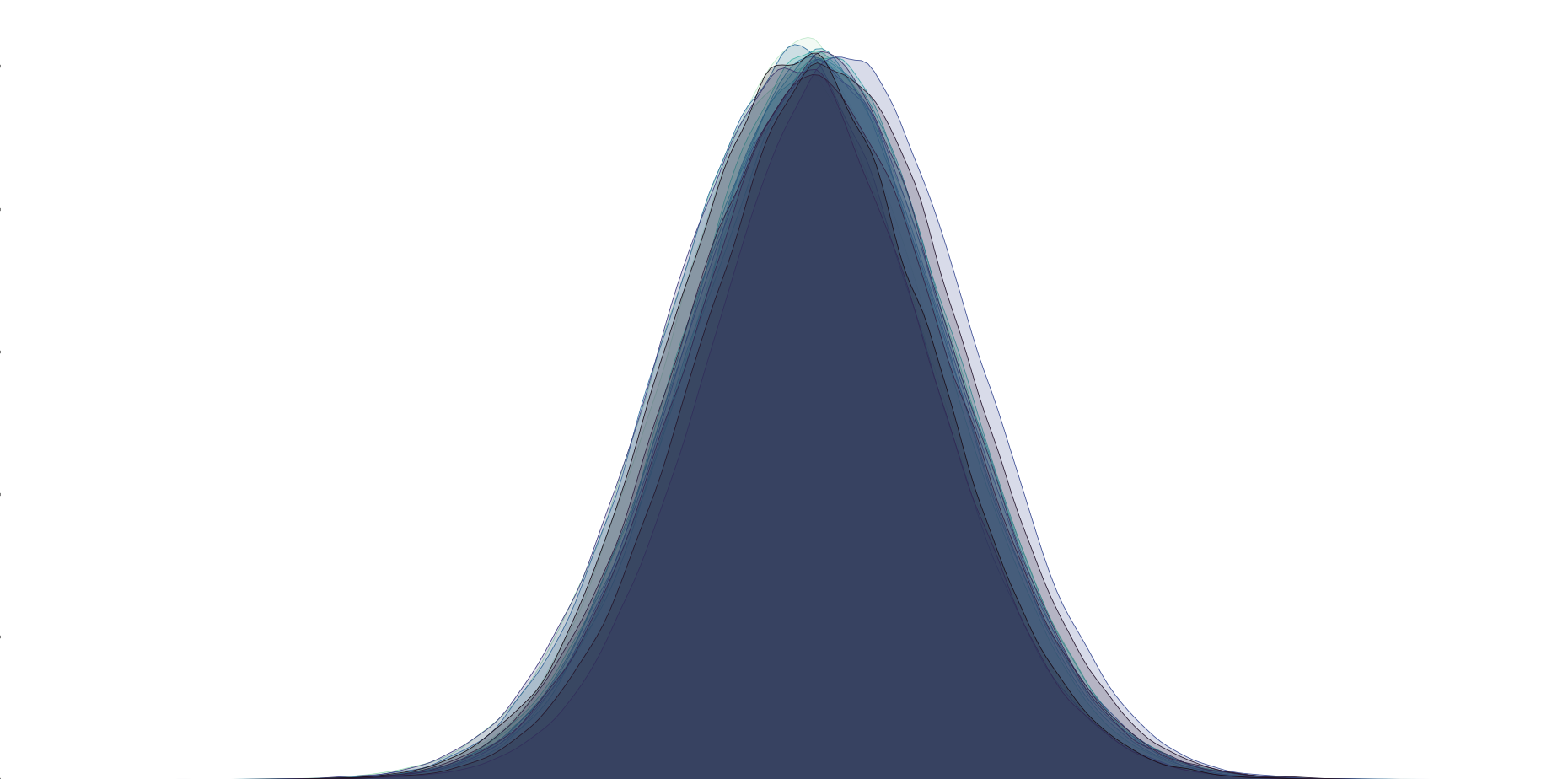}}                                  \\ \hline
GloVe                         & Word Embeddings                & 1,183,514  & 10,000  & 50         & \raisebox{-.4\height}{\includegraphics[width=0.08\textwidth]{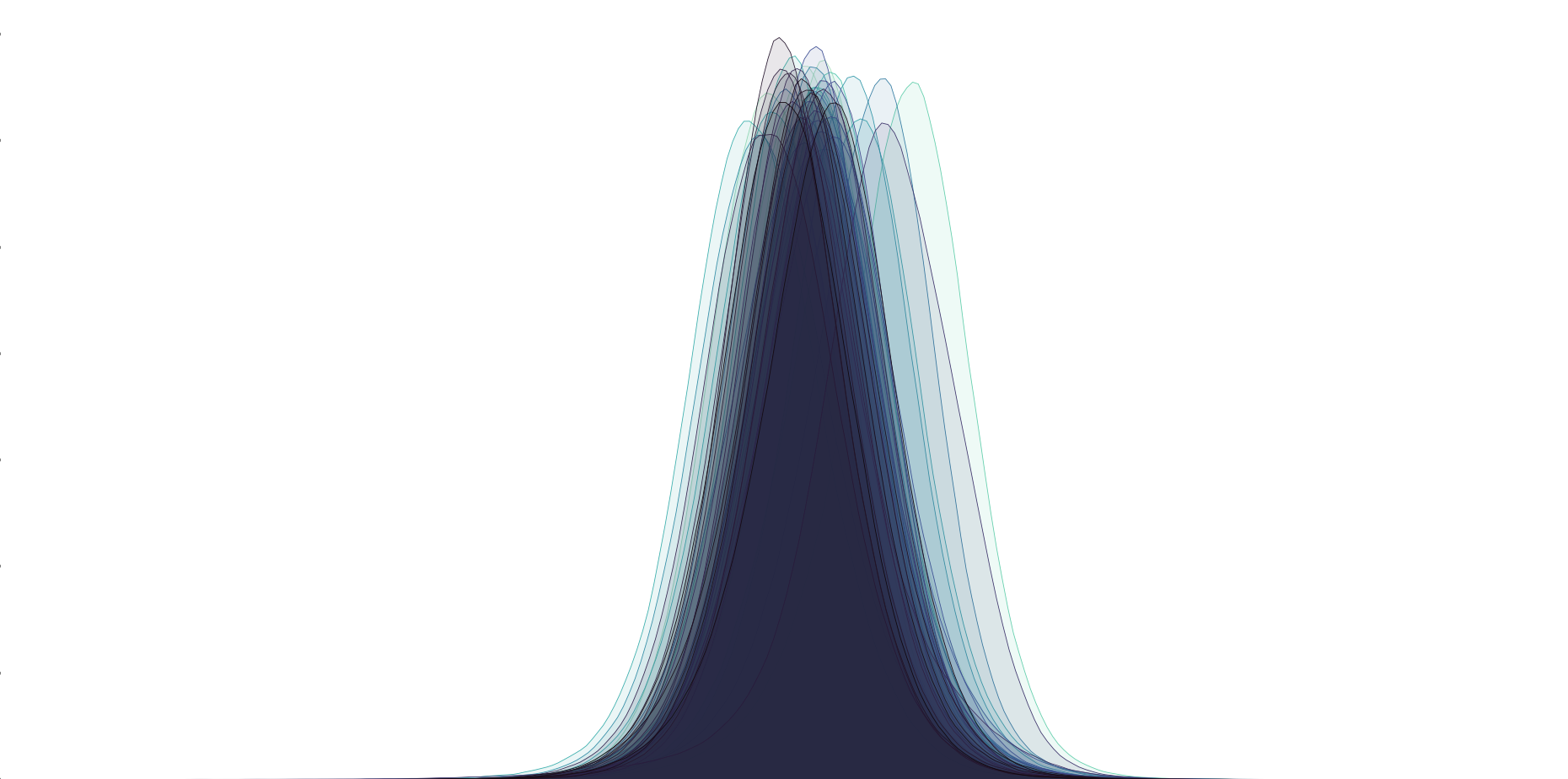}}                                  \\ \hline
DEEP                          & Image Embeddings               & 9,990,000 & 10,000 & 96         &  \raisebox{-.4\height}{\includegraphics[width=0.08\textwidth]{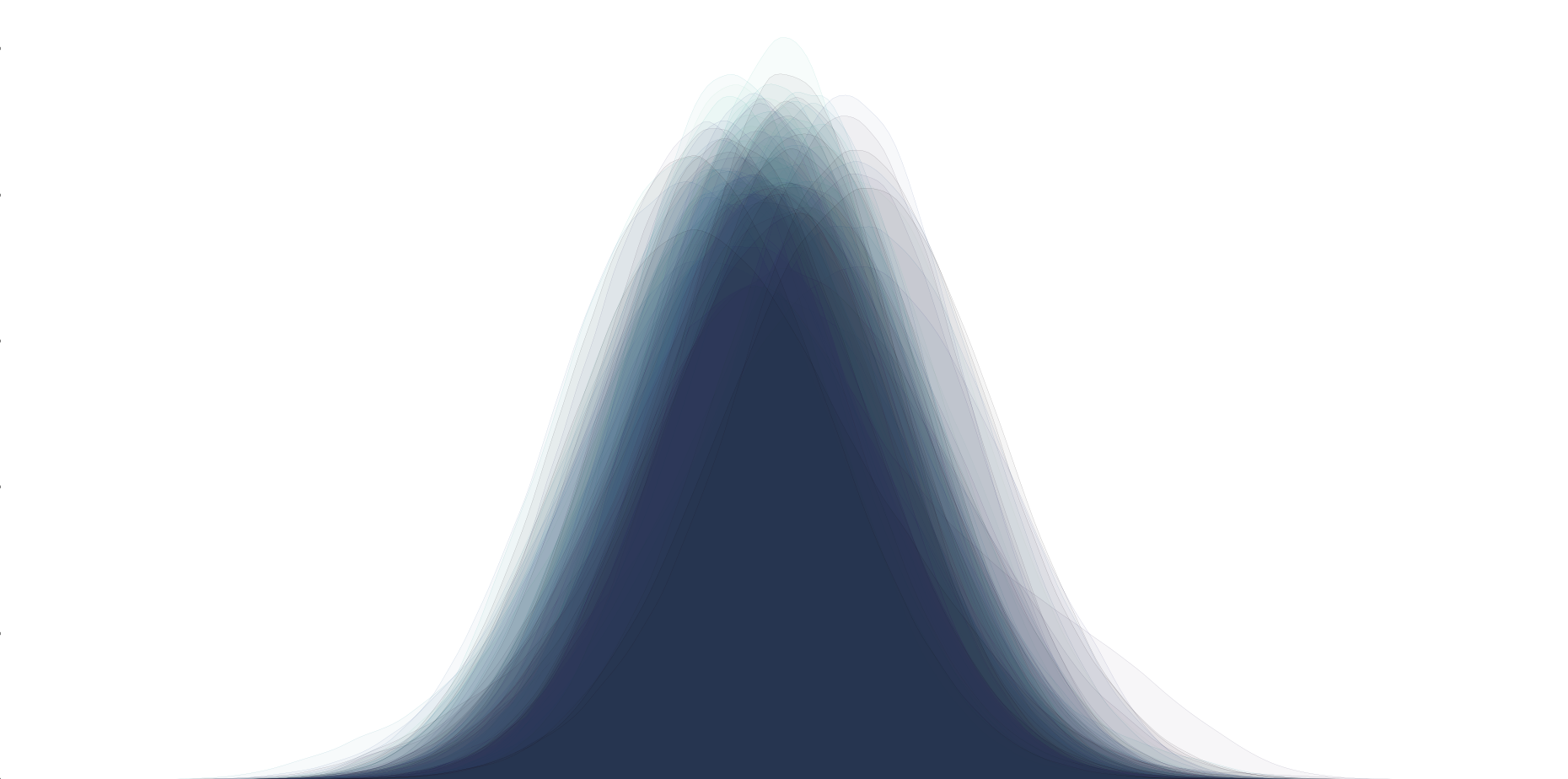}}                                 \\ \hline 
SIFT                          & Image Features                 & 1,000,000  & 10,000  & 128        &  \raisebox{-.4\height}{\includegraphics[width=0.08\textwidth]{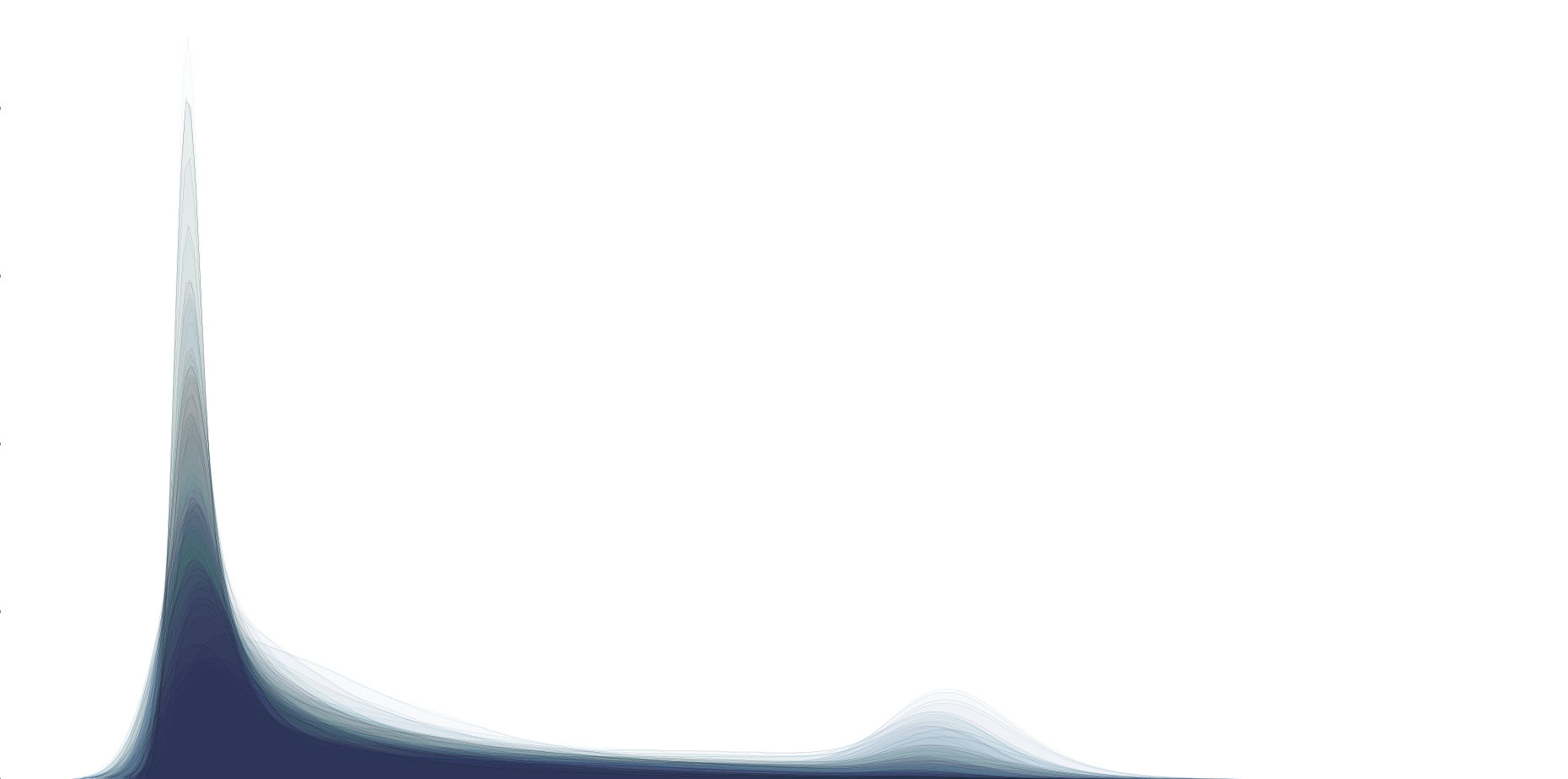}}                                 \\ \hline
GloVe                         & Word Embeddings                & 1,183,514  & 10,000  & 200        & \raisebox{-.4\height}{\includegraphics[width=0.08\textwidth]{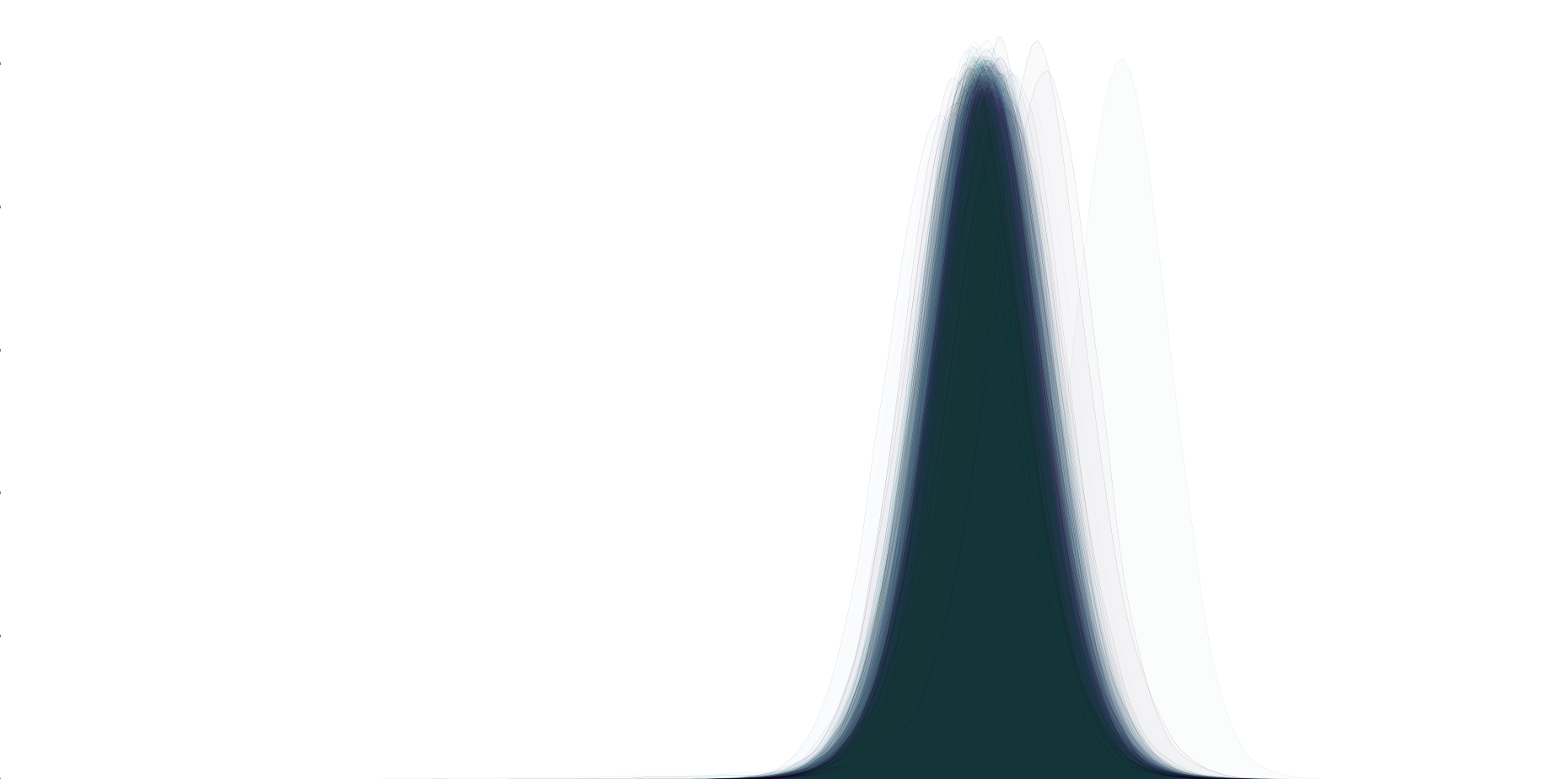}}                                  \\ \hline
MSong                         & Audio Features                 & 983,185    & 1,000    & 420        & \raisebox{-.4\height}{\includegraphics[width=0.08\textwidth]{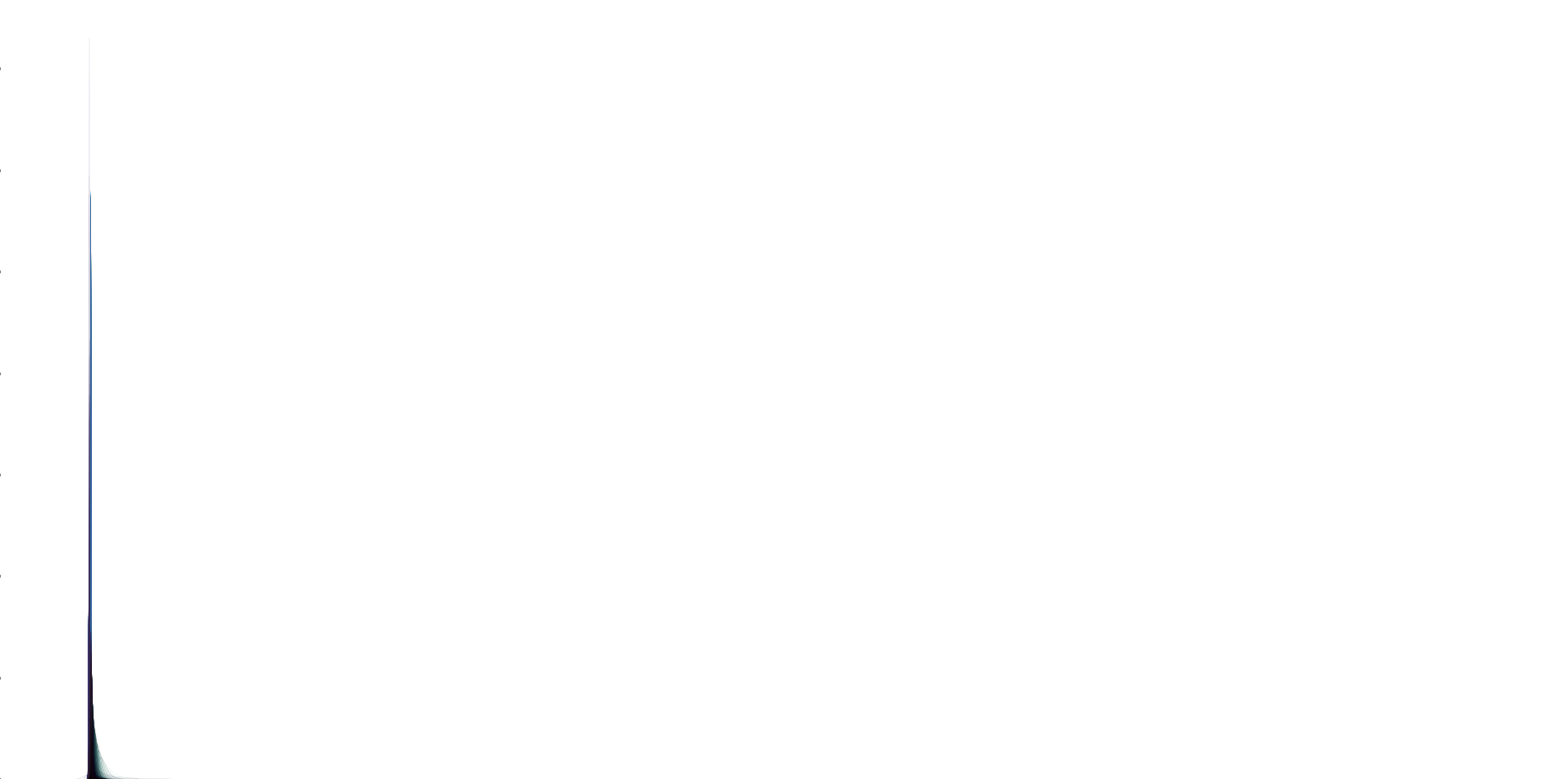}}                                   \\ \hline
Contriever                    & Word Embeddings                & 990,000  & 10,000  & 768        &  \raisebox{-.4\height}{\includegraphics[width=0.08\textwidth]{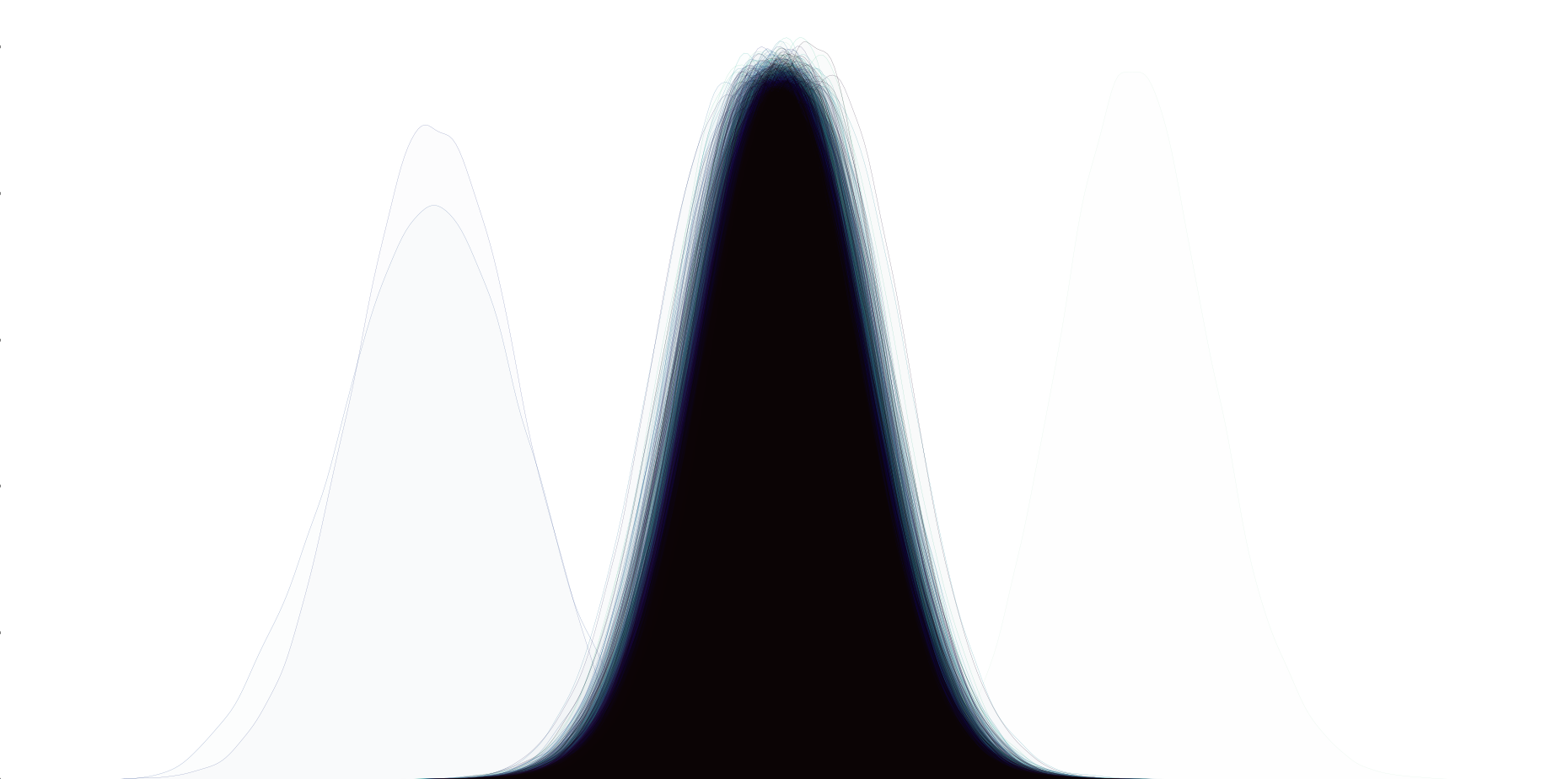}}                                 \\ \hline
arXiv                         & Text Embeddings                 & 2,253,000 & 1,000    & 768        & \raisebox{-.4\height}{\includegraphics[width=0.08\textwidth]{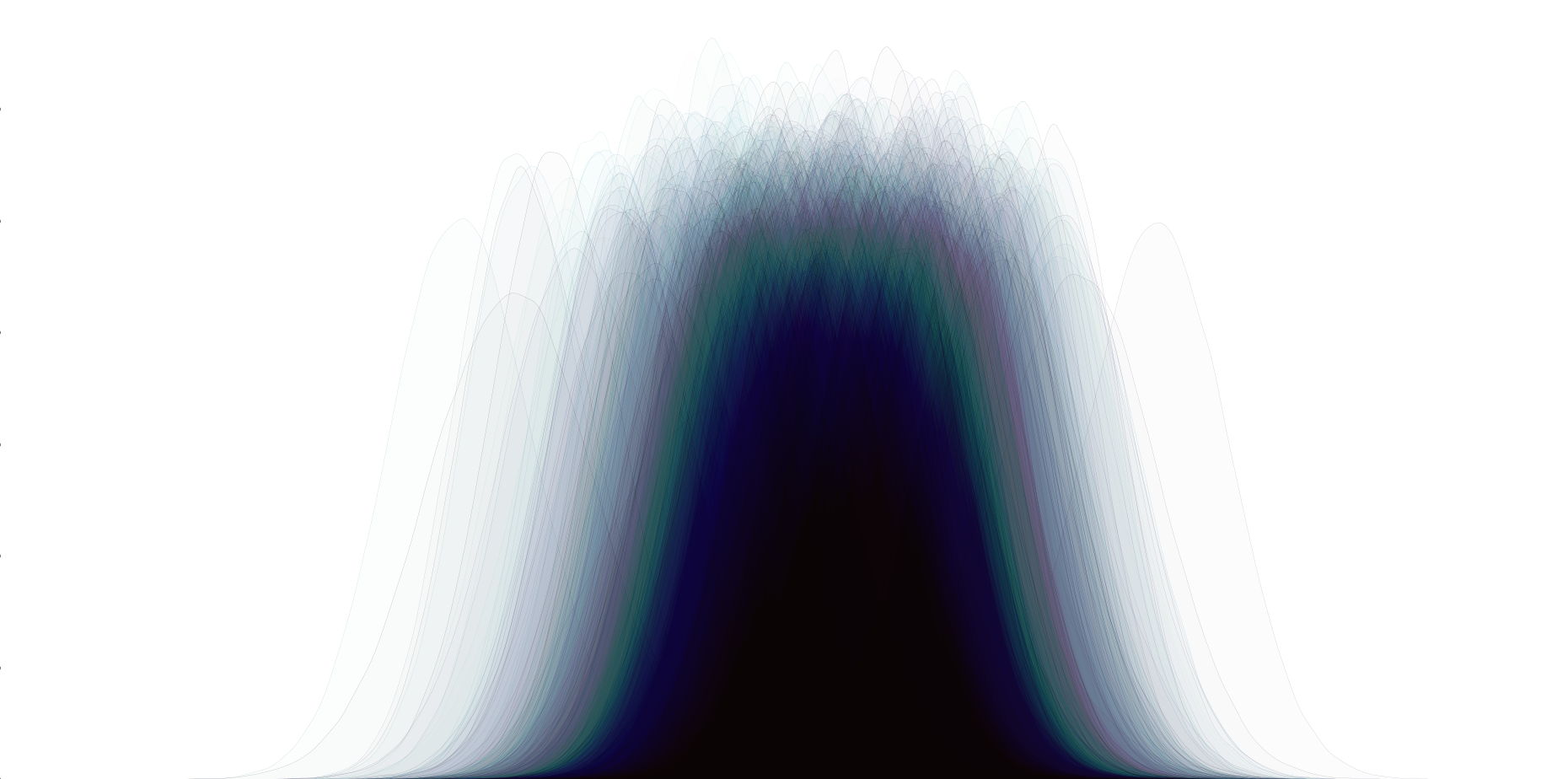}}                                  \\ \hline
% F-MNIST                       & Image Pixels                   & 60,000 & 10,000     & 784        &  \raisebox{-.4\height}{\includegraphics[width=0.08\textwidth]{paper/images/distributions/kde-fashion-mnist-784-euclidean.png}}                                 \\ \hline
%MNIST                         & Image Pixels                   & 60,000     & 784        &  \raisebox{-.4\height}{\includegraphics[width=0.08\textwidth]{paper/images/distributions/kde-mnist-784-euclidean.png}}                                 \\ \hline
GIST                          & Image Features                 & 1,000,000  & 1,000  & 960        &  \raisebox{-.4\height}{\includegraphics[width=0.08\textwidth]{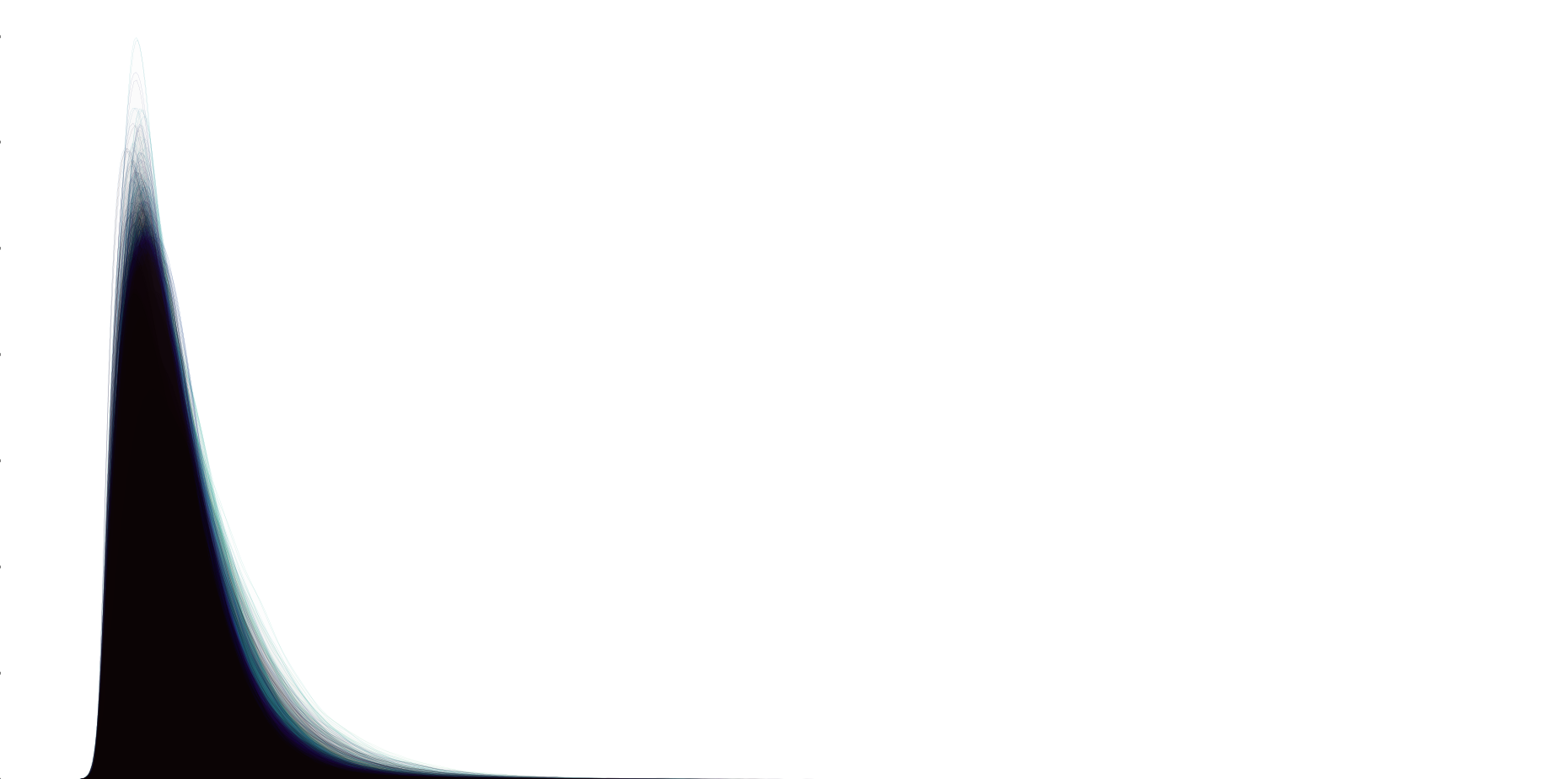}}                                 \\ \hline 
OpenAI                         & Text Embeddings                 & 999,000    & 1,000    & 1536        & \raisebox{-.4\height}{\includegraphics[width=0.08\textwidth]{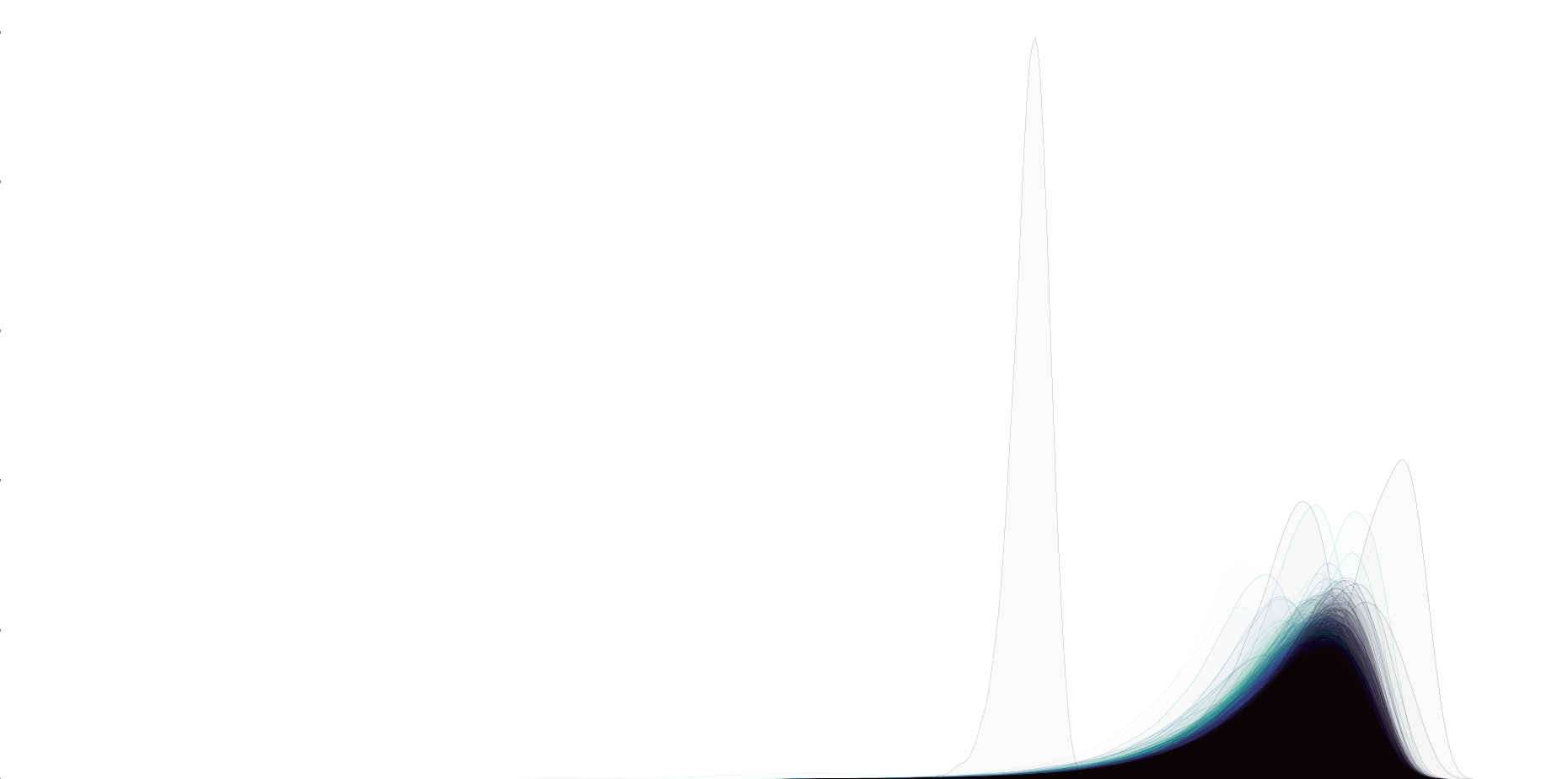}}                                  \\ \hline
%Trevi                         & Image Pixels                   & 91,120     & 4,096      &  \raisebox{-.4\height}{\includegraphics[width=0.08\textwidth]{paper/images/distributions/kde-trevi-4096.png}}                                 \\ \hline
%STL                           & Image Pixels                   & 90,000    & 9,216      & \raisebox{-.4\height}{\includegraphics[width=0.08\textwidth]{paper/images/distributions/kde-stl-9216.png}}                                  \\ \hline

\end{tabular}
}\vspace*{-3mm}
\end{table}

\subsection{Vector Datasets}\label{sec:datasets}
Indexes and quantization techniques for VSS can fail to achieve desirable recall depending on the collection they are applied to~\cite{rabitq}. As such, we have chosen for our analysis and evaluation 10 datasets (see Table~\ref{tab:datasets}) that exhibit various characteristics, some commonly used to evaluate vector similarity search techniques (e.g., SIFT, GIST, GloVe, MSong, DEEP). From these collections, 3 represent vectors from image data (SIFT, GIST, DEEP), 6 from text (NYTimes, GloVe variants, Contriever, arXiv, OpenAI), and 1 from audios (MSong). One dataset has an {\small \tt int} datatype (SIFT); the rest are {\small \tt float32}. From our observations, we classify these datasets based on (i) their dimensionality $D$ and (ii) the distributions of their dimensions. 

\vspace*{3mm}
\noindent{\bf Dimensionality.} As the dimensionality of a collection increases, every extra vector evaluated with the distance function adds more computational overhead and memory consumption (as also memory footprint and CPU cost increases by D). Also, an effect known as \textit{the curse of dimensionality} appears, in which the distance difference between a vector's farthest and nearest neighbor becomes indiscernible~\cite{satrees, surveysystems}. This can occur as low as $D=10$, given that dimensions are identically distributed and independent. The latter makes it harder to construct a good enough set $\hat{R}$. Table~\ref{tab:datasets} shows that vector collections are always of high dimensionality ($D > 10$), and the ones stemming from LLMs (e.g., OpenAI/1536) usually exhibit an even higher dimensionality. Interestingly, these collections have shown higher resilience to the curse of dimensionality~\cite{meaningfulcurse}. 

\vspace*{3mm}
\noindent{\bf Value Distributions.} In the last column of Table \ref{tab:datasets}, we show a plot depicting the shape of the distribution of each dimension for every collection. Here, we observe two types: normal (DEEP, NYTimes, arXiv, Contriever, GloVe variants) and skewed (SIFT, GIST, MSong, OpenAI). These distributions are of importance for the pruning power of algorithms which prune dimensions at search time.

\subsection{The Power of Pruning}\label{sec:pruningpower}
ADSampling~\cite{adsampling} and BSA~\cite{bsa} propose reducing the D-complexity of each distance evaluation by pruning dimension that are no longer needed to determine if a vector will make it into the KNN candidates list of a query (usually a max-heap~\cite{faisspaper}). These algorithms were motivated by the observation that most of a query runtime is spent evaluating vectors that never made it into the KNN candidate list (>84\% in IVF, >63\% in HNSW)~\cite{adsampling}. However, they are not the first algorithms to pursue dimensions pruning per vector, as previous studies tried to do so in an exact fashion by computing exact \textit{bounds} on the distance metric. These algorithms find efficacy since there is a concentration inequality on the distance between two vectors~\cite{distanceconcentration}. 

\vspace*{3mm}\noindent{\bf Exact bounds} to the Euclidean distance have been proposed in \cite{bond, fnn}. A common idea of these is to compute the best-case scenario (a lower-bound) of the distance between $v$ and $q$ after only having inspected a few dimensions. If this best-case scenario distance is higher than an existing threshold (usually the current best $k^{th}$ exact distance), then $v$ cannot make it to the k-nearest neighbors of $q$ (as the distance is monotonically increasing), resulting in the pruning of the dimensions of $v$, which have not been evaluated yet. The simplest of lower-bounds is the partially computed distance itself, which does not incur additional latency to obtain but may lack the power to prune early~\cite{adsampling}.

BOND~\cite{bond} (Branch-and-bound ON Decomposed data) proposes computing lower- \textit{and} upper-bounds to the Euclidean distance. An upper-bound is an estimation of the worst-case scenario of the distance between $v$ and $q$ by only having inspected a few dimensions. Thanks to this upper-bound, BOND can define a pruning threshold without ever visiting all the dimensions of any vector. This allows the data to be vertically decomposed (dimensions are stored together), enabling the search to happen dimension-by-dimension (instead of vector-by-vector) without incurring random access. This vertical decomposition is key to BOND's main idea: to visit dimensions in an order that more rapidly increases the distance metric towards the lower-bound, thus pruning vectors earlier. BOND criteria to prioritize dimensions is to first visit the dimension with the highest value in the query vector (\textit{decreasing} order). We refer to this as a query-aware order to visit dimensions. This achieved a power-law pruning behavior on skewed datasets. However, the speedup of BOND on KNNS (1.6x faster) was limited by the upper and lower bounds computation latency. 

%\vspace*{3mm}
\noindent{\bf Approximate pruning} techniques try to prune vectors with a low probability of getting into the KNN candidates list of a query after having inspected a few of their dimensions. The approximate nature of these techniques makes pruning more efficient and reduces the complexity of evaluating whether a vector can be pruned. ADSampling~\cite{adsampling} is the first of its kind. ADSampling performs a random orthogonal projection on the entire collection. This allows one to take random samples from a vector projected at different dimensions just by sequentially scanning it. At search time, for every vector in the collection, ADSampling reads a subset of its dimensions. The size of this subset is controlled with a parameter fixed for a dataset ($\Delta d$). Then, it evaluates the partial distance metric and estimates if it is already unlikely that the vector will make it into the resulting KNNs of the query. This evaluation is done via hypothesis testing by comparing the partial distance with a threshold (the current best $k^{th}$ exact distance) using a fixed error bound ($\epsilon_0$). Similar to $\Delta d$, this error bound is also fixed for a dataset. If the hypothesis test cannot prune the vector, it continues to read its following $\Delta d$ dimensions and repeats the process. ADSampling prunes 96\% of dimension values in GIST, achieving a speedup of ~3.0x in a brute-force search with >0.99 recall. ADSampling also introduced a data layout that separates the vectors into two blocks: One with the first $\Delta d$ dimensions (fully scanned first) and the other one with the rest of the dimensions (scanned only on the vectors not pruned with the first hypothesis testing). This layout speeds up searches thanks to the first block being cached more efficiently.  

BSA~\cite{bsa} followed ADSampling by transforming the vectors using a PCA projection on the D-dimensional space instead of ADSampling's random orthogonal projection. The latter minimizes the error distribution of the distance approximation. Furthermore, BSA introduces a framework in which the probability of a vector being part of the KNN candidates list is evaluated via error quantiles (given by the Cauchy‐Schwarz inequality~\cite{bhatia1995cauchy}) rather than hypothesis testing. BSA also proposes a learned approach in which the error bounds of the PCA projection at each dimension are learned at preprocessing time using multiple linear regression models. The latter alleviates the user from configuring the significance value for the hypothesis testing. However, it is expensive, as a model has to be trained for every dimension in the collection, and their effectiveness has yet to be proven under distribution shifts in the collection~\cite{candy}. 
BSA reported searches 1.6x faster than ADSampling, limited due to its more expensive data transformation and higher latency of the error quantile evaluation. Like ADSampling, BSA adopted the dual-block layout and pruning every $\Delta d$ steps. 

The effectiveness of these algorithms is dependent on their \textit{pruning power} during a search. We define \textit{pruning power} as the percentage of individual dimensions not used in distance calculations in a KNNS (either exact or approximate). Note that pruning power does not directly translate to speedup, as these algorithms perform additional work to evaluate bounds. Unfortunately, these studies (ADSampling and BSA) have analyzed their pruning power as a final averaged metric. In the next section, we perform a comprehensive analysis of the behavior of pruning to uncover potential bottlenecks and missed opportunities of these novel pruning approaches. 

\begin{table}[]
\centering
\caption{Best, $p^{50}$, $p^{25}$, and worst pruning power of ADSampling when trying to prune at every dimension ($\Delta d$=1). The darker area indicates the portion of values not pruned at that dimension (x-axis). The number inside the plot indicates the total percentage of avoided values. }
\vspace*{-3mm}
\label{tab:pruning}
\resizebox{0.99\columnwidth}{!}{%
\begin{tabular}{l|l|l|l|l|l|l|l|l|}\cline{2-9}
              & \multicolumn{8}{c|}{Datasets}                                                      \\ \hline
\multicolumn{1}{|l|}{Pruning} & \rotatebox[origin=c]{90}{\raisebox{-1.5\normalbaselineskip}[0pt][0pt]{GIST/960}} & \rotatebox[origin=c]{90}{\raisebox{-1.5\normalbaselineskip}[0pt][0pt]{ MSong/420 }} & \rotatebox[origin=c]{90}{\raisebox{-1.5\normalbaselineskip}[0pt][0pt]{NYTimes/16}} & \rotatebox[origin=c]{90}{\raisebox{-1.5\normalbaselineskip}[0pt][0pt]{GloVe/50}} & \rotatebox[origin=c]{90}{\raisebox{-1.5\normalbaselineskip}[0pt][0pt]{DEEP/96}} & \rotatebox[origin=c]{90}{\raisebox{-1.5\normalbaselineskip}[0pt][0pt]{Contriever/786}} & \rotatebox[origin=c]{90}{\raisebox{-1.5\normalbaselineskip}[0pt][0pt]{OpenAI/1536}} & \rotatebox[origin=c]{90}{\raisebox{-1.5\normalbaselineskip}[0pt][0pt]{SIFT/128}} \\ \hline\hline
\multicolumn{1}{|l|}{Best}  &  \raisebox{-.4\height}{\includegraphics[width=0.055\textwidth]{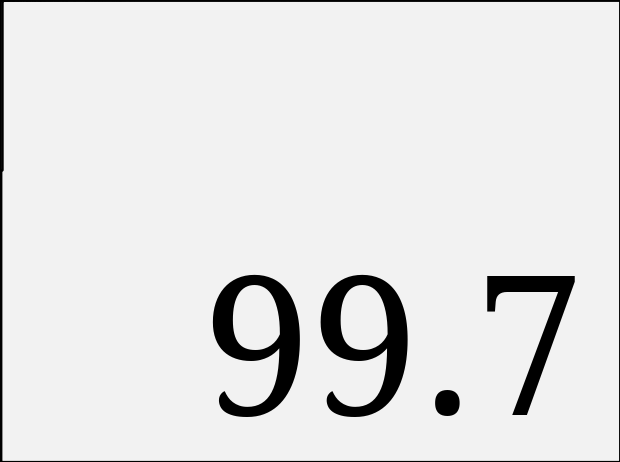}}  & \raisebox{-.4\height}{\includegraphics[width=0.055\textwidth]{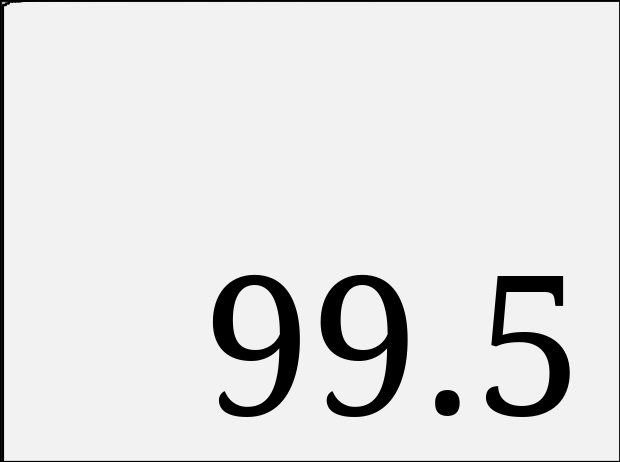}} & \raisebox{-.4\height}{\includegraphics[width=0.055\textwidth]{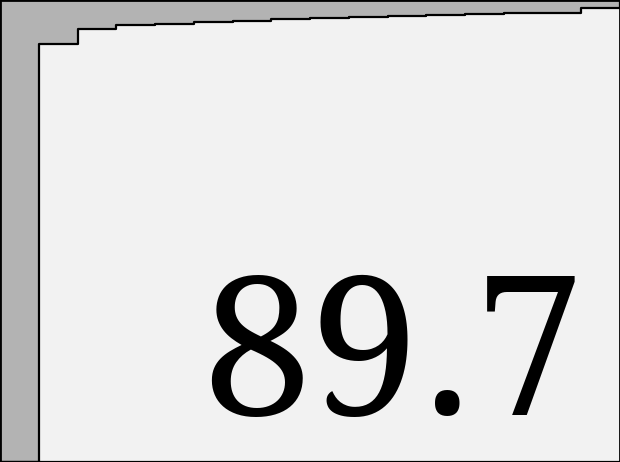}} & \raisebox{-.4\height}{\includegraphics[width=0.055\textwidth]{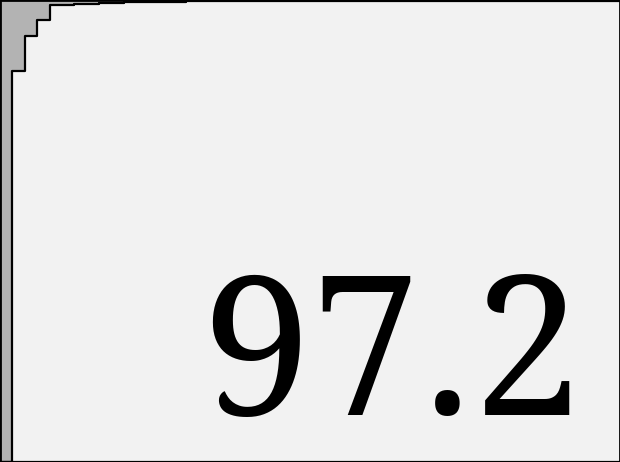}} & \raisebox{-.4\height}{\includegraphics[width=0.055\textwidth]{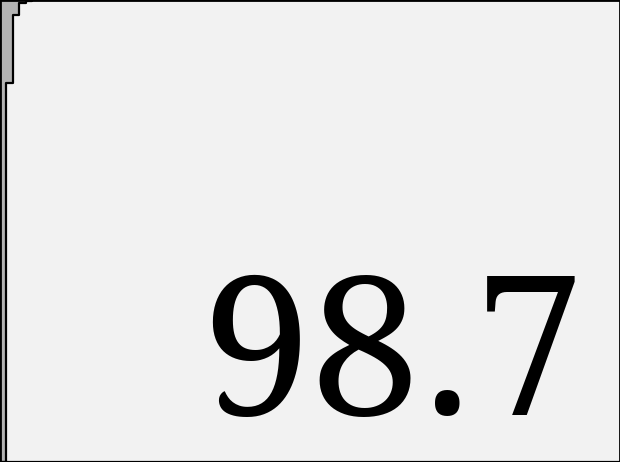}} & \raisebox{-.4\height}{\includegraphics[width=0.055\textwidth]{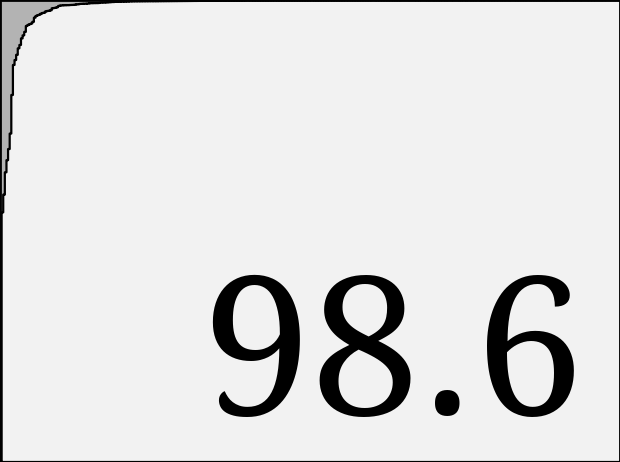}}                 &  \raisebox{-.4\height}{\includegraphics[width=0.055\textwidth]{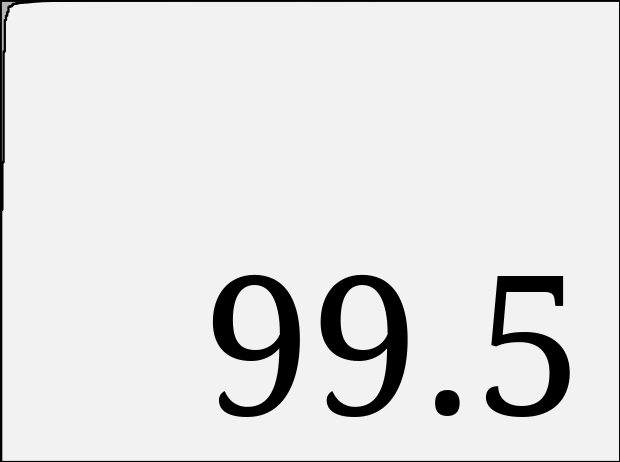}} &  \raisebox{-.4\height}{\includegraphics[width=0.055\textwidth]{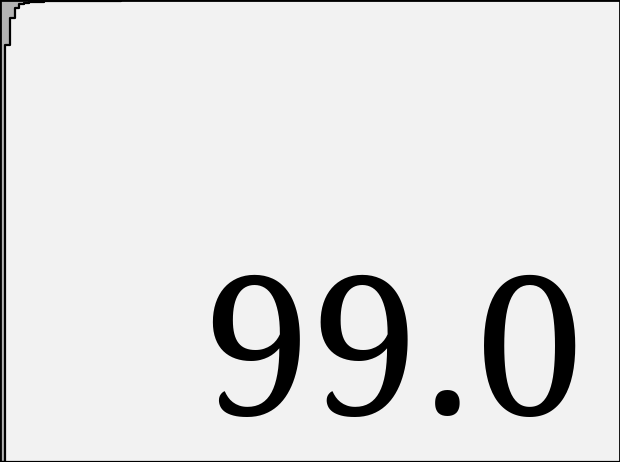}} \\
\hline
\multicolumn{1}{|l|}{$p^{50}$} & \raisebox{-.4\height}{\includegraphics[width=0.055\textwidth]{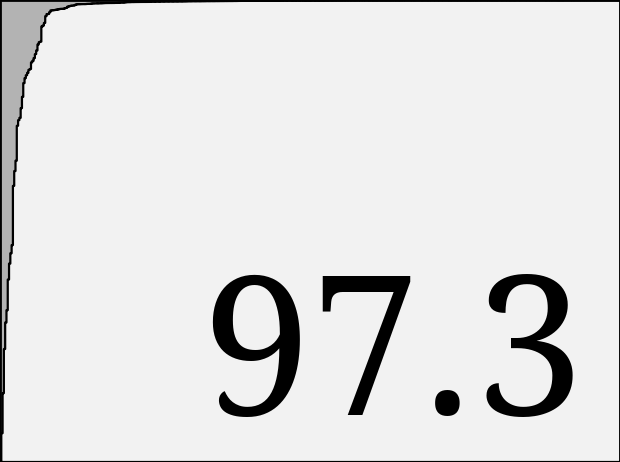}}   & \raisebox{-.4\height}{\includegraphics[width=0.055\textwidth]{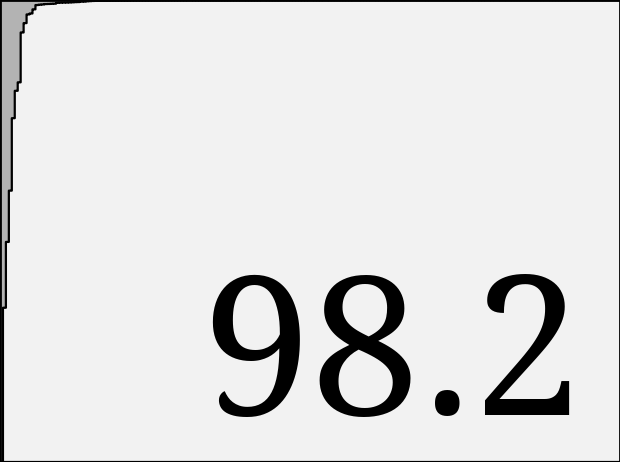}} & \raisebox{-.4\height}{\includegraphics[width=0.055\textwidth]{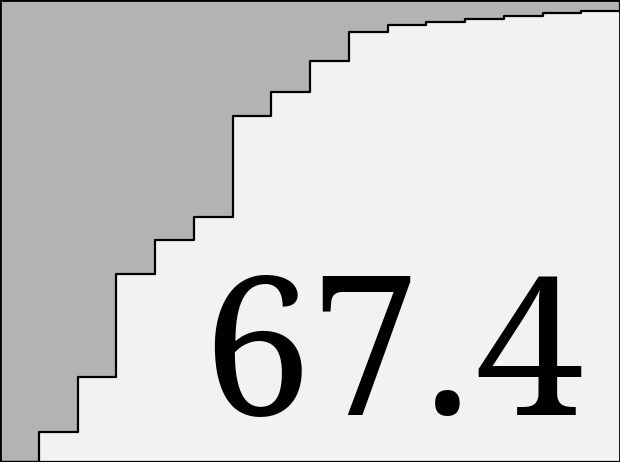}} & \raisebox{-.4\height}{\includegraphics[width=0.055\textwidth]{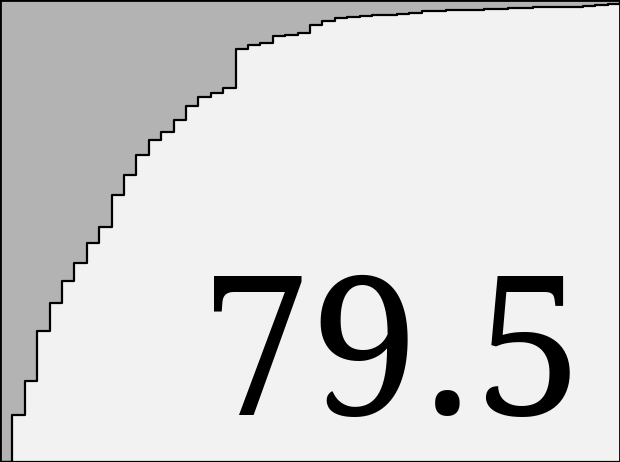}} & \raisebox{-.4\height}{\includegraphics[width=0.055\textwidth]{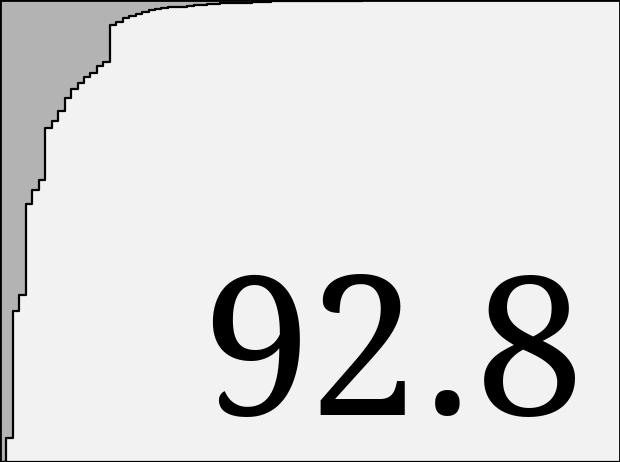}} &  \raisebox{-.4\height}{\includegraphics[width=0.055\textwidth]{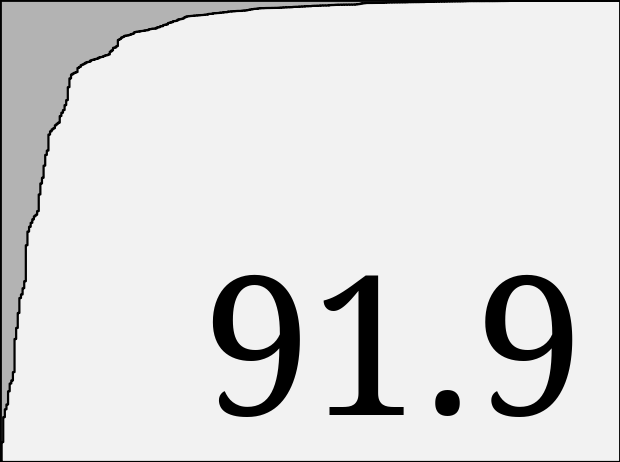}}   &  \raisebox{-.4\height}{\includegraphics[width=0.055\textwidth]{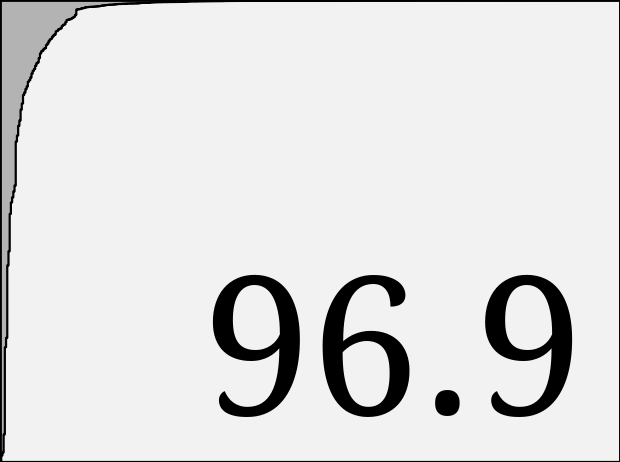}} &  \raisebox{-.4\height}{\includegraphics[width=0.055\textwidth]{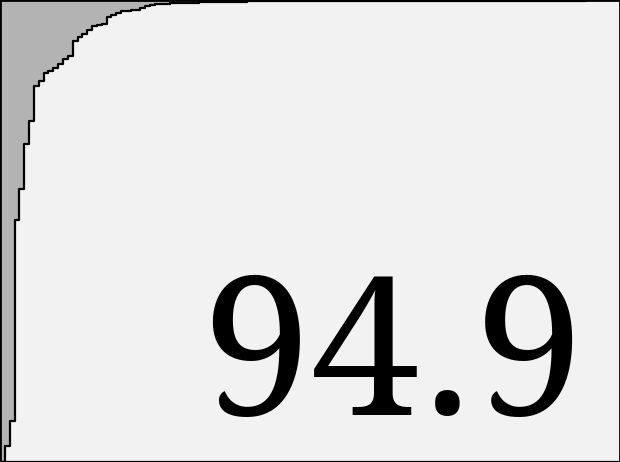}} \\
\hline
\multicolumn{1}{|l|}{$p^{25}$} &  \raisebox{-.4\height}{\includegraphics[width=0.055\textwidth]{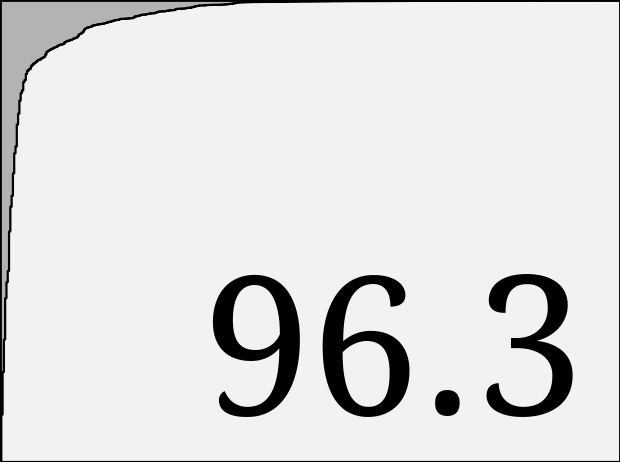}}        &  \raisebox{-.4\height}{\includegraphics[width=0.055\textwidth]{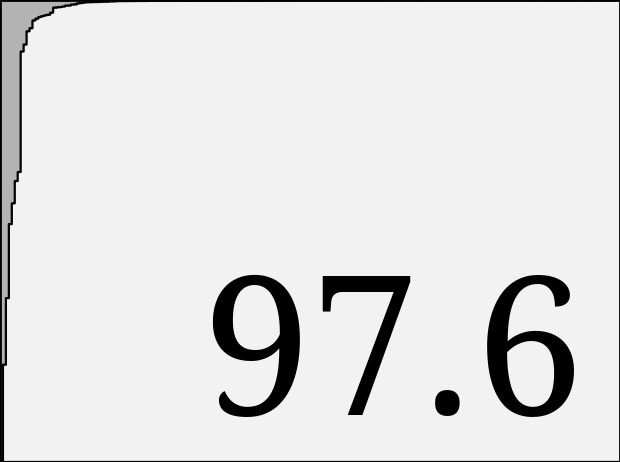}} & \raisebox{-.4\height}{\includegraphics[width=0.055\textwidth]{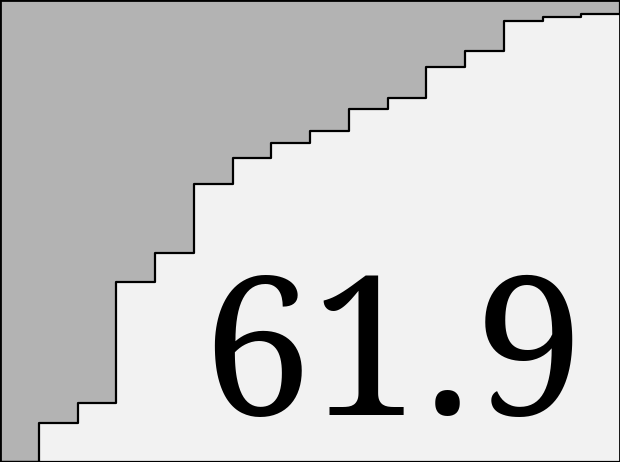}} & \raisebox{-.4\height}{\includegraphics[width=0.055\textwidth]{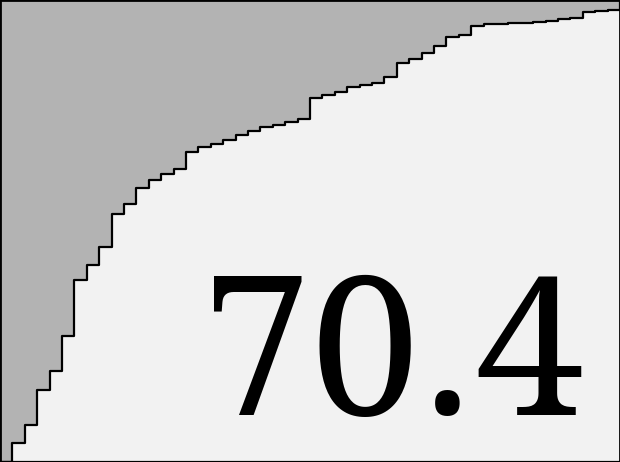}} & \raisebox{-.4\height}{\includegraphics[width=0.055\textwidth]{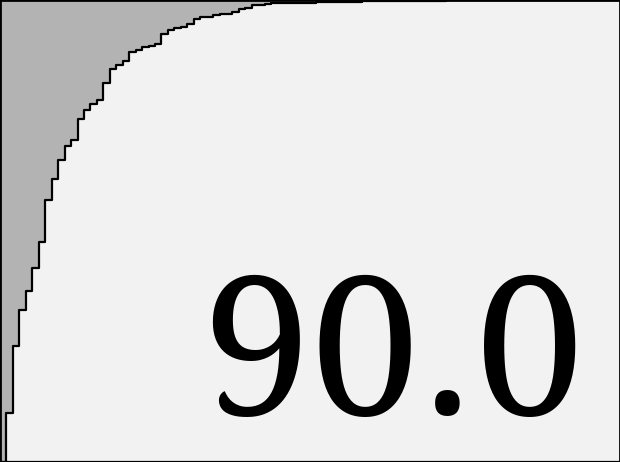}} &  \raisebox{-.4\height}{\includegraphics[width=0.055\textwidth]{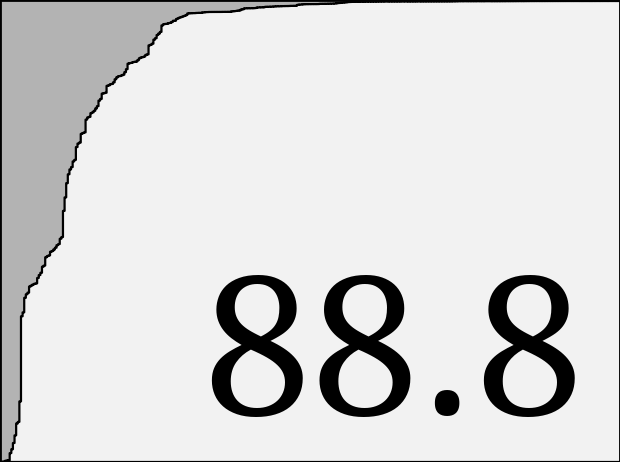}}   &  \raisebox{-.4\height}{\includegraphics[width=0.055\textwidth]{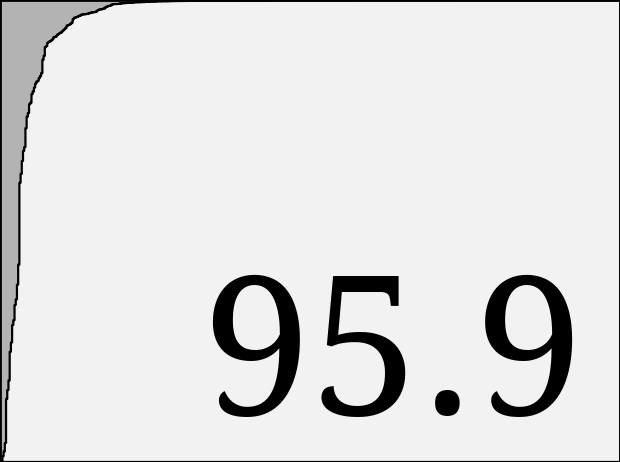}} &  \raisebox{-.4\height}{\includegraphics[width=0.055\textwidth]{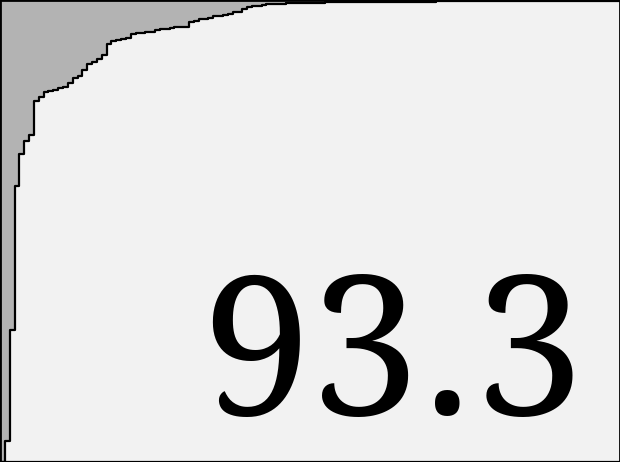}} \\
\hline
\multicolumn{1}{|l|}{Worst} &  \raisebox{-.4\height}{\includegraphics[width=0.055\textwidth]{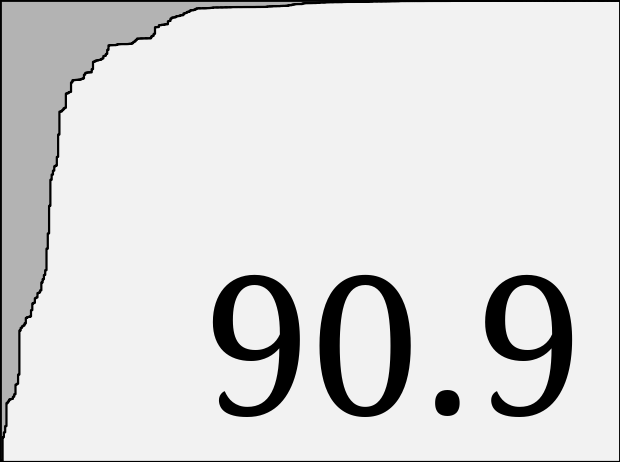}}        &  \raisebox{-.4\height}{\includegraphics[width=0.055\textwidth]{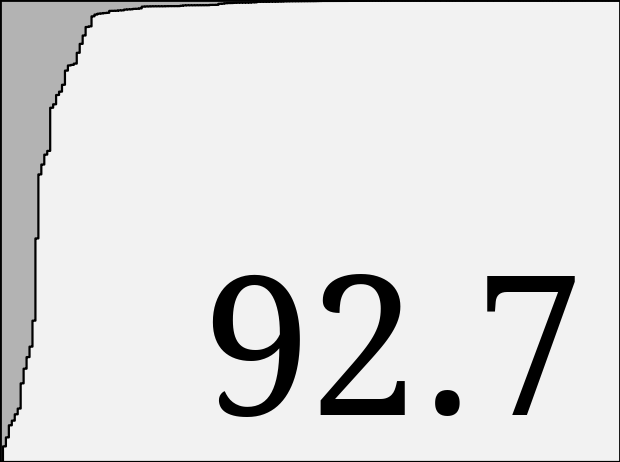}}  & \raisebox{-.4\height}{\includegraphics[width=0.055\textwidth]{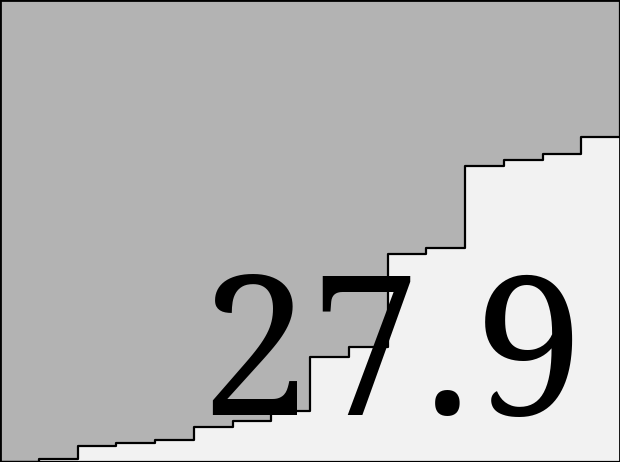}} & \raisebox{-.4\height}{\includegraphics[width=0.055\textwidth]{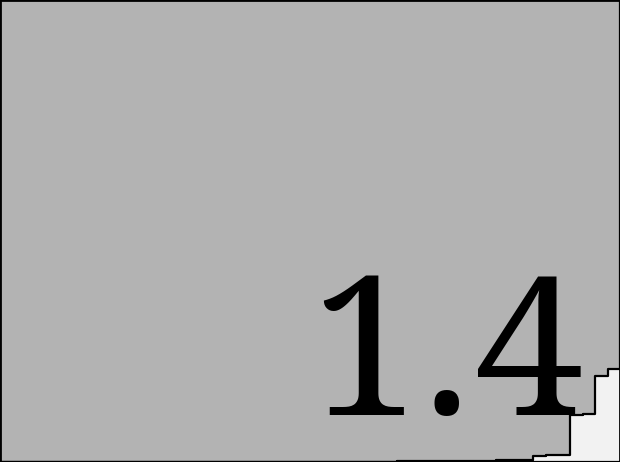}} & \raisebox{-.4\height}{\includegraphics[width=0.055\textwidth]{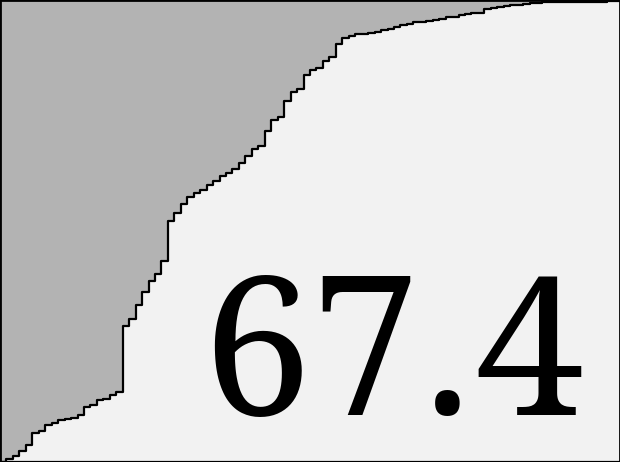}} &  \raisebox{-.4\height}{\includegraphics[width=0.055\textwidth]{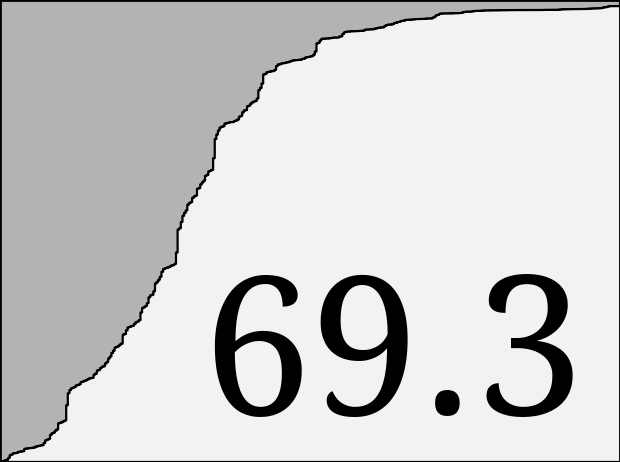}}   &  \raisebox{-.4\height}{\includegraphics[width=0.055\textwidth]{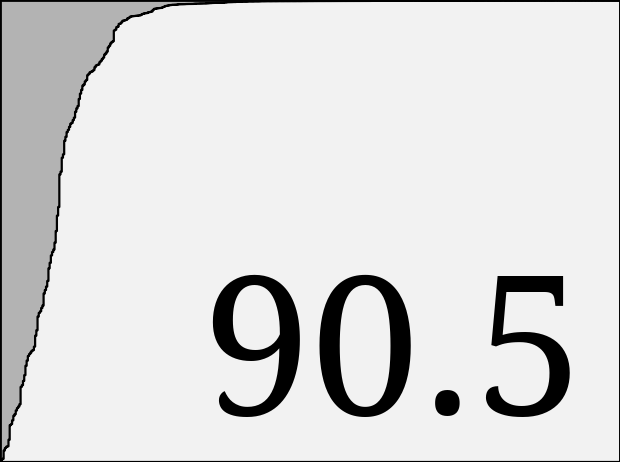}} &  \raisebox{-.4\height}{\includegraphics[width=0.055\textwidth]{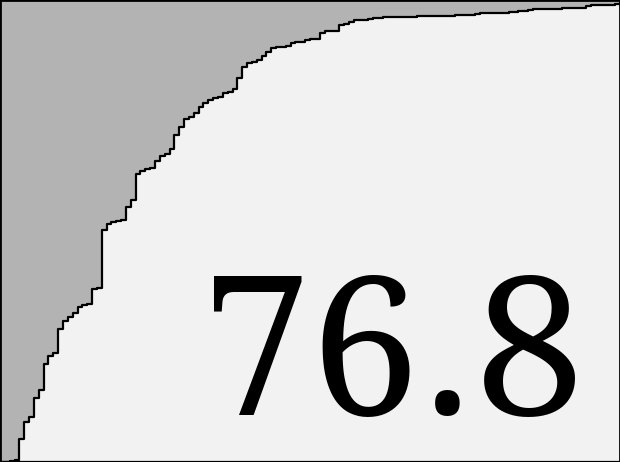}} \\
\hline
\end{tabular}
}\vspace*{-2mm}
\end{table}

\subsection{Pruning Behavior}
Table~\ref{tab:pruning} shows the behavior of the best, $p^{50}$,  $p^{25}$, and worst pruning power with K=10 on ADSampling within eight datasets: Four skewed (GIST, MSong, SIFT, OpenAI) and four normal (NYTimes, GloVe, DEEP, Contriever). We choose ADSampling for this demonstration as the pruning behaviors of BSA\footnote{When referring to BSA, we would be referring to BSA$_{res}$--the version of BSA without the learned instance approach (named BSA$_{pca}$), as the latter incurs a huge preprocessing cost and does not improve query latency at high recalls~\cite{bsa}.} 
are very similar (both in shape and pruning power), and BOND cannot prune normally distributed datasets. The plots show the percentage of vectors from the collection (y-axis) pruned at each scanned dimension (x-axis). The darker area represents values that were not pruned. Here, we visit dimensions one by one ($\Delta d=1$), doing the hypothesis testing at each step to show the real potential of pruning. 

We see that normally distributed datasets are more challenging to prune than skewed datasets. We can also see how the pruning power within the same dataset changes on a query basis. For instance, in Contriever/768 the best pruning power prunes 98.6\% of values while the worst prunes 69.3\%. In NYTimes, half of the queries ($p^{50}$) can prune less than 67\% of the values. From these pruning behaviors, we make the following \textbf{key observations}: (i) Effective pruning has a query-dependent starting point. (ii) For most queries and datasets, once pruning starts, it keeps pruning exponentially fast on the following dimensions (note the power-law behavior of the plots). (iii) A small step size is beneficial when pruning starts; however, benefits are only seen at much bigger steps at later dimensions. From these key observations, we detect where pruning algorithms are missing opportunities to optimize their search strategy.

\vspace*{3mm}
\noindent{\bf Issue \#1: Pruning at fixed steps.} In ADSampling and BSA, pruning happens every 32 dimensions ($\Delta d$). $\Delta d = 32$ was determined by doing an exhaustive parameter search~\cite{adsampling}. However, the optimal step size depends on both the query and the dataset (Table \ref{tab:pruning}). Therefore, the step size must be adaptive, starting low and increasing exponentially. This would also reduce the number of evaluations of the pruning predicate and alleviate the user from finding an optimal $\Delta d$ parameter for their data. 
%\pagebreak

\vspace*{3mm}
\noindent{\bf Issue \#2: Keep pruning at late dimensions:} The hypothesis testing no longer brings benefits after a certain dimension (pruning no longer happens). As such, there is no need to keep doing hypothesis tests; instead, the distance over the rest of the dimensions should be computed. The latter would lower the cost of  hypothesis testing and open opportunities for SIMDizing the distance kernel. 

\vspace*{3mm}\noindent{\bf Issue \#3: Not processing multiple vectors at-a-time.} Given the correct data layout, a dimension-by-dimension search would improve the efficiency of pruning algorithms. A dimension-by-dimension search can evaluate the pruning bounds on multiple vectors at-a-time~\cite{superscalar,monetdb} in a loop separated from the distance calculations. Also, it would allow compilers to seamlessly vectorize the distance calculations as every distance evaluation aggregate into a different result.
%\footnote{One can force vectorization of floating-point reductions (e.g., -ffast-math flag in gcc). However, this incur undefined floating-point behavior. Instead, systems implement SIMD kernels~\cite{usearch, bsa}--which are also faster~\cite{usearch}, for different ISAs~\cite{milvus, weaviate}.}. 
The access patterns to the data would also improve as the CPU cache is not polluted with dimension values that are not needed, and frequently accessed dimensions are cached more efficiently. Both approximate and exact search algorithms can benefit from this vectorized processing, as evaluating the distance of \textit{multiple} vectors is unavoidable.

% It is important to note how ADSampling is already achieving almost 100\% pruning in some datasets (MNIST, MSong, F-MNIST, SIFT, Trevi, GIST, Contriever), only leaving substantial room for improvement in a few datasets (GloVe variants and NYTimes). On these datasets with almost 100\% pruning, BSA preprocessing (especially the learned instance), the extra storage overhead and query transformation overhead could be over-engineering. In fact, on the ones in which ADSampling struggles, BSA with non-learned instances performs just slightly better. This hints to us that, for most datasets, developing tighter bounds would not bring as many benefits as a better pruning strategy would. 

\section{The PDX Data Layout}\label{sec:pdx}
We introduce the \textbf{PDX (Partition Dimensions Across)} data layout for vector similarity search (Figure \ref{fig:pdx}), which finds a balance between a vertically decomposed layout~\cite{bond} and the traditional vector-by-vector layout. PDX stores together the values of each dimension within a {\small \tt block}. Blocks define a subset of vectors within the collection (e.g., IVF/LSH buckets or horizontal partitioning). The motivation of blocks is to maintain all the dimensions of the same vector close by in the storage (analogous to rowgroups in modern file formats such as Parquet~\cite{parquet}, DuckDB~\cite{duckdbpaper}, and FastLanes~\cite{fastlanes}). The PDX layout allows for a dimension-by-dimension search that operates on multiple vectors at-a-time instead of the traditional vector-by-vector search. Furthermore, partitioning dimensions allow pruning algorithms to be adaptive regarding the number of dimensions explored. Thus, tackling the shortcomings uncovered in the previous section. 

\renewcommand{\lstlistingname}{Algorithm}

\lstset{frame=tb,
    language=C++,
    aboveskip=0mm,
    belowskip=4mm,
    showstringspaces=false,
    columns=flexible,
    basicstyle={\small\ttfamily},
    numbers=none,
    numberstyle=\tiny\color{gray},
    keywordstyle=\color{blue},
    stringstyle=\color{mauve},
    breaklines=true,
    breakatwhitespace=true,
    tabsize=3
}
\definecolor{mygreen}{rgb}{0,0.5,0}
\definecolor{mygray}{rgb}{0.5,0.5,0.5}
\definecolor{byzantium}{rgb}{0.44, 0.16, 0.39}

\lstdefinestyle{myCstyle}{
language       = C++,
basicstyle     = \linespread{0.8}\footnotesize\def\fvm@Scale{.80}\fontfamily{fvm}\selectfont,
numbers        = left,
numberstyle    = \color{mygray},
commentstyle   = \color{gray},
keywordstyle   = \bfseries\color{byzantium},
keywordstyle   =[2]\bfseries,
keywordstyle   =[3]\bfseries\color{blue},
stringstyle    = \color{blue},
morekeywords   = {},
directivestyle = \color{violet}\bfseries,
keywords=[2]{ N, D, data, query, FFOR,PATCH_EXCEPTIONS, DECODE, UNFFOR, READ_VECTOR_HEADER, PATCH, GLUE, BITUNPACK, BITUNPACK_DECODEDICT, READ_ROWGROUP_HEADER, CUT, BITPACK, SKEWDICT_BITPACK},
keywordstyle=[2]\color{NavyBlue}\bfseries,
keywords=[3]{ L1, L2, IP },
keywordstyle=[3]\color{ForestGreen}\bfseries,
keywords=[4]{ PDX_BLOCK_SIZE },
keywordstyle=[4]\color{Maroon}\bfseries,
keywords=[5]{ nothing },
keywordstyle=[5]\color{Orange}\bfseries,
keywords=[6]{ p },
keywordstyle=[6]\color{Brown}\bfseries,
caption          = ePWM.c,
}

%\begin{algorithm}\captionsetup{labelfont={sc,bf}, labelsep=newline}
\begin{lstlisting}[
    floatplacement=t,
    float,
    language=C++,
    style=myCstyle,
    caption={L2, L1, IP distance kernels on the PDX Layout},
    captionpos=t,
    morekeywords={STORE, LOAD, AND, AND_RSHIFT, AND_LSHIFT, OR, XOR, ADD, uint, FLMM1024, uint8, SET, U, int64, int16, uint32, MaxHeap, size_t},
    mathescape=true,
    aboveskip=0mm,
    belowskip=-2mm,
    label={algo:pdx}]
const PDX_BLOCK_SIZE = 64;
float[PDX_BLOCK_SIZE] distances;
for (d = 0; d < D; ++d){ // Dimensions loop
    size_t offset_to_dimension = d * PDX_BLOCK_SIZE;
    float query_dim = query[d];
    for (n = 0; n < PDX_BLOCK_SIZE; ++n){ // Vectors loop 
        L2:
            float to_mul = query_dim - data[offset_to_dimension + n];
            distances[n] += to_mul * to_mul;
        L1:
            float to_abs = query_dim - data[offset_to_dimension + n];
            distances[n] += std::fabs(to_abs)
        IP: 
            distances[n] += query_dim * data[offset_to_dimension + n];
    }   
}   
\end{lstlisting}

% PDX_BOND::GetDimensionsOrder(DIMENSION_MEANS, QUERY);

% // START
% float threshold = PDXEARCH::FullScan(VECTORGROUPS[0]);
% for (vectorgroup in VECTORGROUPS[1:]) {

%     // WARMUP
%     while (selectivity > SELECTIVITY_THRESHOLD) {
%         dims = PDXEARCH::FetchNextDimensions(); // Exponentially growing
%         PDXEARCH::FullScanDimensions(dims, query);
%         selectivity = PDX_BOND::GetSelectivity(distances, threshold);
%     }

%     // PRUNE
%     while (selectivity > PLATEAU_THRESHOLD) {
%         dims = PDXEARCH::FetchNextDimensions();
%         PDXEARCH::PrunedScanDimensions(dims, query, positions);
%         selectivity = PDX_BOND::GetSelectivityAndUpdatePositions(distances, positions, threshold);
%     }

%     // PLATEAU
%     for (position in positions) {
%         distance = PDXEARCH::ScanVector(position);
%         if (distance < threshold) {
%             knn.merge(distance);
%         }
%     }
%     threshold = knn.top();
% }

\begin{figure}[t]
\includegraphics[width=1\linewidth]{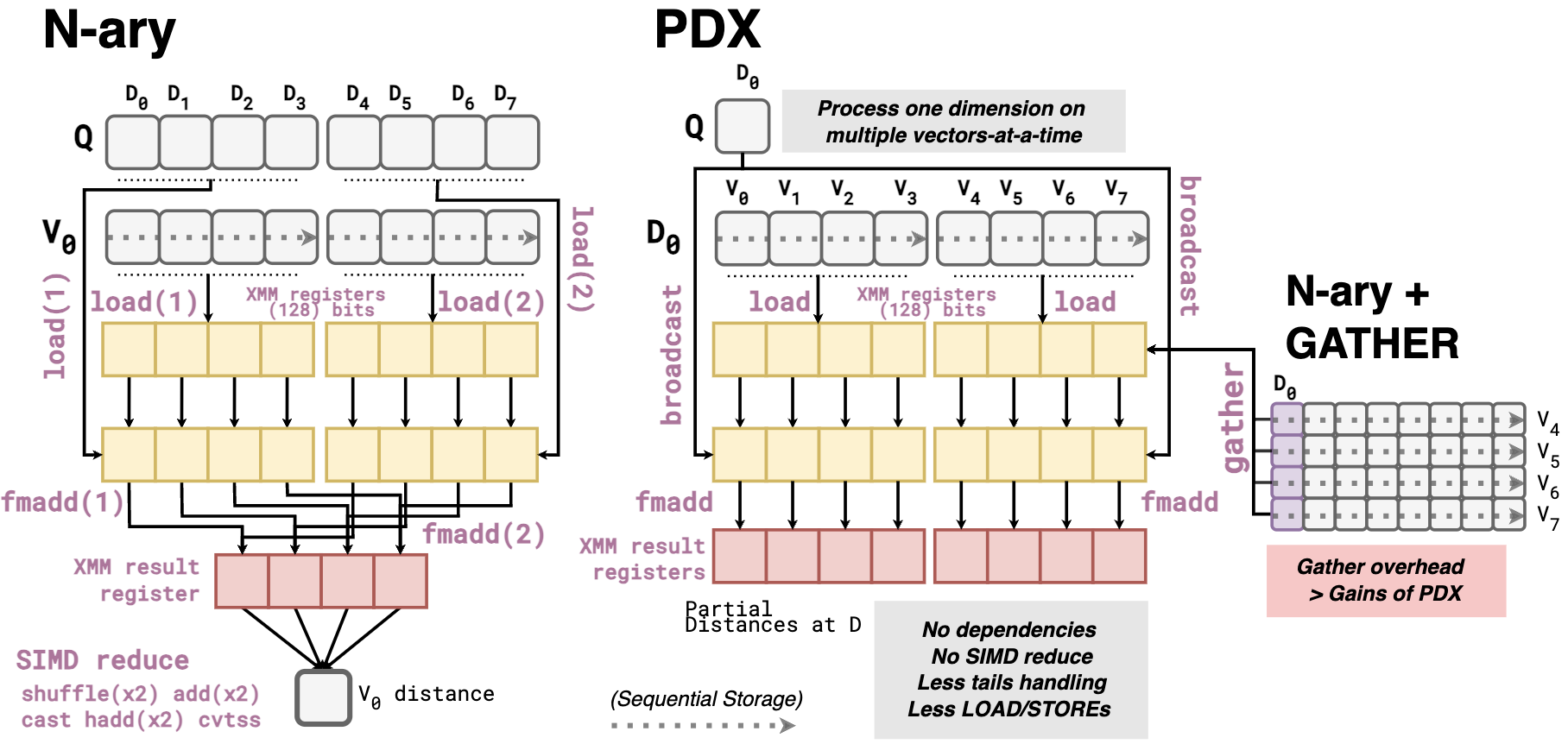}
\centering
\vspace*{-7mm}
\caption{An Inner Product calculation on the horizontal layout (N-ary) and the PDX layout with 128-bit SIMD registers. The PDX kernel does not have dependencies (the distances of {\em different} vectors are aggregated in different SIMD lanes), is unaffected by dimensionality, and avoids the register \textit{reduce} step. Constructing the PDX layout on-the-fly from the N-ary layout for calculations introduces a non-negligible overhead (N-ary + Gather), as discussed in section~\ref{sec:discussion}.}
\label{fig:simdvspdx}
\vspace*{-6mm}
\end{figure}

\vspace*{3mm}
\noindent{\bf Distance kernels that auto-vectorize.} In modern systems, distance kernels are optimized using explicit SIMD intrinsics or manual loop unrolling tailored for every major ISA, processor family, and datatype being used~\cite{milvus, faisspaper,usearch,weaviate}. This increases code size and is not future-proof, as new CPU architectures with different register widths and capabilities are in constant development. More importantly, these approaches have degraded performance in datasets with a dimensionality smaller than the available SIMD register width. The latter is not friendly towards pruning approaches as they would (ideally) inspect only a few dimensions. The PDX layout addresses these issues using distance kernels that process multiple vectors-at-a-time~\cite{superscalar,monetdb}.
 
Algorithm~\ref{algo:pdx} shows pseudo-code for the Euclidean Distance (L2), Manhattan Distance (L1), and Inner Product (IP) kernels in the PDX layout. Here, the inner loop that processes multiple vectors at-a-time is a natural fit for auto-vectorization, where the distances of different vectors are aggregated in different SIMD lanes without dependencies (see PDX in Figure~\ref{fig:simdvspdx}). SIMDizing over vectors rather than dimensions (in the default horizontal approach) is, therefore, no longer affected by dimensionality. Further, the reduction of the SIMD register that must happen at the end of every vector (see the last step of N-ary in Figure~\ref{fig:simdvspdx}) is eliminated. Leftover handling at the end of a vectorized loop (usually handled with masked instructions) also gets reduced since it happens only at the end of every dimension rather than at the end of every vector. Note that these kernels also alleviate technical software debt (absence of intrinsics, as they auto-vectorize efficiently in any architecture when vectors are {\small \tt float32}). %and are future-proof to wider SIMD registers and newer ISAs. 
The resulting code is most efficient if loops are tight enough such that the entire {\small \tt distances} array fits into the available SIMD registers (loaded once before processing and stored once after finishing the processing). From our experiments, processing \textit{64 vectors at-a-time} achieve the highest performance improvement in all major ISAs (NEON, AVX2, and AVX512). In section~\ref{sec:eval-kernels}, we present an in-depth study of the effect of using different block sizes on different architectures. Finally, in section~\ref{sec:discussion} we study the alternate approach to use Algorithm~\ref{algo:pdx} on N-ary storage, by performing an on-the-fly gather (depicted in the rightmost part of Figure~\ref{fig:simdvspdx}). % Note that to do our 64 vector at-a-time processing, the vectors within a block are sub-grouped into \textit{mini-blocks} of 64. 

\vspace*{3mm}
\noindent{\bf Metadata per block. } The notion of blocks allows for storing \textit{metadata} that can aid a search. Metadata can be defined according to the needs of an algorithm. For instance, BSA may store the variance for each dimension of a block. The latter could be used to tune the pruning process per block while being adaptive to potential distribution shifts when the variance of the vectors changes. This idea is not novel, as modern systems, such as DuckDB~\cite{duckdbpaper}, store metadata (min/max) per column in a rowgroup to perform skipping for filter predicates pushed down into the scan.   

%\vspace*{3mm}
\noindent{\bf Inserts and Updates}. Within vector databases, the typical workloads are bulk load, append, or complete rewrite (e.g., when the underlying model that produces the vectors changes). Despite updates being less common, vector systems like Weaviate~\cite{weaviate} and Milvus~\cite{milvus} support individual vector updates. PDX currently does not use compression/quantization, which makes it trivial to update-in-place if data is memory-resident. Otherwise, PDX can implement the same well-known strategies to perform updates in PAX: merge-on-read if updates are frequent or copy-on-write otherwise~\cite{paxupdates}.

\begin{figure}[t!]
\includegraphics[width=1\linewidth]{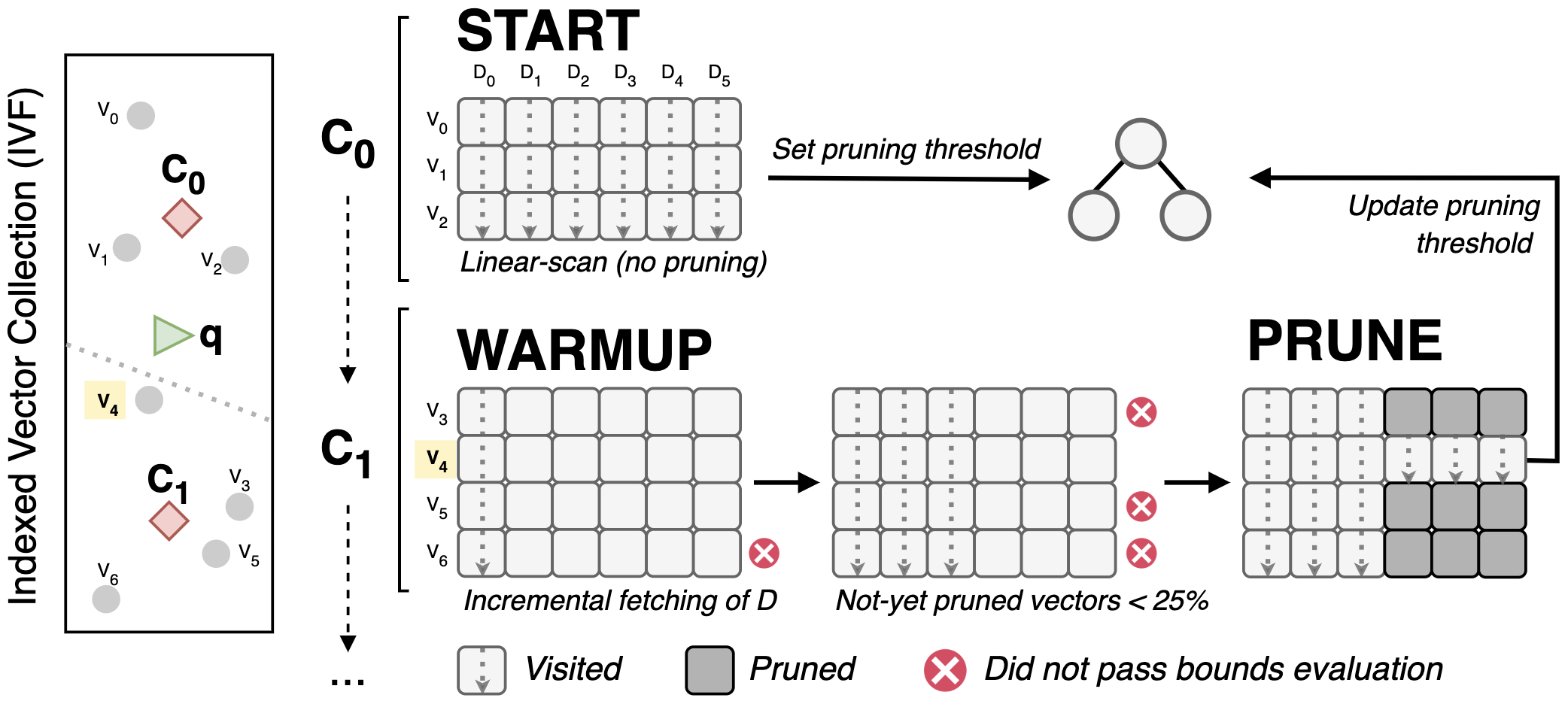}
\centering
\vspace*{-6.5mm}
\caption{The PDXearch framework within an IVF index: A search happens dimension-by-dimension per block (bucket). A linear scan is done in the first block ($C_0$ in the figure) to set a pruning threshold. In the following blocks ($C_1$), the search has two phases: WARMUP (keep scanning all vectors at incremental steps of D) and PRUNE (scan only the not-yet pruned vectors once they are few).}
\label{fig:pdxearch}
\vspace*{-4mm}
\end{figure}

\section{The PDXearch Framework}\label{sec:pdxearch}
We introduce \textbf{PDXearch}, a framework for efficient dimension-by-dimension pruned search on VSS workloads (exact or approximate) powered by the PDX layout. PDXearch is meant to be used by algorithms that prune dimensions at search time~\cite{bond, fnn, adsampling, bsa}. On PDXearch, a search happens block-by-block, propagating the threshold found in one block to the following ones. In each block, vectors are inspected dimension-at-a-time on incremental steps, and distance computations are avoided only when the amount of \textit{not-yet} pruned vectors is low. An example of PDXearch on an IVF index is presented in Figure \ref{fig:pdxearch}. It is important to note that the framework preserves the correctness and recall levels of the underlying pruning algorithm. It only changes how many dimensions to inspect at each step and when to break off computations to maximize efficiency.

\vspace*{3mm}\noindent{\bf PHASE 0: START. } When visiting the first block (start of the search), we do not yet have a threshold to use for pruning. Therefore, we compute the distances without pruning dimensions (a linear scan) to find a threshold for the later search stages. This is a small overhead as one block is only a small percentage of all data. Furthermore, trying to prune on the first block does not bring many benefits, as early in the search is when pruning is least effective. Once a threshold is defined, every following block will start in the WARMUP phase.

\vspace*{3mm}
\noindent{\bf PHASE 1: WARMUP. } In the WARMUP phase, we incrementally fetch dimensions from the block of vectors (we first fetch 2 dimensions, then the following 4 dimensions, then the next 8, and so on). In Figure \ref{fig:pdxearch}, we first fetch 1 dimension, then 2, and then the rest. At each fetching step, we calculate the partial distances, perform the pruning predicate evaluation (e.g., hypothesis testing on ADSampling, bounds evaluation in BOND), and keep track of the number of pruned vectors. In our code implementation, the pruning predicate evaluation is done in a loop separated from the distance calculations to avoid {\small \tt if-then-else} control structures (code is vectorized). 
In the WARMUP, we do not yet break-off distance computation of the pruned vectors (note in our example that $V_6$ is still visited in the second step of the WARMUP, despite being discarded as a candidate on the first step), as when the number of pruned vectors is still low, it would make the following partial distance computation slower due to random access~\cite{sancaaccess, bond}. Once the number of remaining vectors is lower than a threshold, we start the PRUNE phase. % Fetching dimensions in an incremental manner is key--recall that effective pruning has a starting point that is query- and dataset-dependent (Table \ref{tab:pruning}), after which the number of vectors pruned increases exponentially. In Section \ref{sec:adaptive} we show the benefits of using our incremental fetching instead of a fixed one.

\vspace*{3mm}\noindent{\bf PHASE 2: PRUNE. } During the PRUNE phase, only a few vectors remain as candidates. As such, we break off distance computations by skipping the already discarded vectors. We do so by maintaining the count of the remaining vectors and their positions within the block. Then, the following distance calculations are performed using this positions array. These random accesses can be optimized on Intel CPUs using a {\small\tt gather} operation. As fetching happens exponentially, the last dimension is reached quickly while still trying to prune vectors. Once the last dimension of the block is reached, we merge the remaining vectors distances on the max-heap (in our example, $V_4$ is merged into the heap, and the pruning threshold becomes tighter). We keep repeating the WARMUP and PRUNE phases for the remaining blocks. In our example, blocks represent buckets on an IVF index, but a block can also represent randomly partitioned vectors for an exact search without an index.

% Furthermore, we know that once effective pruning has started, it exponentially prunes vectors (Figure \ref{fig:sift-prune}). Accordingly, we stop increasing the dimensions fetched size and maintain the last one used in the WARMUP phase. Finally, we start the PLATEAU phase once the pruning predicate selection percentage is lower than another threshold. 

% \vspace*{3mm}\noindent{\bf PHASE 3: PLATEAU. } In the PLATEAU phase pruning can not happen anymore or does not bring substantial benefits. As such, we iterate over the rest of the dimensions of the not-pruned vectors using the positions array and evaluate their exact distance with the query. Finally, we merge these evaluated distances to the KNN max-heap and update the threshold for the next blocks. Here, the max-heap and threshold could remain unchanged. 

% \input{paper/code/pdx-bond}

\section{PDX-BOND: Our DCO optimizer}\label{sec:pdx-bond}

BOND's~\cite{bond} capabilities to speed up KNNS were limited due to its inability to fully evaluate the distance between the query and a vector until the last dimension was visited at the end of the search. Furthermore, the BOND strategy to prioritize dimension access (from biggest to smallest value in the query) is only effective if the values of the query are \textit{outliers} relative to the dimensions of the collection.  

We propose \textbf{PDX-BOND}, a follow-up to BOND~\cite{bond} that uses PDXearch, in which dimensions are accessed in terms of how far their mean is to the values in the query. Thanks to the START phase of the PDXearch framework, a tight-enough lower-bound is found early in the search (tackling the main shortcoming of BOND). To improve latency, PDX-BOND only uses the partially computed distance to determine whether a vector can be pruned. As a result, PDX-BOND is an exact DCO optimizer without extra latency to compute bounds. Contrary to ADSampling and BSA, PDX-BOND does not require any data transformation. Hence, it is a plug-and-play technique to quickly accelerate KNNS on any collection of vectors (assuming vectors are stored with PDX). 

\begin{figure}[t!]
\includegraphics[width=1\linewidth]{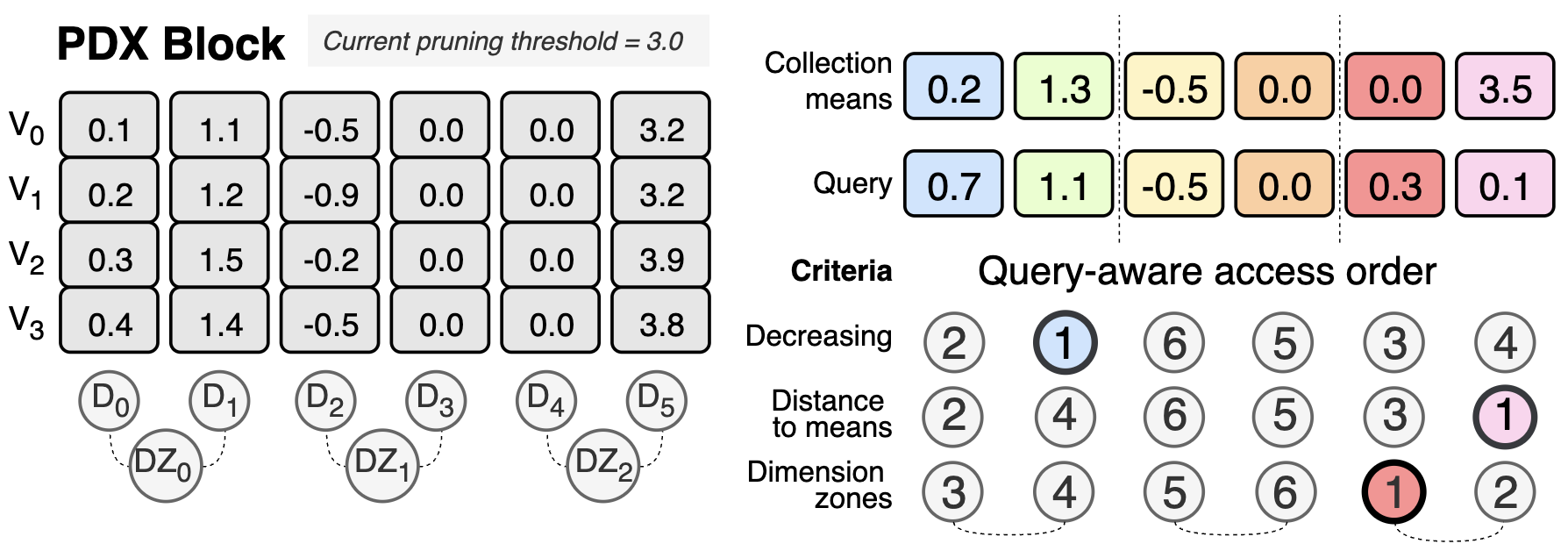}
\centering
\vspace*{-6mm}
\caption{Example of three query-aware access order criteria: i) Decreasing: the dimension with the highest query value is accessed first ($D_1$ in the figure), ii) Distance to means: the dimension of the query with the largest distance to the collection means is accessed first ($D_5$), iii) Dimension zones: the subset of consecutive dimensions with the highest distance to the collection means are accessed first ($DZ_2$ in the figure).}
\label{fig:dz}
\vspace*{-4mm}
\end{figure}

Figure \ref{fig:dz} shows an example of how different query-aware strategies to determine the order in which dimensions are visited lead to different access patterns to the collection. In this example, we use L1 as the distance metric. The decreasing criteria (used in BOND \cite{bond}) solely leverage the query values, accessing first the dimension with the highest query value ($D_1$). As $D_1$ is not enough to reach the pruning threshold in any vector, we have to explore the block dimensions further. However, by leveraging the block statistics (mean), one can prioritize $D_5$ instead, as its mean in the collection is the farthest from the query value. In this case, $D_5$ is the only dimension needed to visit to discard all the vectors.

A high pruning power cannot be achieved without prioritizing dimensions~\cite{adsampling}. However, both the decreasing and distance-to-means criteria affect the memory access efficiency of the algorithm: by making more jumps and accessing shorter memory stretches, {\em automatic prefetching} by the Memory Management Unit (MMU) of a CPU, that is triggered on detecting sequential access, becomes less efficient. This can make a sequential access of dimensions more performant than a query-aware order despite its lower pruning power in certain microarchitectures. Note that we cannot re-order the collection as the access order depends on the incoming query. To make up for that, when blocks are small (as buckets in an IVF index will be), we divide dimensions into \textit{dimension zones}. Dimension zones are multiple dimensions residing sequentially in storage, providing longer sequential stretches. When a query arrives, we rank each zone based on the "distance-to-means" criteria of its dimensions and first visit the most promising zones. In Figure \ref{fig:dz}, the most promising dimension-zone is $DZ_2$. As such, we visit $D_4$ and then $D_5$ to maximize sequential access while still visiting the most promising dimensions for pruning. Note that the fetching size of PDXearch still rules the amount of dimensions visited. 

\section{Evaluation}\label{sec:eval}
We now experimentally evaluate the following research questions:
\begin{questions}
    \item How do the auto-vectorized distance kernels on the PDX layout compare to the current best SIMD kernels for the major ISAs (AVX512, AVX2, and NEON)?\label{q:kernels}
    \item How does the performance of PDXearch compare against a pruned search in the horizontal vector-by-vector layout?\label{q:pdxearch}
    \item Is a search that prunes vectors always faster than a linear scan on modern vector systems?\label{q:isfaster}
    \item How does PDX-BOND compare to ADSampling and BSA in terms of its pruning power and speed?\label{q:bond}
    \item What is the performance of an exact search on the PDX layout, and how does it compare to other systems (USearch, Milvus, FAISS, and, Scikit-Learn)?\label{q:exact}
\end{questions}

\begin{table}[t!]
\renewcommand{\tabcolsep}{1.5pt}
\centering
\caption{Hardware Platforms Used}
\vspace*{-4mm}
{\small
\resizebox{0.99\columnwidth}{!}{%
\begin{tabular}{lccccccc}
\hline
{\bf Architecture}  & {\bf Scalar ISA} & {\bf Best SIMD ISA} & {\bf CPU Model} & {\bf Freq.}   \\ \hline
Intel Sapphire Rapids  & x86\_64 & AVX512 & Gold 6455B & 3.9 GHz \\ 
AMD Zen4  & x86\_64 & AVX512 & Ryzen9 7900 & 3.7 GHz \\ 
AMD Zen3  & x86\_64 & AVX2 (256-bits) & EPYC 7R13 & 3.6 GHz \\ 
%Apple M1  & ARM64 & NEON (128-bits) & Apple M1 & 3.2 GHz \\ \hline
AWS Graviton4  & ARM64 & NEON (128-bits) & Neoverse-V2 & 2.7 GHz  \\ \hline
\end{tabular}}
}
\label{tab:machine_details}
\vspace*{-6mm}
\end{table}

% \begin{figure*}[t!]
% \includegraphics[width=1\linewidth]{paper/images/pdx_kernels.png}
% \centering
% \vspace*{-8mm}
% \caption{Speedup of the L2 Euclidean Distance calculation of auto-vectorized PDX vs. horizontal kernels with explicit SIMD intrinsics on vector collections of different sizes and dimensionality. White space represent  1.0 $\leq$ speedup $<$ 1.2. Over all scenarios, PDX is on average 1.4-1.7x faster on the various architectures. Crucial for pruning algorithms is that with few ($\leq$ 32) dimensions, PDX is {\em much} faster than horizontal kernels and gives a 3-8x speedup (darker green vertical patterns on the left). }
% \label{fig:kernels}
% \vspace*{-3mm}
% \end{figure*}

\subsection{Setup}\label{sec:setting}
We start by evaluating the performance of the auto-vectorized distance kernels on the PDX layout in subsection \ref{sec:eval-kernels} ~\ref{q:kernels}. For this, we used the architectures presented in Table ~\ref{tab:machine_details}. These cover the major ISAs (AVX512, AVX2, NEON) and popular processors (Intel, AMD, and Graviton). Here, we compare the auto-vectorization of our kernels produced by LLVM (C++) against the state-of-the-art SIMD distance kernels ~\cite{simsimd,faisspaper,milvus} of three distance metrics: L2-euclidean, L1-manhattan, and Inner Product. 
In subsection \ref{sec:eval-pdxearch}, we experimentally evaluate the PDXearch framework by performing queries on an IVF index on our presented datasets, optimizing the distance calculation with ADSampling. Here, we compare the query throughput of the algorithm when using PDXearch against the search on the horizontal layout~\ref{q:pdxearch}. FAISS~\cite{faisspaper} and Milvus~\cite{milvus} IVF indexes are used as baselines~\ref{q:isfaster}. All the IVF indexes are constructed using the same parameters.   
Next, in subsection \ref{sec:eval-pdxbond}, we evaluate PDX-BOND against ADSampling and BSA, also within an IVF index search~\ref{q:bond}. Finally, in subsection \ref{sec:eval-exact}, we evaluate the end-to-end performance of PDX-BOND on exact queries against three vector systems (Milvus, FAISS, and USearch~\cite{usearch})~\ref{q:exact}.

\vspace*{3mm}\noindent{\bf Parameters. } We set the parameter $\Delta d$ of ADSampling and BSA to 32, as recommended by the authors. On the datasets with less than 128 dimensions, we set $\Delta d = D/4$, as using $\Delta d = 32$ on these datasets would be unfair. The $\epsilon_0$ parameter on ADSampling, which tunes the recall, is set to 2.1, as recommended by the authors. The multiplier $m$ parameter on BSA is set to achieve a recall similar to the one of ADSampling. Furthermore, we adopt the dual-block layout proposed by ADSampling, splitting the vectors into two blocks at $\Delta d$. For this experiment, we show a \textit{Recall vs QPS} curve. The recall is tuned with the \textit{nprobe} parameter of the IVF index, which determines how many buckets are visited. A higher nprobe increases recall and reduces QPS as more vectors are explored. Finally, on the PDX version of the algorithms, we set to 20\% the selection percentage threshold to advance through the PRUNE phase. An in-depth study of this parameter is presented in subsection \ref{sec:eval-select}. 

\vspace*{3mm}
\noindent{\bf Implementation and Hardware. } PDXearch was implemented in C++ and compiled with the {\small \tt Release} (CMake) and {\small \tt -O3} compiler flags alongside the recommended {\small \tt -march} or {\small \tt -mtune} for each architecture. For ADSampling and BSA algorithms, we used an optimized version of the original implementation that improves the performance of the query transformation phase. These implementations were adapted to work with PDXearch. Furthermore, in our codebase, we also SIMDized the original implementation of ADSampling to compare it fairly to PDXearch. We used the available software of Milvus, FAISS, and USearch. We used machines with 64GB of RAM (enough to fit every dataset in memory) and Ubuntu 24.01 as OS. We deactivated any multi-threading capabilities in all benchmarks to compare raw performance between different approaches without introducing possible parallelization artifacts.

% For our evaluations within an IVF index, we decided to not take into consideration the time to find the nearest buckets to the query. We removed this time from our QPS metric since we want to measure the  effectiveness of pruning methods. Furthermore, there are multiple ways to optimize the finding of the most promising buckets (e.g., creating an HNSW index on the centroids~\cite{spann}). Depending on which is chosen, the effectiveness of the pruning methods could be overshadowed, especially when the amount of probed vectors is magnitudes lower than the number of buckets in the index. 

\begin{table}[t]
%\small
\centering
\caption{Speedup of the distance calculation (L2, IP and L1) of auto-vectorized PDX vs. horizontal kernels with explicit SIMD intrinsics on a variety of {\small \tt float32} vector collections. PDX is, on average, 2.0x faster across architectures. Crucial for pruning algorithms is that with few ($\leq$ 32) dimensions, PDX is {\em much} faster than horizontal kernels (1.5-7.4x speedup).}
\vspace*{-3mm}
\label{tab:kernels}
\resizebox{1.00\columnwidth}{!}{%
\begin{tabular}{|l|cccc|cccc|cccc|}
\hline
\multicolumn{1}{|c|}{\multirow{2}{*}{\textbf{Arch.}}} & \multicolumn{4}{c|}{\textbf{\begin{tabular}[c]{@{}c@{}}Euclidean\\Distance L2\end{tabular}}}                               & \multicolumn{4}{c|}{\textbf{\begin{tabular}[c]{@{}c@{}}Inner Product\\ IP\end{tabular}}}                                     & \multicolumn{4}{c|}{\textbf{\begin{tabular}[c]{@{}c@{}}Manhattan\\Distance L1\end{tabular}}}                               \\ \cline{2-13}  
\multicolumn{1}{|c|}{}                                                       & \multicolumn{1}{c|}{\footnotesize D=8} & \multicolumn{1}{c|}{\footnotesize D=16,32} & \multicolumn{1}{c|}{\footnotesize D$>$32} & \footnotesize All & \multicolumn{1}{c|}{\footnotesize D=8} & \multicolumn{1}{c|}{\footnotesize D=16,32} & \multicolumn{1}{c|}{\footnotesize D$>$32} & \footnotesize All & \multicolumn{1}{c|}{\footnotesize D=8} & \multicolumn{1}{c|}{\footnotesize D=16,32} & \multicolumn{1}{c|}{\footnotesize D$>$32} & \footnotesize All \\ \hline 
\begin{tabular}[c]{@{}l@{}}Intel S.R. \\ AVX512\end{tabular} & \multicolumn{1}{c|}{\cellcolor[HTML]{C3DEBF}5.8} & \multicolumn{1}{c|}{\cellcolor[HTML]{C3DEBF}2.4}                           & \multicolumn{1}{c|}{1.3}               & 1.8 & \multicolumn{1}{c|}{\cellcolor[HTML]{C3DEBF}5.6} & \multicolumn{1}{c|}{\cellcolor[HTML]{C3DEBF}2.4}                           & \multicolumn{1}{c|}{1.2}               & 1.7 & \multicolumn{1}{c|}{\cellcolor[HTML]{C3DEBF}5.3} & \multicolumn{1}{c|}{\cellcolor[HTML]{C3DEBF}2.5}                           & \multicolumn{1}{c|}{1.2}               & 1.7 \\ 
\begin{tabular}[c]{@{}l@{}}Zen 4\\ AVX512\end{tabular}             & \multicolumn{1}{c|}{\cellcolor[HTML]{C3DEBF}7.4} & \multicolumn{1}{c|}{\cellcolor[HTML]{C3DEBF}2.7}                           & \multicolumn{1}{c|}{1.4}               & 2.0 & \multicolumn{1}{c|}{\cellcolor[HTML]{C3DEBF}6.6} & \multicolumn{1}{c|}{\cellcolor[HTML]{C3DEBF}2.5}                           & \multicolumn{1}{c|}{1.4}               & 2.0 & \multicolumn{1}{c|}{\cellcolor[HTML]{C3DEBF}6.7} & \multicolumn{1}{c|}{\cellcolor[HTML]{C3DEBF}2.8}                           & \multicolumn{1}{c|}{1.4}               & 2.0 \\ 
\begin{tabular}[c]{@{}l@{}}Zen 3\\ AVX2\end{tabular}               & \multicolumn{1}{c|}{\cellcolor[HTML]{C3DEBF}6.2} & \multicolumn{1}{c|}{\cellcolor[HTML]{C3DEBF}3.3}                           & \multicolumn{1}{c|}{1.7}               & 2.3 & \multicolumn{1}{c|}{\cellcolor[HTML]{C3DEBF}5.9} & \multicolumn{1}{c|}{\cellcolor[HTML]{C3DEBF}3.1}                           & \multicolumn{1}{c|}{1.5}               & 2.1 & \multicolumn{1}{c|}{\cellcolor[HTML]{C3DEBF}7.4} & \multicolumn{1}{c|}{\cellcolor[HTML]{C3DEBF}3.5}                           & \multicolumn{1}{c|}{1.4}               & 2.2 \\ 
\begin{tabular}[c]{@{}l@{}}Grav. 4\\ NEON\end{tabular}          & \multicolumn{1}{c|}{\cellcolor[HTML]{C3DEBF}2.7} & \multicolumn{1}{c|}{1.5}                           & \multicolumn{1}{c|}{1.8}               & 1.8 & \multicolumn{1}{c|}{\cellcolor[HTML]{C3DEBF}3.1} & \multicolumn{1}{c|}{1.8}                           & \multicolumn{1}{c|}{1.9}               & 2.0 & \multicolumn{1}{c|}{\cellcolor[HTML]{C3DEBF}2.6} & \multicolumn{1}{c|}{1.5}                           & \multicolumn{1}{c|}{1.9}               & 1.9 \\ \hline \hline
\textbf{Avg.}                       & \multicolumn{1}{c|}{5.5}                                            & \multicolumn{1}{c|}{2.5}                                                                         & \multicolumn{1}{c|}{1.5}                                                          & 2.0 & \multicolumn{1}{c|}{5.3}                                            & \multicolumn{1}{c|}{2.4}                                                                         & \multicolumn{1}{c|}{1.5}                                                          & 2.0 & \multicolumn{1}{c|}{5.5}                                            & \multicolumn{1}{c|}{2.6}                                                                         & \multicolumn{1}{c|}{1.5}                                                          & 2.0 \\ \hline
\end{tabular}
}\vspace*{-2mm}
\end{table}

\subsection{Distance Kernels on the PDX Layout}\label{sec:eval-kernels}
We measured the raw performance of three distance kernels (L2, L1, and Inner Product) in vector collections of different sizes ({from 64 to 131K}) and dimensionalities ({8, 16, 32, 64, 128, 192, 256, 384, 512, 768, 1024, 1536, 2K, 4K, 8K}) consisting of standardly distributed randomly generated {\small \tt float32}. Here, we do \textit{not} perform a KNNS. The only work measured is the distance calculation between one query and the entire collection in the N-ary (horizontal) and the PDX layout. Note that in PDX, we process blocks of 64 vectors at-a-time. The L2 and IP N-ary kernels were taken from SimSIMD~\cite{simsimd} (used by USearch~\cite{usearch}), and the L1 kernel was taken from FAISS~\cite{faisspaper}. 

The auto-vectorized PDX kernels perform never worse and generally better than the horizontal kernel with SIMD intrinsics in all scenarios across all architectures~\ref{q:kernels}. Table~\ref{tab:kernels} highlights the average speedup at four granularities of dimensionalities: D$=$8, 8$<$D$\leq$32, D$>$32 and throughout all values of D. The L2 PDX kernel outperforms the horizontal explicit SIMD kernels when $D \leq 32$ ($\approx$4-10x faster in Zen4, $\approx$3-9x faster in Zen3, $\approx$3-8x faster in Intel and $\approx$2-3x faster in Graviton4 depending on the size of the collection). Similar speedups are obtained in the IP and L1 kernels. 

These speedup benefits at low D are due to the kernels in the PDX layout pipelining loops over the number of vectors instead of the number of dimensions (recall Figure~\ref{fig:simdvspdx}), thus always fully utilizing the SIMD registers. In contrast, in the horizontal kernels, the entire vector fits in one register when D is low (in AVX512, not even one full register is utilized when D=8). Note that efficiency on distance computations with limited dimensions ($\leq$ 32) is crucial for pruning algorithms, which only fully scan the first dimensions and then break off full computation thanks to pruning.

The speedups are not limited to these cases of low D, as all kernels are 1.5x faster, averaging all architectures if we do not consider the low ($\leq$ 32) dimensionalities. These benefits are thanks to eliminating the SIMD register reduction step at the end of each vector, the absence of dependencies, and better loop pipelining as the distances of different vectors are aggregated in different SIMD lanes (recall Figure~\ref{fig:simdvspdx}). There are also performance benefits when $D \geq 4096 $ (up to 2.1x faster in Zen4, 2.1x in Zen3, and 2.6x faster in Graviton4 for the L2 kernel). These are due to the tight loops (64 at a time) avoiding LOAD/STORE instructions from/to the {\small \tt distances} array at every iteration of the outer loop, as the entire array can fit in the available registers (red registers in Figure~\ref{fig:simdvspdx}). When we are not memory-bound (data fits in L2), the PDX kernel is consistently 1.5-2.0x faster. % hinting that on \textit{quantized} vectors (not done here yet), the PDX kernels could find even more benefits.

The differences between architectures are due to their different set of instructions with different latencies, SIMD register widths, and cache sizes. For instance, in Graviton4, the gains of PDX kernels at $D\leq32$ are less evident compared to the gains in Intel/Zen4/Zen3 due to the register width in NEON being 128 bits (fitting four float32), hence using two full SIMD registers at D=8. In contrast, in Zen4 (512-bit registers), not even one register can be filled at D=8, which results in masked instructions and, thus, under-utilization of the registers.

\begin{table}[t!]
\centering
\caption{Average speedups (higher is better) of the L2 PDX kernel against the N-ary kernel using different block-sizes for the vectors on the PDX layout. A block size of 64 achieves the highest speedups across all architectures}
\vspace*{-3mm}
\label{tab:block}
\resizebox{0.73\columnwidth}{!}{%
\begin{tabular}{lcccccc}
\hline
\multicolumn{1}{c}{}                                        & \multicolumn{6}{c}{\textbf{PDX Block Size}}                                                                                                                                                                                           \\ \cline{2-7} 
\multicolumn{1}{c}{\multirow{-2}{*}{\textbf{Architecture}}} & \multicolumn{1}{l}{\textbf{16}} & \multicolumn{1}{l}{\textbf{32}} & \multicolumn{1}{l}{\textbf{64}}                 & \multicolumn{1}{l}{\textbf{128}} & \multicolumn{1}{l}{\textbf{256}} & \multicolumn{1}{l}{\textbf{512}} \\ \hline
Intel S.R. (AVX512)                                                & \multicolumn{1}{c}{1.5}         & \multicolumn{1}{c}{1.6}         & \multicolumn{1}{c}{\cellcolor[HTML]{C3DEBF}1.8} & \multicolumn{1}{c}{1.8}          & \multicolumn{1}{c}{1.7}          & 1.6                                  \\ 
Zen 4 (AVX512)                                                         & \multicolumn{1}{c}{1.6}         & \multicolumn{1}{c}{1.9}         & \multicolumn{1}{c}{\cellcolor[HTML]{C3DEBF}2.0} & \multicolumn{1}{c}{2.0}          & \multicolumn{1}{c}{1.8}          & 1.5                             \\ 
Zen 3 (AVX2)                                                         & \multicolumn{1}{c}{1.7}         & \multicolumn{1}{c}{2.2}         & \multicolumn{1}{c}{\cellcolor[HTML]{C3DEBF}2.3} & \multicolumn{1}{c}{2.0}          & \multicolumn{1}{c}{1.5}          & 1.6                               \\
Graviton 4 (NEON)                                                    & \multicolumn{1}{c}{1.6}         & \multicolumn{1}{c}{1.7}         & \multicolumn{1}{c}{\cellcolor[HTML]{C3DEBF}1.8} & \multicolumn{1}{c}{1.5}          & \multicolumn{1}{c}{1.4}          & 1.4                               \\ \hline
\end{tabular}
}\vspace*{-5mm}
\end{table}

\vspace*{3mm}\noindent{\bf Study on PDX block sizes}. Table~\ref{tab:block} shows how different block sizes affect the speedup of the PDX L2 distance kernel over the N-ary kernel. Blocks of 64 vectors perform best across all architectures as the SIMD registers holding the resulting distances are recycled through the entire block processing without intermediate LOAD/STORE instructions. When increasing the block size beyond 64, NEON and AVX2 performance is diminished as this effect is not achieved, leading to intermediate LOAD/STORE instructions. The latter also happens when increasing the block size beyond 128 in AVX512 (Sapphire Rapids and Zen4). On the other hand, reducing the block size hinders performance across all architectures due to the under-utilization of the available registers. Note that gains are still present at any block size due to the advantage of PDX kernels at lower dimensionalities and the elimination of the reduction step.

\begin{figure*}[t!]
\includegraphics[width=1\linewidth]{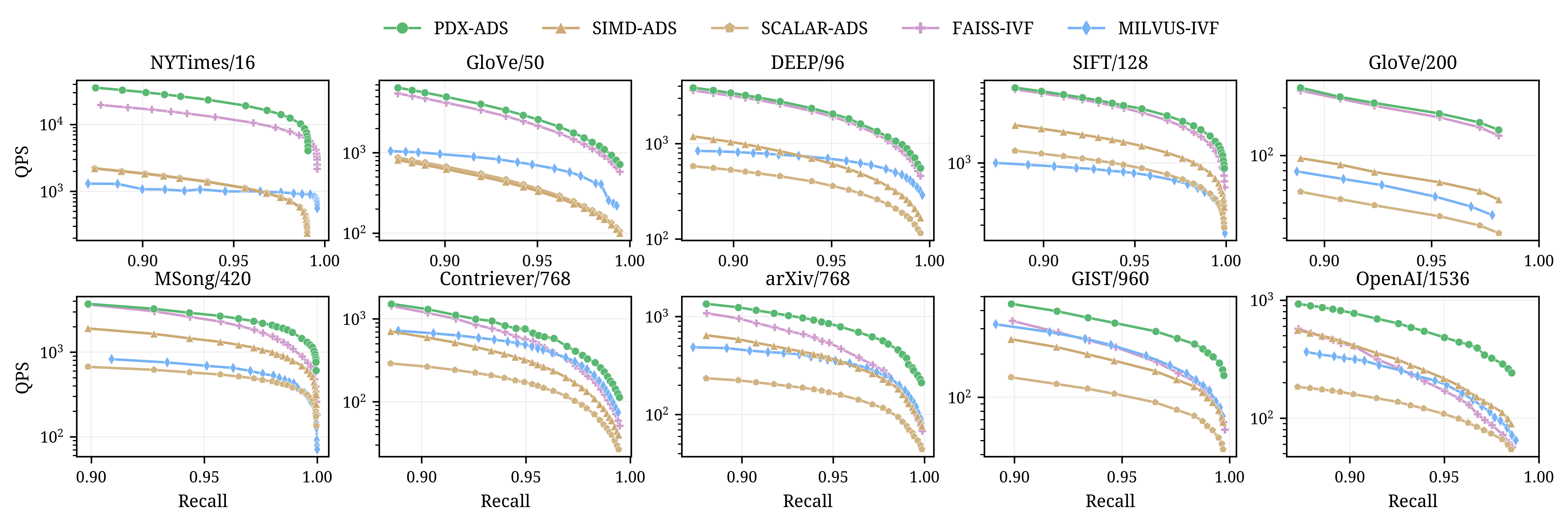}
\centering
\vspace*{-8mm}
\caption{QPS on an IVF index search with KNN=10 in the AMD Zen4 architecture. Three versions of ADSampling are compared: vanilla (scalar), SIMDized, and using auto-vectorized PDXearch. Only the latter  is always superior to the baselines, especially in the high dimensional datasets of the bottom row (3.1x and 3.5x faster than FAISS and Milvus, respectively). Contrary to the horizontal versions of ADSampling, PDX-ADS is never worse than FAISS and Milvus.}
\label{fig:eval-pdxearch}
\vspace*{-3mm}
\end{figure*}

\subsection{PDXearch Framework}\label{sec:eval-pdxearch}
Figure \ref{fig:eval-pdxearch} shows the higher QPS obtained by ADSampling paired with PDXearch (PDX-ADS) in contrast to the SIMDized and vanilla versions of ADSampling on the horizontal layout (SIMD-ADS and SCALAR-ADS)~\ref{q:pdxearch}, while also being faster than both FAISS and Milvus, especially in the vector collections stemming from LLMs (arXiv/768, OpenAI/1536). Here, all the approaches perform queries within a space-partitioning index (IVF) using L2 as the distance metric. This index is referred to as {\small\tt IVF\_FLAT} in the FAISS and Milvus documentation. In this experiment, the recall is controlled by the \textit{nprobe} parameter of the IVF index (at higher recalls, more buckets/blocks are accessed). The highest \textit{nprobe} we used is 512. FAISS and Milvus are doing a linear scan of the IVF buckets. A linear scan is a search (exact or approximate) that fully explores vectors without pruning dimensions. 

Averaging all datasets at the highest recall, PDX-ADS achieves a 4.6x speedup against SIMD-ADS and a 2.2x and 3.5x speedup over FAISS and Milvus {\small\tt IVF\_FLAT} index, respectively. Higher gains are seen on datasets of higher dimensionalities and when targeting high recalls as the pruning strategy can avoid more work. For instance, in OpenAI/1536 and arXiv/768, PDX-ADS is 4.3x and 3.1x faster than FAISS. FAISS cuts close to our pruned search in PDX at low recalls (when the number of visited buckets is low) and for datasets of lower dimensionalities (top row of Figure~\ref{fig:pdxearch}).  

Note how only with PDXearch can a pruned search surpass the performance of FAISS and Milvus~\ref{q:isfaster}, as these are faster than SIMD-ADS (2.7x and 1.4x resp. on average). The latter is due to the pruned vector-by-vector search having few opportunities to parallelize work as the distance must be evaluated every 32 dimensions, incurring 4x more branch mispredictions that stall the CPU. On low-dimensional datasets with low pruning power (NYTimes/16, GloVe/50), SIMD-ADS struggles more due to being unable to use the available registers at each step fully. 

While FAISS and Milvus are superior to the horizontal pruning algorithms in Zen4, the same is not true in the other microarchitectures, where the SIMD performance of ADSampling is not outperformed so heavily. For instance, in Intel Sapphire Rapids, SIMD-ADS is, on average, 2.0x faster than FAISS, and PDX-ADS comes on top, being 3.5x faster than FAISS and 5.3x faster when $D \geq 420$, with a remarkable 7.2x speedup on the OpenAI dataset. Thanks to PDX, pruning methods become the clear winners regardless of the architecture and dimensionality of the data. In Section \ref{sec:eval-across}, we present a summary of our results across architectures.

It is important to mention that all competitors share the same IVF index created by FAISS (identical buckets), except Milvus. Milvus uses its own algorithm to train and build the index. As such, these results are not evidence of better raw performance, as Milvus could be evaluating fewer vectors (or vice-versa). Furthermore, Milvus uses a \textit{dynamic batching} mechanism that executes queries at intervals ($\approx$1ms); thus, the asymptotic behavior at $\approx10^3$ QPS.

\begin{figure}[t]
\includegraphics[width=0.90\linewidth]{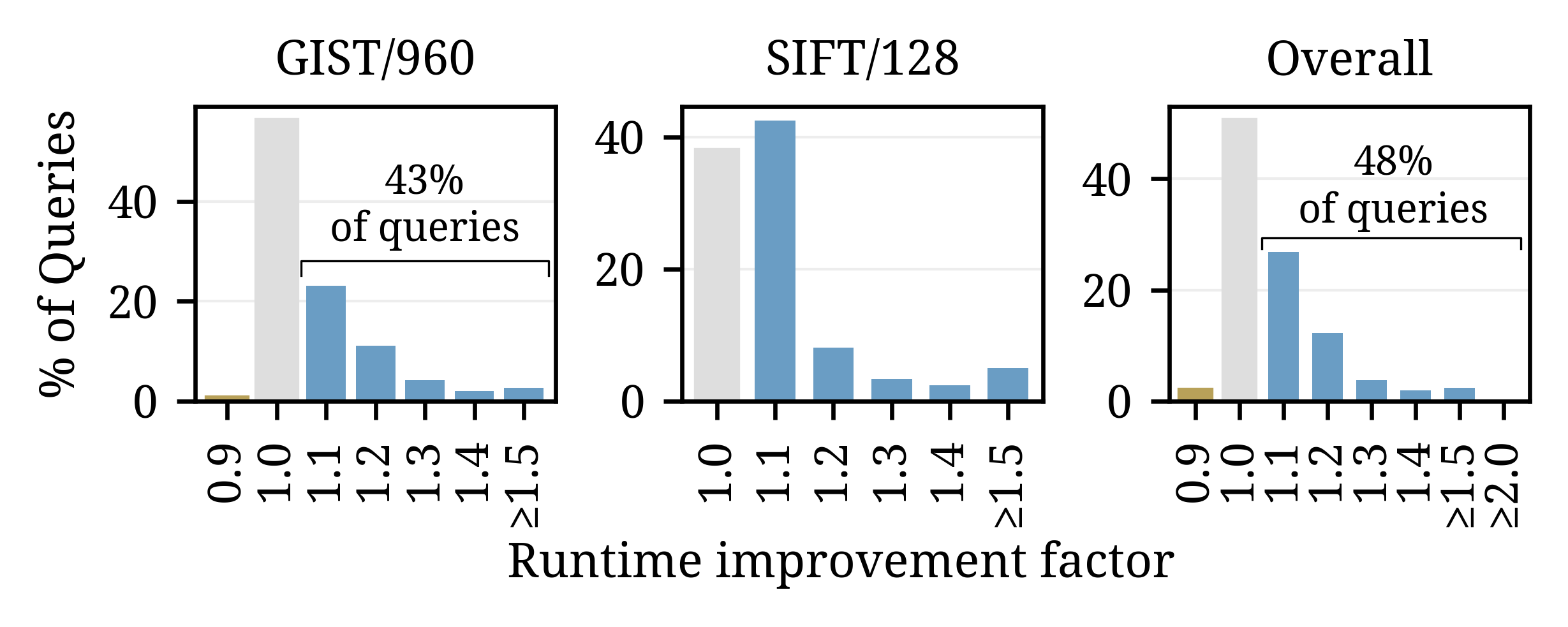}
\centering
\vspace*{-4mm}
\caption{The effect of using adaptive steps on PDXearch vs. a fixed one. On GIST--dataset in which $\Delta d$ was determined with an exhaustive parameter search\cite{adsampling}, 43\% of queries improved their runtime, 3\% being $\geq$1.5x faster.}
\label{fig:adaptive}
\vspace*{-8mm}
\end{figure}

\vspace*{3mm} 
\noindent{\bf Adaptive vs fixed steps}. Figure \ref{fig:adaptive} shows the frequency of runtime improvements on individual queries by using our proposed incremental steps vs a fixed $\Delta$d=32 on the Intel CPU. The only difference between both experiments is the number of dimensions explored at each step in the PDXearch algorithm. On GIST, 43\% of queries see a speedup, with 3\% being $\geq$ 1.5x faster and less than 1\% seeing a 10\% slower performance (queries in which the needed granularity for most vectors is exactly 32). Remarkably, the adaptive threshold finds gains even on GIST/960 (dataset used in \cite{adsampling} to determine the value of $\Delta$d=32 by doing an exhaustive parameter search). These speedups happen when only 4, 8, or 16 dimensions are enough to prune, especially useful at late stages in a search when most vectors do not make it into the KNN. Also, in PDX, bounds are evaluated \textit{much} faster, making these evaluations before the 32nd dimensions have little overhead on the runtime. Across all datasets, almost half of queries benefit from having an adaptive threshold, which also alleviates users to find this parameter in the first place. % While a new $\Delta d$ can be trained for this dataset, a $\Delta d$ smaller than 32 will always lose to the PDX kernels (recall our distance kernels experiment on Section \ref{sec:eval-kernels}). 

\vspace*{3mm} 
\noindent{\bf PDX vs N-ary disabling vectorization}. We performed an experiment disabling the compiler vectorization ({\small \tt -fno-vectorize} \linebreak {\small \tt -fno-tree-vectorize} {\small \tt -fno-slp-vectorize} flags). Here, PDX-ADS is still faster by 1.8x on average, mainly due to better data access patterns, fewer branch mispredictions, and better cache utilization. In this experiment, the CPU profiling of AMD shows that PDXearch executes twice as many instructions per cycle, incurs 4x fewer branch mispredictions (retiring 3x fewer branch instructions), and hits twice as much the L2 cache when compared to the vector-by-vector search~\ref{q:pdxearch}. Note that PDXearch code is branchless, contrary to the horizontal versions that interleave the distance calculations and the bounds evaluation.

\begin{figure*}[t!]
\includegraphics[width=1\linewidth]{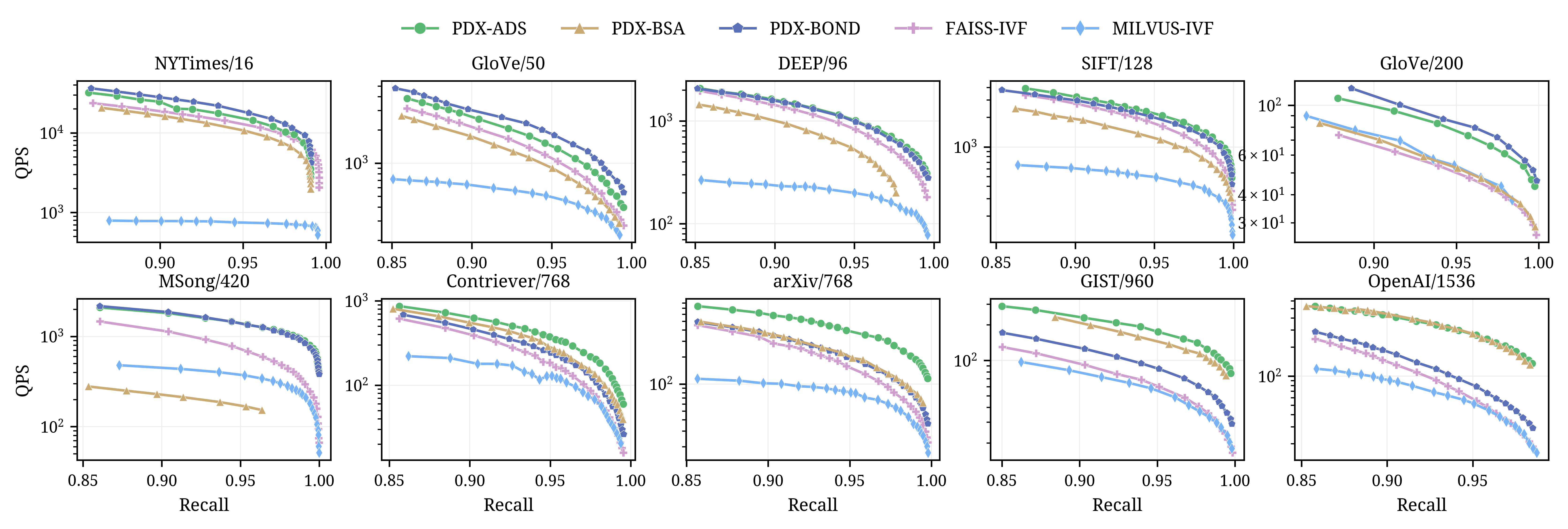}
\centering
\vspace*{-8mm}
\caption{QPS on an IVF index search with KNN=10 in the Intel architecture. All pruning algorithms are compared in the PDX layout. PDX-BOND is 2.1x faster than FAISS, with a performance comparable to ADSampling (1.7x and 1.2x slower at 0.99 and 0.90 recall resp.). The performance of PDX-BOND is still remarkable, given that it does not trade-off recall and works on raw vectors without preprocessing.}
\label{fig:eval-pdxbond}
\vspace*{-3mm}
\end{figure*}

\begin{table}[t]
\centering
\caption{Best, $p^{50}$, $p^{25}$, and worst pruning behaviors of PDX-BOND when trying to prune at every dimension. The darker area indicates the portion of values not pruned at that dimension (x-axis). The number inside the plot indicates the total percentage of avoided values. }
\vspace*{-3mm}
\label{tab:pruning-pdxbond}
\resizebox{1\columnwidth}{!}{%
\begin{tabular}{l|l|l|l|l|l|l|l|l|l|}\cline{2-9}
              & \multicolumn{8}{c|}{Datasets}                                                      \\ \hline
\multicolumn{1}{|l|}{Pruning} & \rotatebox[origin=c]{90}{\raisebox{-1.5\normalbaselineskip}[0pt][0pt]{GIST/960}} & \rotatebox[origin=c]{90}{\raisebox{-1.5\normalbaselineskip}[0pt][0pt]{ MSong/420 }} & \rotatebox[origin=c]{90}{\raisebox{-1.5\normalbaselineskip}[0pt][0pt]{NYTimes/16}} & \rotatebox[origin=c]{90}{\raisebox{-1.5\normalbaselineskip}[0pt][0pt]{GloVe/50}} & \rotatebox[origin=c]{90}{\raisebox{-1.5\normalbaselineskip}[0pt][0pt]{DEEP/96}} & \rotatebox[origin=c]{90}{\raisebox{-1.5\normalbaselineskip}[0pt][0pt]{Contriever/786}} & \rotatebox[origin=c]{90}{\raisebox{-1.5\normalbaselineskip}[0pt][0pt]{OpenAI/1536}} & \rotatebox[origin=c]{90}{\raisebox{-1.5\normalbaselineskip}[0pt][0pt]{SIFT/128}} \\ \hline\hline
\multicolumn{1}{|l|}{Best}  &  \raisebox{-.4\height}{\includegraphics[width=0.055\textwidth]{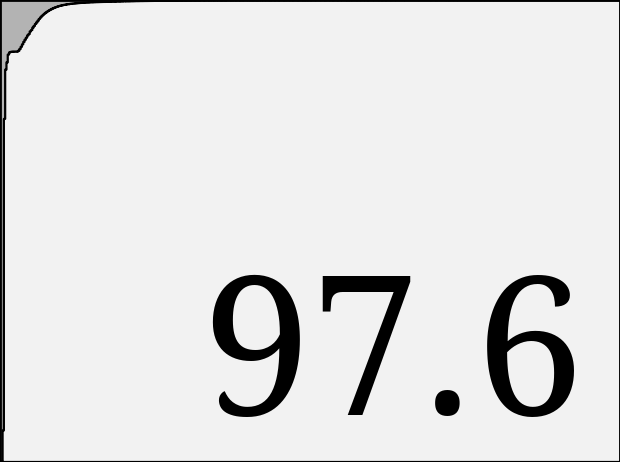}}  & \raisebox{-.4\height}{\includegraphics[width=0.055\textwidth]{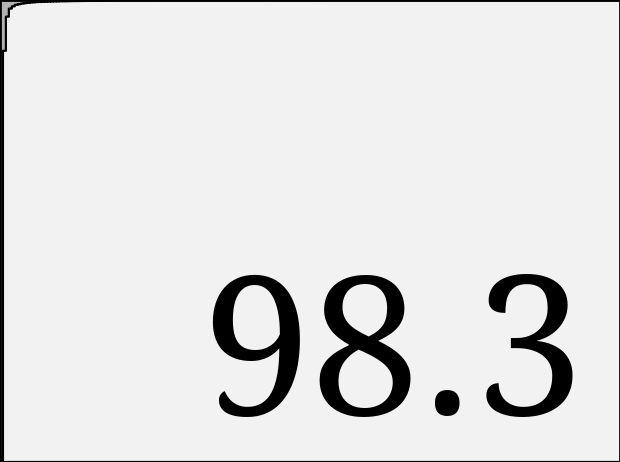}} & \raisebox{-.4\height}{\includegraphics[width=0.055\textwidth]{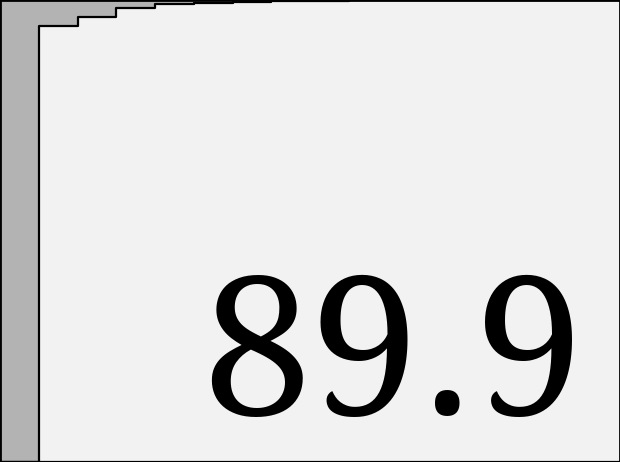}} & \raisebox{-.4\height}{\includegraphics[width=0.055\textwidth]{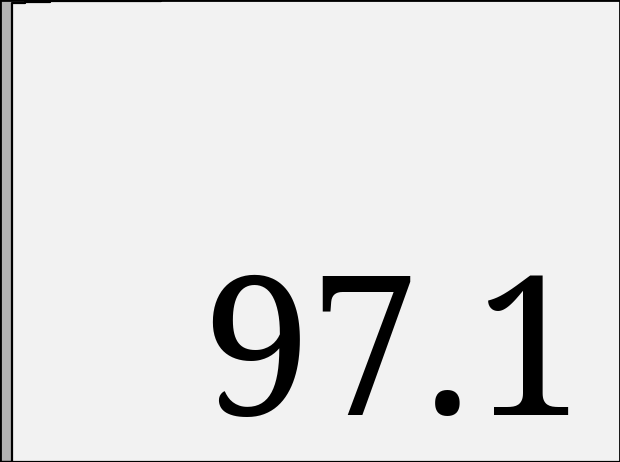}} & \raisebox{-.4\height}{\includegraphics[width=0.055\textwidth]{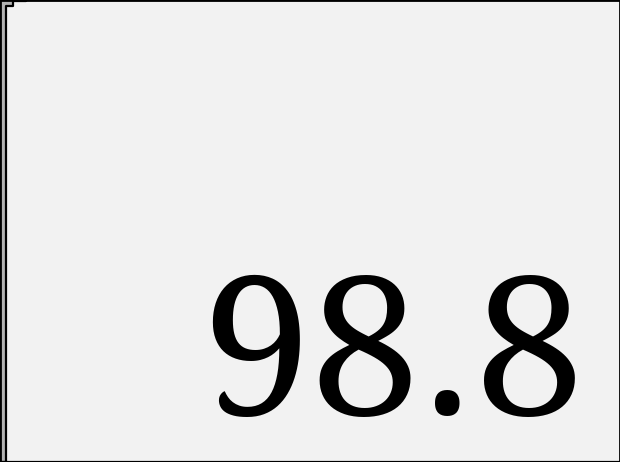}} & \raisebox{-.4\height}{\includegraphics[width=0.055\textwidth]{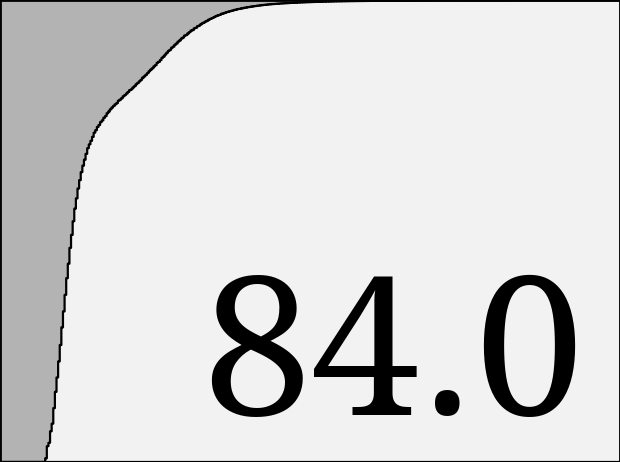}}                &  \raisebox{-.4\height}{\includegraphics[width=0.055\textwidth]{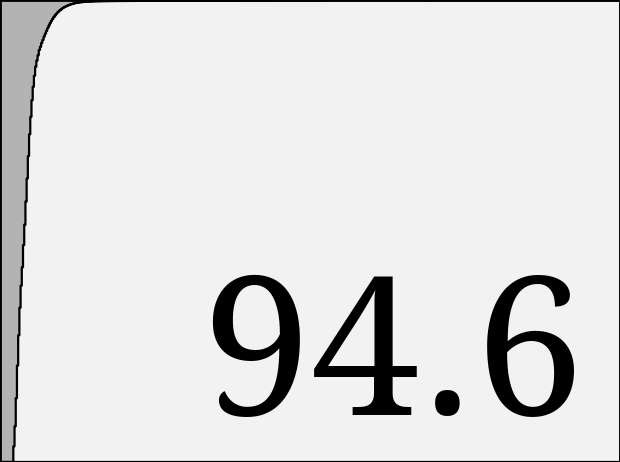}} &  \raisebox{-.4\height}{\includegraphics[width=0.055\textwidth]{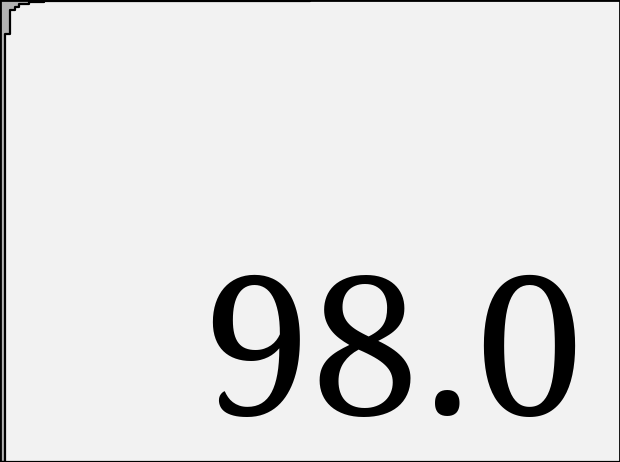}} \\
\hline
\multicolumn{1}{|l|}{$p^{50}$} & \raisebox{-.4\height}{\includegraphics[width=0.055\textwidth]{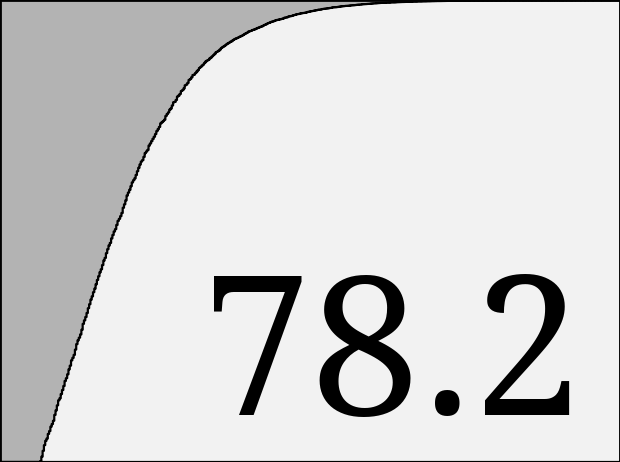}}   & \raisebox{-.4\height}{\includegraphics[width=0.055\textwidth]{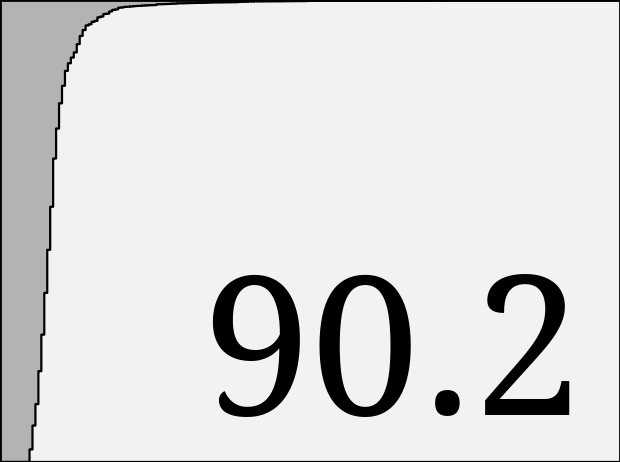}} & \raisebox{-.4\height}{\includegraphics[width=0.055\textwidth]{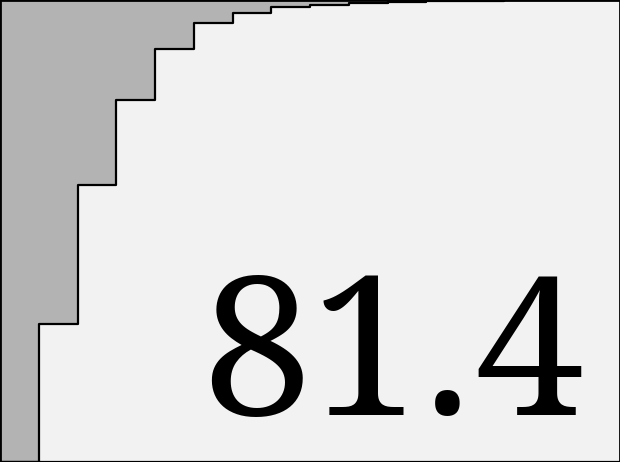}} & \raisebox{-.4\height}{\includegraphics[width=0.055\textwidth]{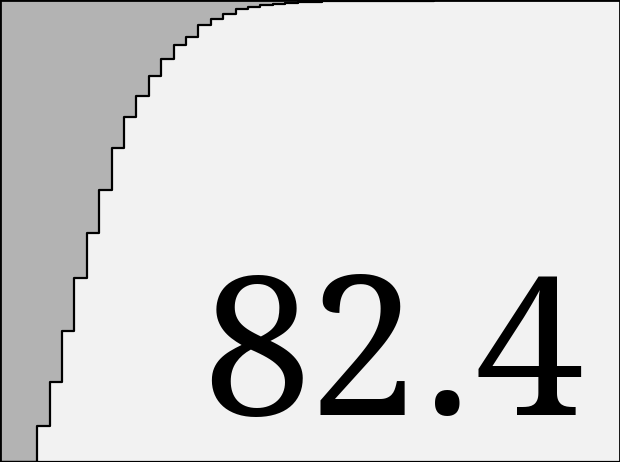}} & \raisebox{-.4\height}{\includegraphics[width=0.055\textwidth]{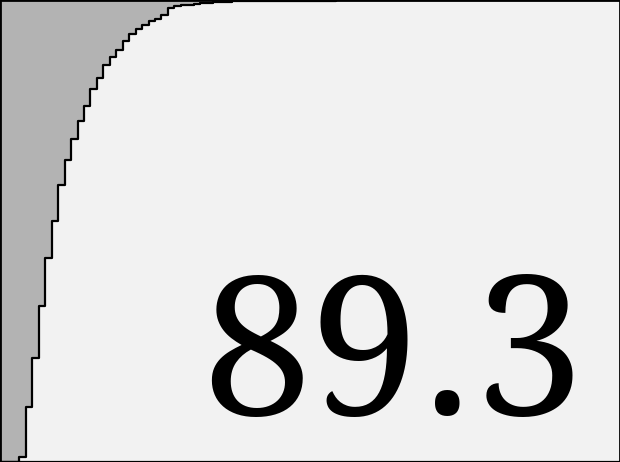}} &  \raisebox{-.4\height}{\includegraphics[width=0.055\textwidth]{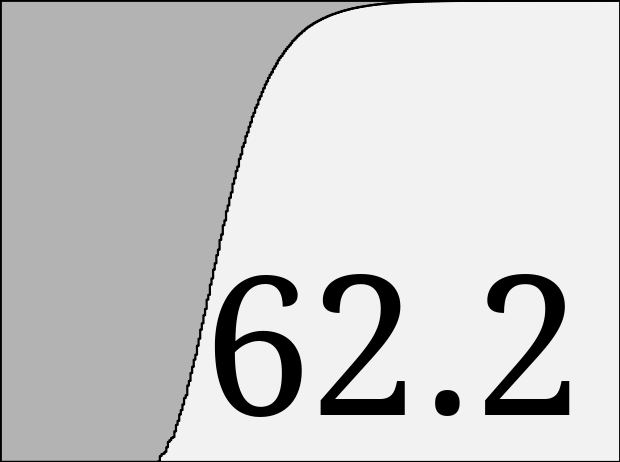}}  &  \raisebox{-.4\height}{\includegraphics[width=0.055\textwidth]{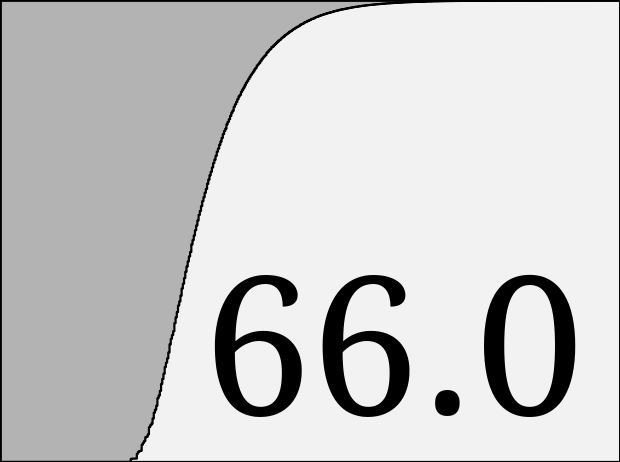}} &  \raisebox{-.4\height}{\includegraphics[width=0.055\textwidth]{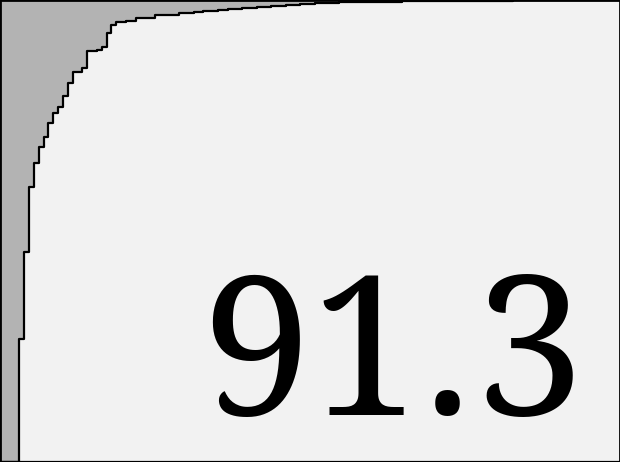}} \\
\hline
\multicolumn{1}{|l|}{$p^{25}$} &  \raisebox{-.4\height}{\includegraphics[width=0.055\textwidth]{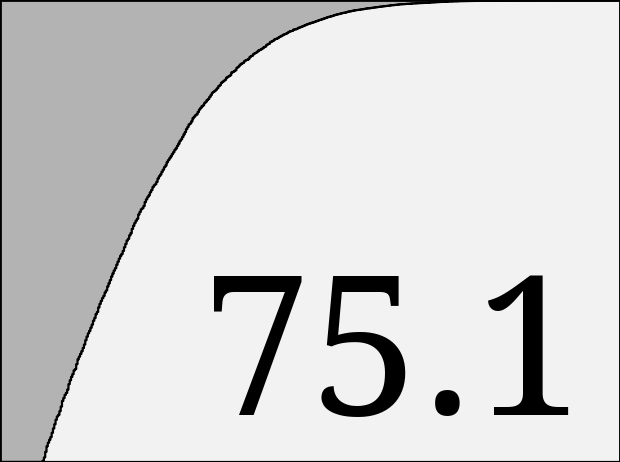}}        &  \raisebox{-.4\height}{\includegraphics[width=0.055\textwidth]{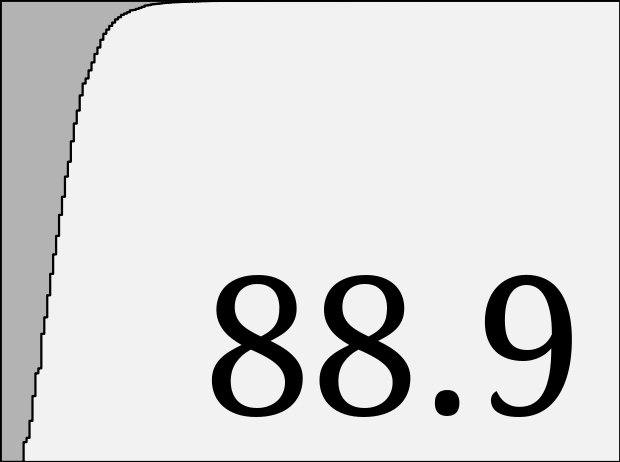}} & \raisebox{-.4\height}{\includegraphics[width=0.055\textwidth]{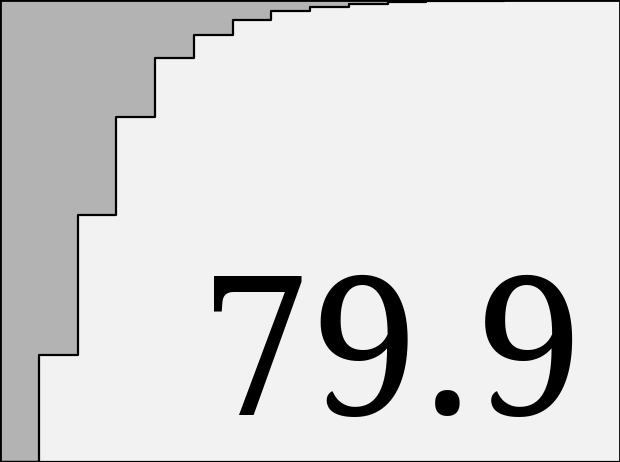}} & \raisebox{-.4\height}{\includegraphics[width=0.055\textwidth]{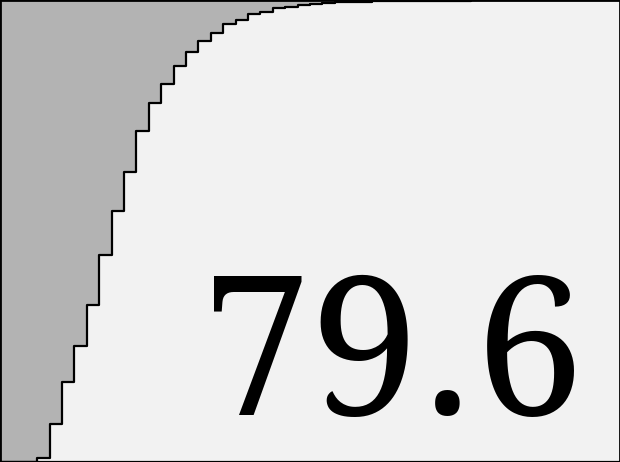}} & \raisebox{-.4\height}{\includegraphics[width=0.055\textwidth]{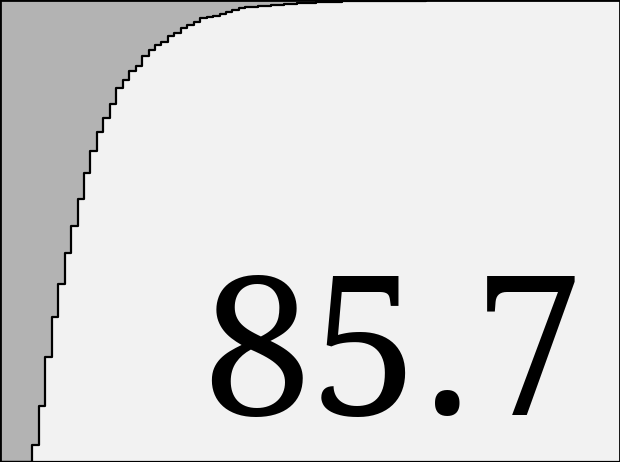}} &  \raisebox{-.4\height}{\includegraphics[width=0.055\textwidth]{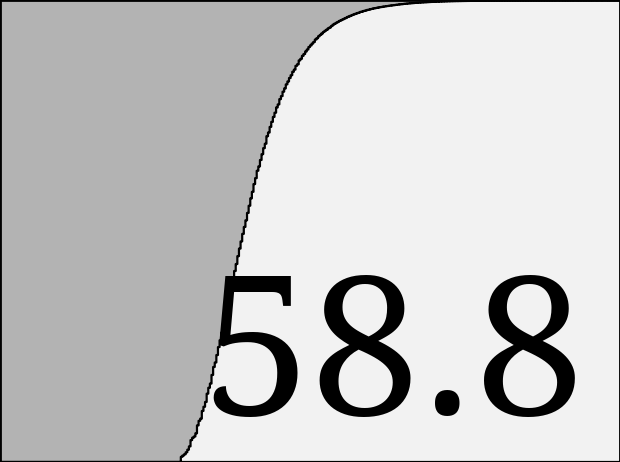}}  &  \raisebox{-.4\height}{\includegraphics[width=0.055\textwidth]{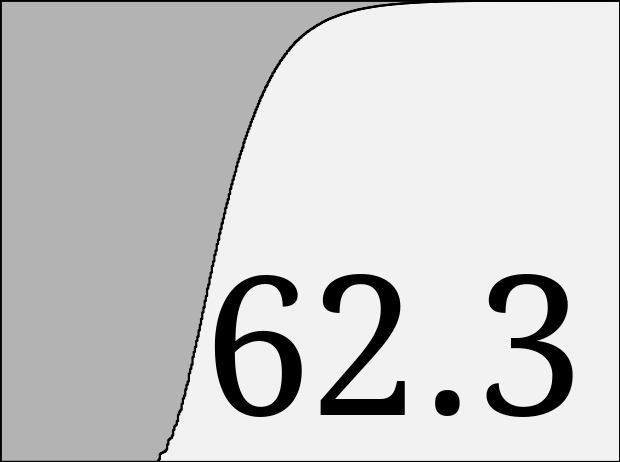}} &  \raisebox{-.4\height}{\includegraphics[width=0.055\textwidth]{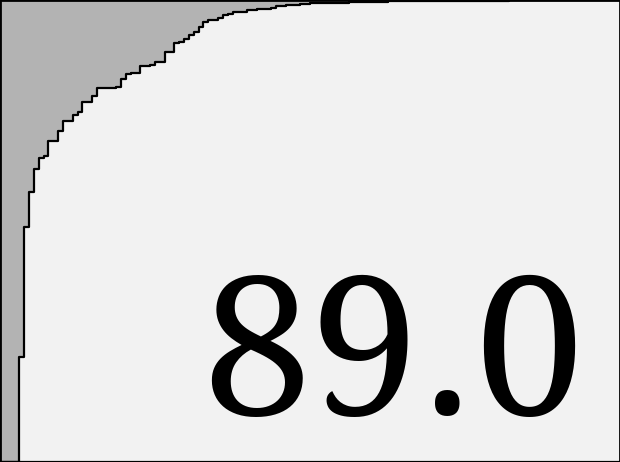}} \\
\hline
\multicolumn{1}{|l|}{Worst} &  \raisebox{-.4\height}{\includegraphics[width=0.055\textwidth]{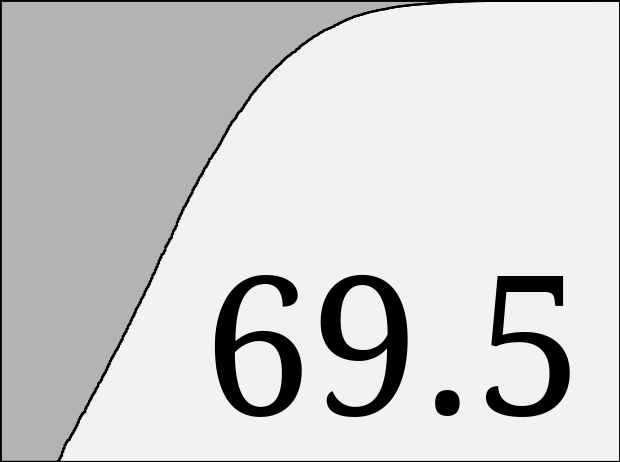}}        &  \raisebox{-.4\height}{\includegraphics[width=0.055\textwidth]{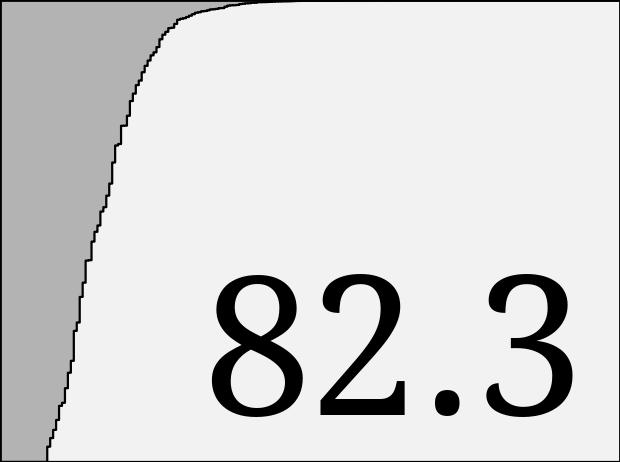}}  & \raisebox{-.4\height}{\includegraphics[width=0.055\textwidth]{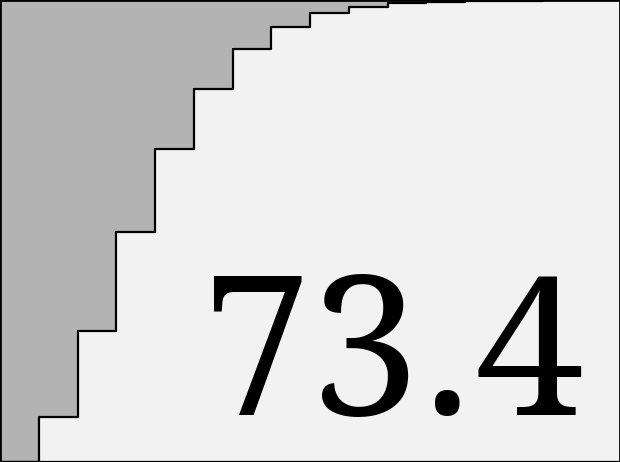}} & \raisebox{-.4\height}{\includegraphics[width=0.055\textwidth]{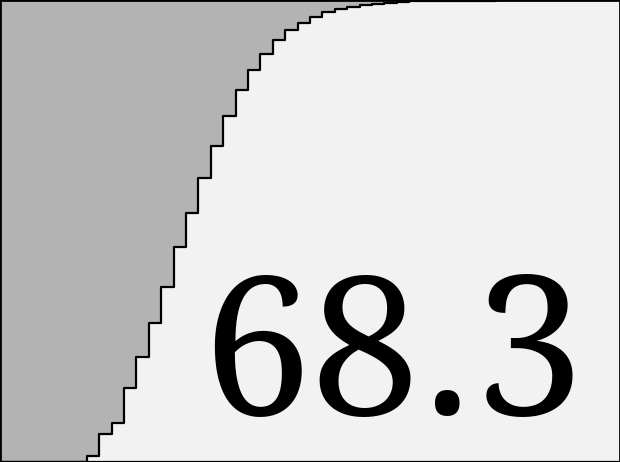}} & \raisebox{-.4\height}{\includegraphics[width=0.055\textwidth]{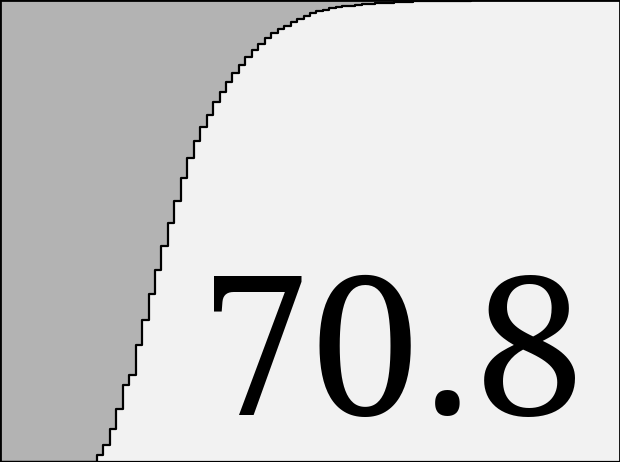}} &  \raisebox{-.4\height}{\includegraphics[width=0.055\textwidth]{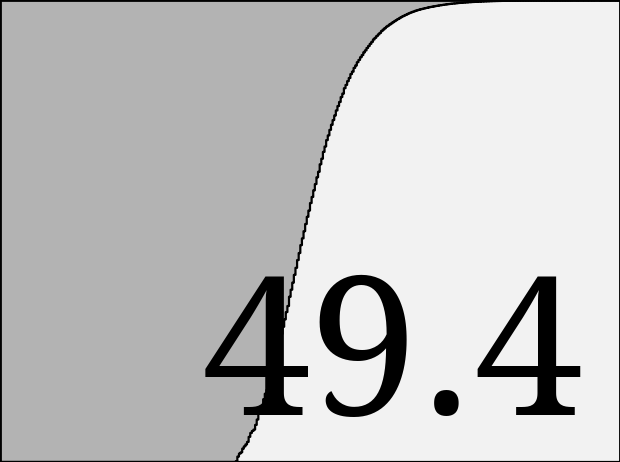}}  &  \raisebox{-.4\height}{\includegraphics[width=0.055\textwidth]{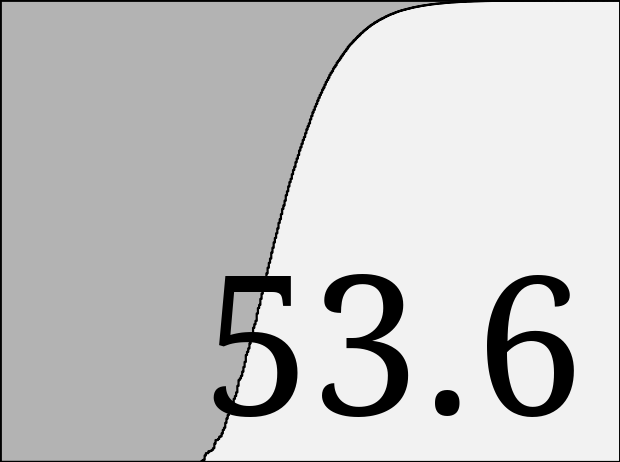}} &  \raisebox{-.4\height}{\includegraphics[width=0.055\textwidth]{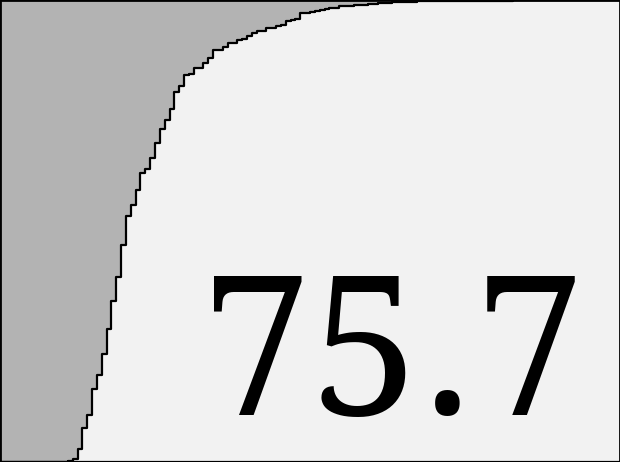}} \\
\hline
\end{tabular}
}
\end{table}

\subsection{PDX-BOND}\label{sec:eval-pdxbond}
Table \ref{tab:pruning-pdxbond} shows PDX-BOND pruning power, and Figure \ref{fig:eval-pdxbond} shows how its speed compares to ADSampling and BSA (both using the PDXearch framework) in Intel Sapphire Rapids. PDX-BOND pruning behavior adapts perfectly to the PDXearch framework: (i) It has a starting point, and (ii) once it starts, it prunes exponentially fast, further demonstrating the versatility of our framework. However, PDX-BOND pruning power is slightly worse than ADSampling (shown in Table \ref{tab:pruning}). 

On average, PDX-BOND is 2.1x and 3.0x faster than FAISS and Milvus at the highest recall level, respectively, and 20\% slower than BSA. However, it is 1.9x slower than ADSampling. When looking at recall levels of $\approx$0.9, PDX-BOND is only 1.3x slower than ADSampling and on par with BSA \ref{q:bond} while still being 30\% faster than FAISS. ADSampling and BSA take the upper hand on datasets of higher dimensionality due to their higher pruning power, thanks to their preprocessing on the vectors. Moreover, ADSampling and BSA benefit from sequential access as dimensions are always accessed sequentially regardless of the query. PDX-BOND main losses are in OpenAI/1536 and arXiv/768 (datasets of high-dimensionality in which the pruning power of PDX-BOND is low). Despite this, PDX-BOND performance is remarkable as it is an exact method (does not compromise recall) and does not require data or query preprocessing (it uses the raw vectors). Therefore, it can be used to increase the throughput of any VSS application just by changing the layout of the stored data. Moreover, the absence of preprocessing means it can fit into systems where data is ingested or updated frequently and at fine granularity. 

Our "dimension zones" approach to trade-off pruning effectiveness for sequential access is 30\% faster on average compared to accessing individual dimensions based on the "distance to means" criteria and 40\% faster than the "decreasing" criteria of BOND~\cite{bond} (recall Figure~\ref{fig:dz}). Another finding of our experiments is that BSA can be slower than ADSampling, especially at datasets of lower dimensionality (top row of Figure~\ref{fig:eval-pdxbond}), where it loses to all the other PDX-competitors. 

\vspace*{3mm}\noindent{\bf Breakdown of end-to-end query execution}. Table \ref{tab:phases} shows how the IVF query runtime of algorithms is distributed into four phases: query preprocessing, finding nearest buckets in the IVF index, bounds evaluation for pruning, and distance calculation. We only show the OpenAI/1536 dataset at 0.99 recall on Intel for presentation simplicity. The PDX versions of the algorithms drastically reduce the time spent evaluating bounds on both ADSampling and BSA, thanks to the branchless code we use to evaluate bounds (code is vectorized), evaluating bounds fewer times (incremental steps), and avoiding the interleaving of distance calculations and bounds evaluation. The PDX versions also spend less time calculating distances than N-ary versions due to our faster auto-vectorizing kernels (see also Table~\ref{tab:kernels}). Similarly, finding the nearest buckets is also sped up with our kernels, because the bucket centroids are also stored with the PDX layout. This phase may in future work also be optimized further by pruning itself. Finally, PDX-BOND query preprocessing (computing the order in which dimensions are accessed) is almost free compared to ADSampling/BSA.

\begin{figure*}[t!]
\includegraphics[width=1\linewidth]{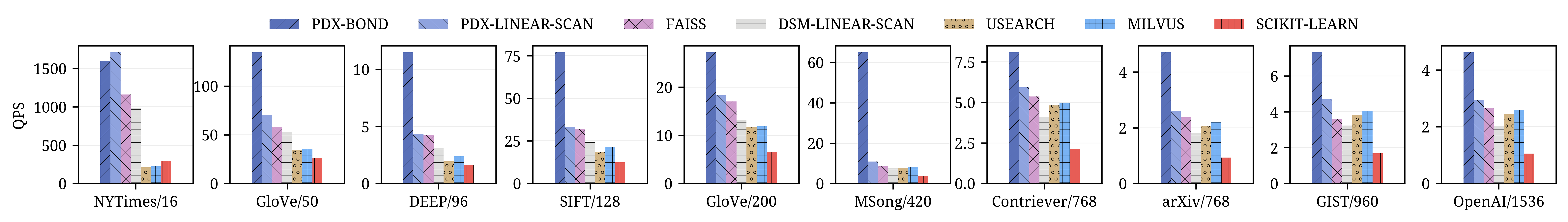}
\centering
\vspace*{-8mm}
\caption{Exact search QPS of all competitors with KNN=10 in the Intel architecture. PDX-BOND is superior to all the other approaches, thanks to pruning. Note that ADSampling and BSA are not exact methods, due to their probabilistic pruning. }
\label{fig:exact}
\vspace*{-3mm}
\end{figure*}

\begin{table}[t]
\centering
\caption{IVF query runtime breakdown into components, for the OpenAI/1536 dataset at 0.95 recall. The PDX version of the algorithms not only decreases Distance Calculation cost, but also the Bounds Evaluation latency, thanks to vectorized execution. It decreases the latency of Finding Nearest Buckets, since centroids are also stored with PDX. %Query Preprocessing is cheapest in PDX-BOND.
}
\vspace*{-3mm}
\label{tab:phases}
\resizebox{1\columnwidth}{!}{%
\begin{tabular}{|l|c|l|l|l|l|}
\hline
\multicolumn{1}{|c|}{\textbf{Algorithm}} & \textbf{\begin{tabular}[c]{@{}c@{}}Query \\ Time\\ (ms)\end{tabular}} & \multicolumn{1}{c|}{\textbf{\begin{tabular}[c]{@{}c@{}}Distance \\ Calculation\end{tabular}}} & \multicolumn{1}{c|}{\textbf{\begin{tabular}[c]{@{}c@{}}Find Nearest \\ Buckets\end{tabular}}} & \multicolumn{1}{c|}{\textbf{\begin{tabular}[c]{@{}c@{}}Bounds \\ Evaluation\end{tabular}}} & \multicolumn{1}{c|}{\textbf{\begin{tabular}[c]{@{}c@{}}Query \\ Preprocessing\end{tabular}}} \\ 
\hline \hline
N-ary ADS    & 17.9                                                                                          & 64.8\% (11.6ms)                                                          & 6.8\% (1.2ms)                                                             & 26.3\% (4.7ms)                                                       & 2.2\% (0.4ms)                                                        \\ 
PDX ADS     & \textbf{4.9}                                                                                          & \textbf{73.2\% (3.3ms)}                                                          & \textbf{18.5\% (0.8ms)}                                                             & \textbf{1.9\% (0.1ms)}                                                        & 6.45\% (0.3ms)                                                        \\ \hline \hline
N-ary BSA           & 25.5                                                                                         & 76.5\% (19.5ms)                                                          & 4.5\% (1.1ms)                                                              & 17.6\% (4.5ms)                                                       & 1.5\% (0.4ms)                                                        \\ 
PDX BSA            & \textbf{3.9}                                                                                          & \textbf{70.1\% (2.7ms)}                                                          & \textbf{17.7\% (0.7ms)}                                                             & \textbf{5.9\% (0.2ms)}                                                        & 6.4\% (0.3ms)                                                        \\ \hline \hline
PDX BOND           & 11.0                                                                                          & 91.9\% (10.1ms)                                                          & \textbf{7.0\% (0.8ms)}                                                              & \textbf{1.0\% (0.1ms)}                                                         & \textbf{0.03\% (0.003ms)}                                                        \\ \hline
\end{tabular}
}
\end{table}

\subsection{Exact Search}\label{sec:eval-exact}
We compare PDX and PDX-BOND capabilities on exact search against FAISS, USearch, Milvus, and a linear-scan on a fully decomposed layout (DSM). Like the ANN-Benchmarks project~\cite{annbench}, we use Scikit-learn~\cite{scikit} as a baseline. In this setting, a {\small \tt block} of PDX-BOND is defined by horizontally partitioning vectors in equally sized partitions (each of, at most, 10K vectors). Despite the bigger block size not allowing  tight loops, it enables longer sequential access per dimension. This allows PDX-BOND to use the "distance to means" criteria to prioritize dimensions (recall Figure~\ref{fig:dz}), which is the one that achieves the highest pruning power. 

Figure \ref{fig:exact} shows the QPS of each competitor on the Intel Sapphire Rapids architecture. In this plot, we also show the QPS of a linear scan on the PDX layout (with our auto-vectorized kernel). Here, PDX-BOND and the linear scan on PDX are the fastest-performing approaches in all datasets \ref{q:exact}, being PDX-BOND on average 2.5x, 3.7x, 4.0x and 6.2x faster than FAISS, USearch, Milvus and Sklearn respectively; being remarkably higher in some datasets (e.g., 6.0x faster than USearch in DEEP/96 and 7.6x faster than FAISS in MSong/420). We can also see how USearch and Milvus are close to the baseline on relatively low-dimensional datasets (GloVe/50, NYTimes/16). The latter again shows that the performance of SIMD kernels on the horizontal layout depends on the high dimensionality of the dataset. Also, in this experiment, the "distance to means" to prioritize dimensions is 40\% faster than the "decreasing" criteria. Finally, note that a linear-scan on PDX is still faster than doing so in a DSM layout (1.5x in avg.). The latter is due to a search on DSM incurring intermediate LOAD/STORE instructions as the {\small \tt distances} array holding the intermediate results prevents tight loops.

PDX-BOND and the linear scan on the PDX layout are the clear winners (also across architectures) without any recall compromise by relying only on auto-vectorization of scalar code, making it superior in code maintainability and portability to different ISAs. 

\vfill

\subsection{Effect of Pruning Percentage Threshold}\label{sec:eval-select}
\begin{figure}[t!]
\includegraphics[width=0.90\linewidth]{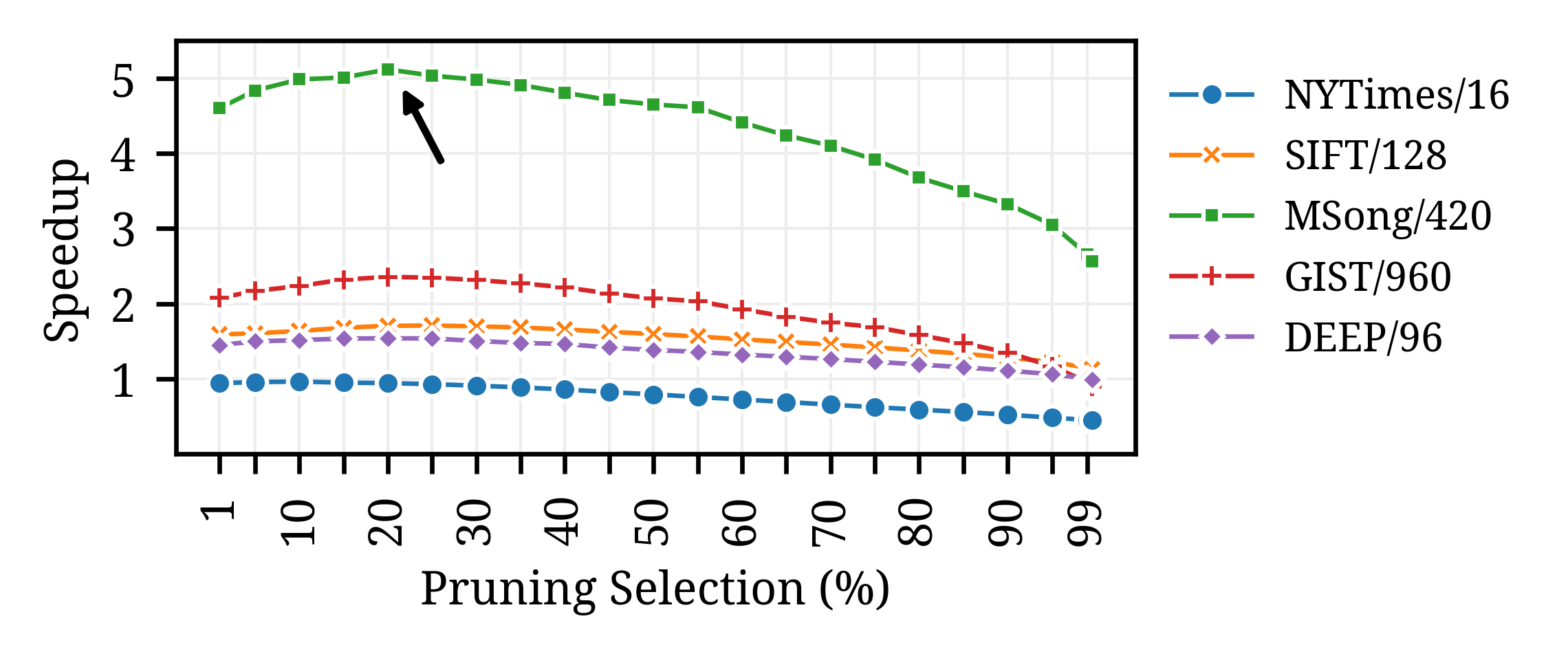}
\centering
\vspace*{-4mm}
\caption{The effect of different selection percentage values on the speedup of PDXearch over a linear scan on PDX which does not prune vectors in Zen4. A sweet spot is found when the amount of \textit{not-yet} pruned vectors is around 20\%. %We have used this threshold for all other experiments.
}
\label{fig:selectivity}
\vspace*{-4mm}
\end{figure}

The PDXearch framework uses the percentage of \textit{not-yet} pruned vectors as a threshold to advance to the PRUNE phase. Figure~\ref{fig:selectivity} shows for 6 datasets how different thresholds effect the speedup of ADSampling using the PDXearch framework against a linear scan on PDX which does not prune vectors (in Zen4). Generally, starting to prune when the selection percentage is too high (>40\%) or too low (<10\%) is detrimental to search speed. The sweet spot is found around 20\%. Interestingly, the difference between 5\% and 20\% is subtle. This is due to the exponential behavior of pruning that makes the difference between reaching 20\% and 5\% have little effect on search speed (as it is probable that both are reached in the same step of dimension scanning). On the other hand, on datasets with low pruning possibilities (NYTimes/16), a linear scan is faster than pruning. This is due to the pruning predicate evaluation turning into an overhead without any gains.  

\subsection{PDXearch Across Architectures}\label{sec:eval-across}
\begin{figure}[t]
\includegraphics[width=0.98\linewidth]{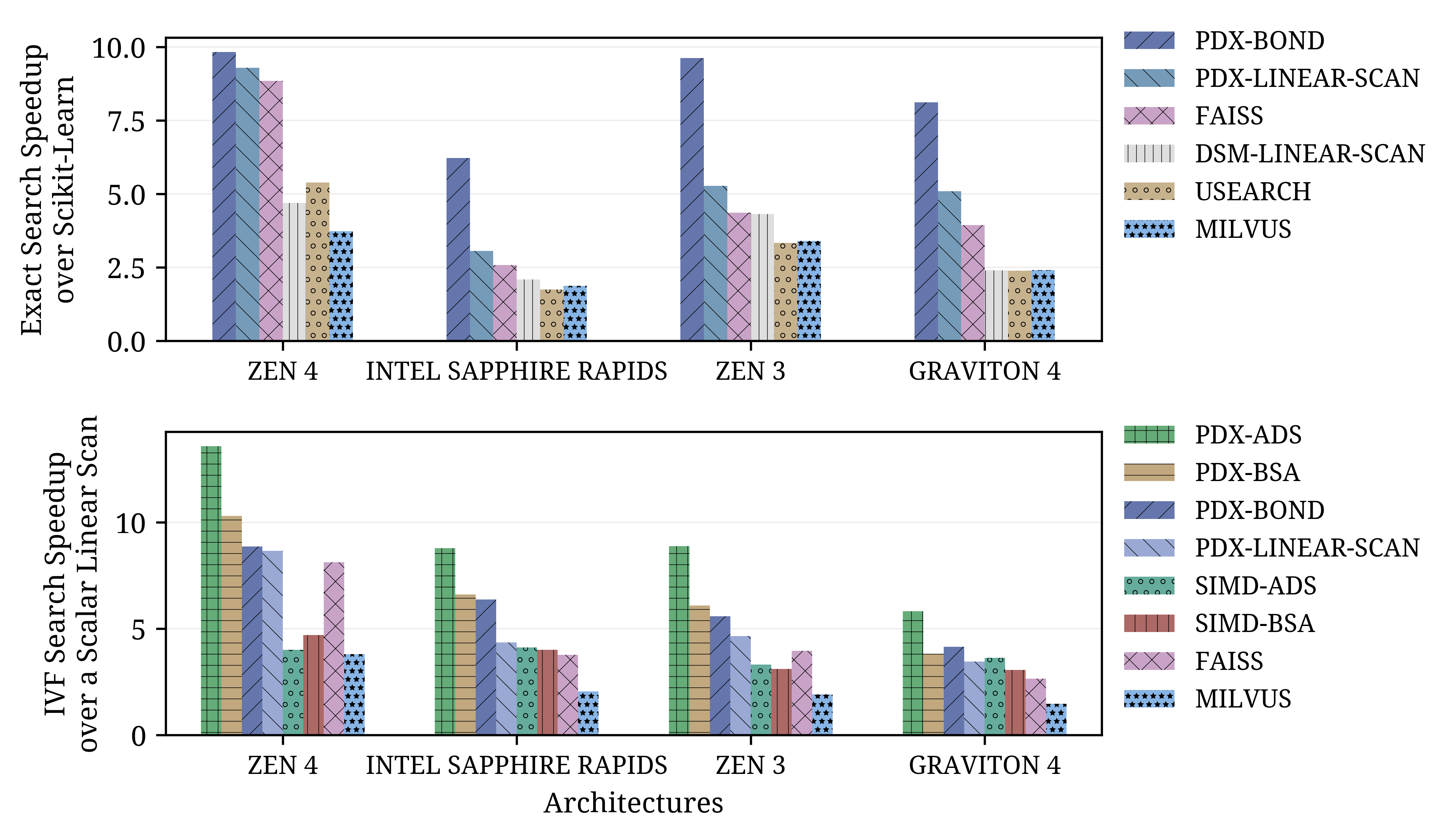}
\centering
\vspace*{-5mm}
\caption{Geometric mean of performance over all datasets, per architecture. PDX-BOND outperforms in exact search  and PDX brings strong benefits to approximate search.}
\label{fig:summary}
\vspace*{-3mm}
\end{figure}

Figure~\ref{fig:summary} summarizes our experiments across all architectures as the geometric mean of speedup obtained across all datasets (at all recall levels in approximate search) against a baseline: Scikit-learn in exact search and a scalar (non-SIMDized) linear scan in the IVF index search. On exact search, PDX-BOND and a linear scan on the PDX layout (PDX-LINEAR-SCAN) are faster than FAISS, Milvus, USearch and a linear-scan on the DSM layout (DSM-LINEAR-SCAN) across all architectures, showing the effectiveness and versatility of our distance kernels and data layout. On the other hand, on an approximate search on an IVF index, PDXearch brings remarkable benefits to existing pruning approaches, which, despite not being exact, still provide guaranteed error bounds that pose little effect on the search recall. Despite PDX-BOND not taking the upper-hand in approximate search, it is still faster than the other non-PDX competitors.

\section{Discussion}\label{sec:discussion}

\noindent{\bf PDXearch on N-ary Storage}. The PDX layout is a transposition of the horizontal/N-ary layout of the vectors. One could question the need to physically store the data using PDX, and rather perform a {\small\tt gather} operation \cite{gathersimd, microfog, tousegather, gathermimdsimd} on-the-fly during search. To test this, we implemented a (N-ary+Gather) kernel, depicted in the rightmost of Figure~\ref{fig:simdvspdx}, that does an on-the-fly transposition of the N-ary layout into blocks of 64 vectors into PDX layout using AVX512 {\small\tt gather} instructions~\cite{tousegather} on Intel Sapphire Rapids and AMD Zen4. This kernel needs multiple scalar loads on NEON, since ARM lacks a {\small\tt gather} instruction. We used the same variety of randomly generated collections from our kernels experiment (Section \ref{sec:eval-kernels}).

In Zen4, this produces an average slowdown of 13x and up to 130x at higher dimensionalities (avg. 33x). In Intel, the slowdowns go from 1.9x to 17x (avg. 4.6x)%, depending on wether the data fits on L1/L2/L3 cache (48KiB, 2MiB and 60MiB resp.)
. In NEON, the slowdowns go from 2.5x to 22x (avg. 8.5x). %In all CPUs there are increased slowdowns when D is a multiple of the cache \textit{critical stride}~\cite{trasposition}, as every iteration of the transposition contend for the same cache lines (a known problem in matrix transposition \cite{trasposition}).
CPU performance counter profiling on Intel and AMD showed that these slowdowns happen for two reasons: (i) increased $\mu$ops of the {\small\tt gather} and (ii) increased memory stalls. 

Regarding (i): One {\small\tt gather} performs 81 $\mu$ops on AMD\footnote{\href{https://web.archive.org/web/20250207181843/https://uops.info/html-instr/VGATHERDPS\_ZMM\_K\_VSIB\_ZMM.html\#ZEN4}{uops.info/html-instr/VGATHERDPS\_ZMM\_K\_VSIB\_ZMM.html\#ZEN4} and \href{https://web.archive.org/web/20250207181843/https://uops.info/html-instr/VGATHERDPS\_ZMM\_K\_VSIB\_ZMM.html\#ADL-P}{\#ADL-P}} and 5-6 on Intel~\cite{uops, instructions}. These benchmarks align with our experiments, which also showed a similarly increased amount of $\mu$ops and instructions retired across architectures. We must stress that the PDX kernel is \textit{extremely} fast: it processes each float in 0.1 CPU cycles when the data fits in L1d. For one AVX512 register with 16 floats, it needs just 3$\mu$ops (one {\small\tt load} [1$\mu$op], one {\small\tt sub} [1$\mu$op], and one {\small\tt fmadd} [1$\mu$op]). While the amount of $\mu$ops does not map 1:1 to runtime, adding 81 (AMD) or even just 5 (Intel) to the original 3 has noticable impact.

\begin{figure}[t]
\includegraphics[width=0.98\linewidth]{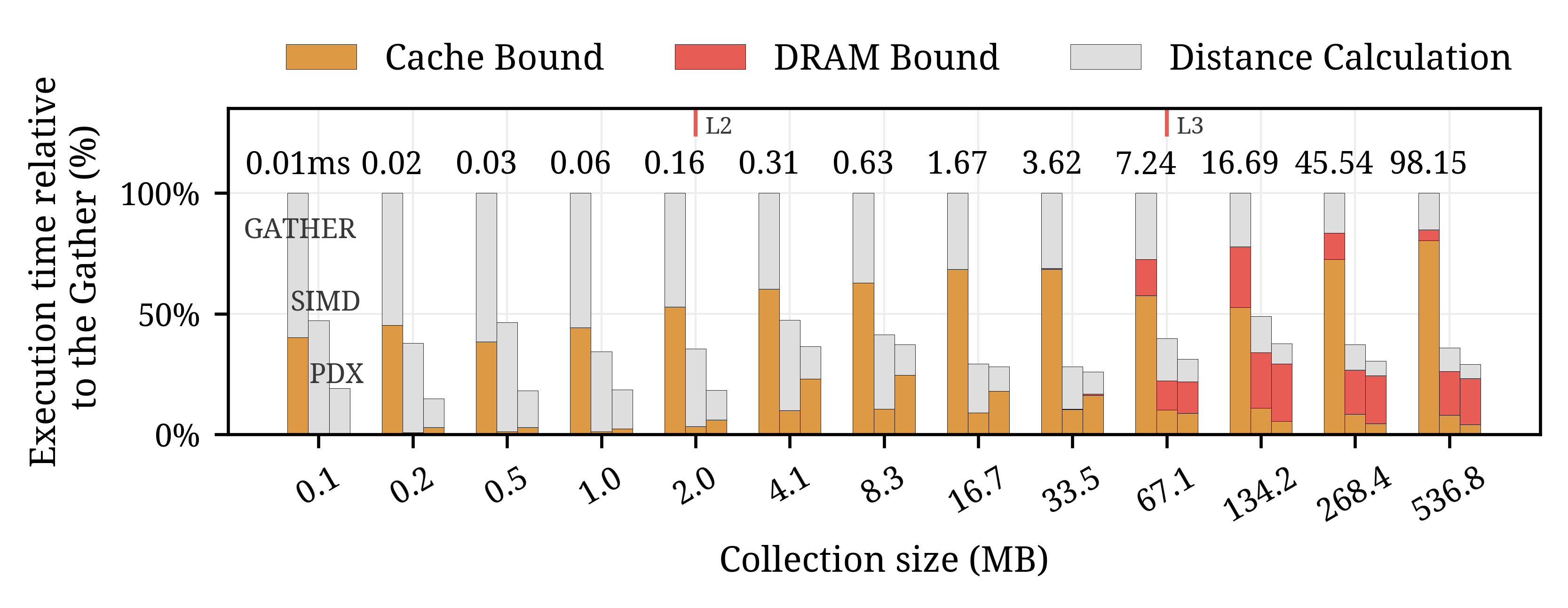}
\centering
\vspace*{-4mm}
\caption{Performance breakdown for three distance kernels: N-ary+Gather, N-ary+SIMD, and PDX (bars in order). The Y-axis shows execution time relative to (N-ary+Gather), which is always slowest, as its gather operation comes with data access cost, even if data fits in L1/L2. When data size (X-axis) exceeds L2, all kernels become data access-bound. As (N-ary+Gather) runs slower, it shows as less DRAM-bound. %PDX is the most affected by memory-bounds as its distance kernel is the fastest.
}\label{fig:gather}
\vspace*{-6.5mm}
\end{figure}

Regarding (ii): Figure~\ref{fig:gather} shows the  execution time of the three kernels (N-ary+Gather, N-ary SIMD, PDX) for different collection sizes, on Intel. The time is shown relative to the first (N-ary+Gather). In each bar, we separate CPU distance calculation cost from  data access cost, as measured with CPU performance counters. Contrary to the other kernels, the {\small\tt gather} kernel is always data access-bound, even when data fits in L1d and L2. When data is bigger than L3, all kernels become data access-bound, but the {\small\tt gather} kernel spends more time on memory stalls. This results from its non-sequential memory access patterns that is less able to profit from automatic hardware prefetching and subsequent cache use. We know of few cases in which a {\small\tt gather} \textit{can} achieve the same performance as a sequential {\small\tt load} on Intel CPUs~\cite{gathermimdsimd, tousegather}. However, these depend on the SIMD register width, L1d cache alignment, and the width of the data type being used. Reportedly, AVX512 with 32-bit data cannot achieve this effect due to size limitations of the L1 cache~\cite{tousegather}. These observations from~\cite{tousegather} align with our experiments. 
% On an important note: \cite{gathermimdsimd, tousegather} recently showed that the memory access patterns of a {\small \tt gather} on Intel CPUs \textit{can} achieve the same performance of a sequential {\small \tt load} when each consecutive gather access the same cache lines pre-fetched from the previous gather. However, this effect can only be achieved with specific datatypes, SIMD registers width and size of L1d cache. A setting in which this effect is not present and (quote) \textit{``high throughputs cannot be achieved"} is AVX512 with 32bit data~\cite{tousegather} (or smaller types) due to size limitations on commercially available L1d's~\cite{tousegather}. These observations from \cite{tousegather} align with our experiments. Note also that vector embeddings will always use float32 or smaller data types (float16, bfloat, int8, int4) which are also not compatible.  

Overall, since (N-ary+Gather) introduces the overhead of the on-the-fly transposition, it is never faster than PDX. Since this overhead is higher than the gains from PDX distance calculation, it is also never faster than (N-ary+SIMD). The performance gap on the other architectures is larger than in Figure~\ref{fig:gather} (which shows Intel), since ARM (resp. AMD) lacks a (fast) {\small\tt gather} instruction. Therefore, we conclude that vectors need to be stored in PDX layout, in order to reap the benefits from the PDX distance calculation.

% The GATHER approach at D=8 is 1.9x faster than the SIMD kernel (thanks to SIMDizing over vectors rather than dimensions). However this pales in comparison to the 5.8x speedup of the PDX kernel at D=8 (Table~\ref{tab:kernels}) and the performance quickly degenerates as soon as D>8. 

%, with a performance of 0.10 CPU cycles needed per value when data fits in the L1 cache (64x64 collection), and 0.85 CPU cycles per value with a collection bigger than the L3 cache. 

%On the other hand, the GATHER based kernel on the 64x64 matrix is 0.61 cycles per tuple (6.1x slower), which is close to the benchmarks on matrix transposition using GATHER reported by Intel\cite{intel}. 

\vspace*{3mm} \noindent{\bf PDX vs DSM}. We also tried a fully decomposed layout (DSM). Note that in an IVF index, data gets (horizontally) partitioned in buckets, so applying vertical decomposition inside buckets yields a PDX layout. Therefore we tested DSM on {\em linear scan}, performed on a dimension-by-dimension partitioning of the complete dataset. For the vertical distance computation kernels we propose, DSM maximizes sequential access, as one complete dimension is typically much larger than one IVF bucket. This makes DSM potentially interesting for secondary storage devices that require large access granularity for efficiency, such as S3 or magnetic drives. However, its column-at-a-time distance calculation needs to sequentially update a result array with all distances many times (=for each dimension), introducing extra load/stores. In our in-memory experiments, this made it slower than PDX-linear-scan, as shown in Figures \ref{fig:exact} and ~\ref{fig:summary}.

\vspace*{3mm} \noindent{\bf PDX on Graph Indexes. } The PDX layout and PDXearch are a perfect fit for bucketing indexes and exact search. However, its effectiveness is yet to be tested on graph indexes like HNSW~\cite{hnsw}. Despite these indexes still finding benefits from DCO optimization~\cite{adsampling, bsa}, the notion of blocks is not intrinsically present in these data structures. Recent studies have proposed optimized data layouts for graph indexes, which allows efficient fetching of \textit{neighborhoods-at-a-time} during a search when the data is not memory-resident~\cite{starling, aisaq}. Here, a block could represent neighborhoods of the graph; consequently achieving the desired benefits of our proposed layout and pruning algorithm. The latter is a  common use case, as the memory requirements to keep an HNSW index on modestly sized datasets are beyond commodity hardware. 
\pagebreak

%\vspace*{3mm} 
\noindent{\bf PDX Storage Designs. }
PDX also opens opportunities to improve speed in memory-constrained environments, as the data only needs to be loaded in memory not only block- but dimension-at-a-time. Here, a hybrid storage on disk using both PDX and the traditional N-ary layout could minimize the random access overhead during the PRUNE phase of the search. Similarly, PDX could benefit VSS workloads performed in a hybrid setting where data needed to perform computations must be fetched through a network. In distributed settings, vectors could also be partitioned by dimensions (e.g., certain dimensions are assigned to a node within a cluster). The latter hints that a follow-up to the PDX layout would be on efficient \textit{compressed} representations of dimensions within blocks. This would reduce even more the memory/network bandwidth needed and bring more benefits to the PDX distance kernels which are memory-bounded (recall Figure~\ref{fig:gather}).

\vspace*{3mm} \noindent{\bf PDX in Data Systems. }
Vector-databases like Weaviate~\cite{weaviate} and Milvus~\cite{milvus} have surged as a new category of data systems, but (relational) database systems have also added capabilities for vector storage, search, and indexing.
Typically, when database systems implement an array data type, their implementation stores each array as one data item (the horizontal vector layout). However, in analytical systems that use row-groups and columnar storage within these, it will be easy to use the row-groups as blocks for the PDX layout. Often, within a row-group, data is partitioned per "vector" (in the meaning of vectorized execution), and these smaller blocks of a few thousand values, might even be more beneficial to PDX. Thus, PDX would be well-suited for integration in database systems such as DuckDB~\cite{duckdbpaper}, which also support floats compression~\cite{alp}.

% Undoubtedly, the novel idea on optimizing DCOs opened by~\cite{adsampling} has a lot of potential and exciting opportunities to explore. Combined with the PDX layout, these could bring the next leap of performance improvement on vector similarity search under different settings such as cloud/hybrid/distributed settings, RAM-constrained environments, or vectors which are casually queried. 
\section{Related Work}\label{sec:related}
The recent surge in research attention into ANNS has resulted in a wide range of studies regarding indexes and quantization which are orthogonal to this work. Therefore, we redirect readers to the surveys done in Vector Databases systems~\cite{surveysystems}, approximate graph-indexes~\cite{surveygraph}, approximate hashing indexing~\cite{surveylshquant,surveylsh}, quantization methods~\cite{surveylshquant, polysemous} and benchmarking frameworks~\cite{candy, annbench}. This section focuses on the scenario in which the PDX layout shines: algorithms that avoid the distance evaluation on every dimension. 

\vspace*{3mm} 
\noindent{\bf Data formats for vectors.} {\small \tt .bvecs}, {\small \tt .fvecs} and {\small \tt .ivecs} are three mainstream formats developed by INRIA to store vectors. These formats store vectors one after the other as a serialized sequence of bytes, floats or integer types respectively. The formats contain a header with the dimensionality of the vectors as a 4 byte unsigned integer. Figure \ref{fig:pdx} shows a visual example of the {\small \tt .fvecs} format. The ANN-Benchmarks project~\cite{annbench} stores vectors separated into two {\small \tt .fvecs} datasets (train and test) which are stored within a {\small \tt .hdf5} file. Here, the ground truth of the test set at a predefined \textit{k} is also stored. Vector systems adopt the {\small \tt .fvecs} format to store raw vectors~\cite{usearch, faisspaper, milvus} either in memory or disk which are usually only accessed if the result set $\hat{R}$ needs a re-ranking phase. 

BOND~\cite{bond} proposed a vertically decomposed layout in which the values of each dimension are stored together. More recently, ADSampling~\cite{adsampling} proposed to divide vectors into two blocks that follow the {\small \tt .fvecs} format. One containing the first $\Delta d$ dimensions of every vector and the other containing the rest of the dimensions. During a search, the first block is always scanned fully for all the vectors, only accessing the second block to inspect the vectors which were not pruned yet. This dual-block layout improved speeds thanks to the more efficient use of cache. However, it falls short, as the optimal number of dimensions to scan are query- and dataset-dependent. Until the PDX layout, there has not been any research to develop a new data format for vectors. The PDX layout speeds up search, makes pruning algorithms more efficient, and can co-exist with indexes and quantization techniques. % Furthermore, it is able to process multiple vectors at-a-time and its scalar code autovectorizes without need for explicit SIMD intrinsics. %by auto-vectorizing distance kernels in any ISA and processor family. 

%\pagebreak
\vspace{3mm}
\noindent{\bf Exact Pruned Search}. BOND~\cite{bond} proposed lower- and upper-bounds for the Euclidean distance to discard vectors that could not make it into the KNN of a query after only partial distance calculations. The novelty of BOND was its vertically decomposed layout to store vectors. This layout allowed to prioritize the order in which the algorithm visited dimensions while still benefiting from sequential access to the data. However, the vertical layout impedes BOND to fully visit a vector until the end of the search. As such, the lower bound lacked the necessary tightness to achieve high pruning powers. Despite being able to prune on skewed datasets, BOND speedup of x1.6 over an exact KNN was limited by the overhead of the computation of the bounds.

\vspace*{3mm} \noindent{\bf Approximate Pruned Search}. The goal of ADSampling~\cite{adsampling}  is to quickly determine when vectors have a low probability of making it to the KNN candidate list of a query. The key idea of ADSampling relies on projecting the vector collection to a different number of dimensions flexibly during the query phase. This is achieved by randomly transforming each vector with a random orthogonal transformation (a random rotation). On these transformed vectors, the level of resolution of the distance metric is given by the number of sampled dimensions. This level of resolution has a guaranteed error bound depending on the number of sampled dimensions. BSA~\cite{bsa} followed up on ADSampling by using learned PCA projections on the D-dimensional space instead of a random orthogonal projection, which turns out to further reduce the error-bound of the approximation and achieve a higher pruning power.

\section{Conclusions}\label{sec:conclusion}
We have presented PDX: a new data layout for vectors that allows vector similarity search to happen dimension-by-dimension. This turned out to be a better layout for existing~\cite{adsampling, bsa} dimension pruning algorithms. Furthermore, we introduced PDXearch, a search framework that allows pruning algorithms to be adaptive to any query and dataset and improve cache efficiency. We showed its effectiveness in improving query throughput  using a wide variety of vector datasets on four mainstream CPUs (Zen4, Intel Sapphire Rapids, Zen3, Graviton4).  We also introduced PDX-BOND: a pruning algorithm that does not need data transformations (it works on raw floats) and does not have any recall trade-off (can be used for exact search). % PDX-BOND outperforms modern VSS libraries  (FAISS, Usearch, Milvus) when performing end-to-end exact queries and is as fast as BSA and not much slower than ADSampling on approximate queries, while not incurring data transformation cost or recall loss. 

As for future work, we think that pruning algorithms can benefit from GPU processing. Furthermore, fusing the idea of PDX-BOND (dimensions reordering in terms of query) and ADSampling (random sampling of projections at different levels) could bring benefits to the pruning power. % We also believe that more data types and more distance/similarity functions can be efficiently auto-vectorized on the PDX layout, including user defined functions (UDFs). As such, a future study showcasing more distance kernels being used within different indexes and vectors quantized on different datatypes ({\small \tt float16}, {\small \tt bfloat}, {\small \tt uint16}, {\small \tt uint8}) is a next step for our research. 

\iffalse
\begin{acks}
...
\end{acks}
\fi

\bibliographystyle{ACM-Reference-Format}
\bibliography{_main}

\appendix

\end{document}